\documentclass[a4paper,10pt]{article}
\usepackage{graphicx,amssymb,bm,latexsym,amsmath}
\allowdisplaybreaks[1]
\makeatletter
\@addtoreset{equation}{section}

\makeatother
\newcommand{\bea}{\begin{eqnarray}}
\newcommand{\ena}{\end{eqnarray}}
\newcommand{\vs}[1]{\vspace{#1 mm}}

\renewcommand{\a}{\alpha}
\renewcommand{\b}{\beta}
\renewcommand{\c}{\gamma}
\renewcommand{\d}{\delta}
\newcommand{\e}{\epsilon}
\newcommand{\s}{\sigma}

\newcommand{\pa}{\partial}

\newcommand{\nn}{\nonumber\\}
\newcommand{\p}[1]{(\ref{#1})}

\begin{document}

\begin{titlepage}

\begin{flushright}
OU-HET 555 \\
WU-AP/240/06\\
hep-th/0602242
\end{flushright}

\vs{10}
\begin{center}
{\Large\bf Inflation from Superstring/M-Theory Compactification with
Higher Order Corrections II\\ \vs{3}
-- Case of Quartic Weyl Terms --}
\vs{10}

{\large
Kenta Akune,$^a$\footnote{e-mail address:
akune@gravity.phys.waseda.ac.jp}
Kei-ichi Maeda$^{a,b,c,}$\footnote{e-mail address:
maeda@waseda.jp} and Nobuyoshi Ohta$^{d,}$\footnote{e-mail
address: ohta@phys.sci.osaka-u.ac.jp.
Address after 31 March 2006: Department of Physics, Kinki University,
Higashi-Osaka, Osaka 577-8502, Japan}}\\
\vs{10}
$^a${\em Department of Physics, Waseda University,
Shinjuku, Tokyo 169-8555, Japan}\\
$^b${\em Advanced Research Institute for Science and Engineering,
Waseda University, Shinjuku, Tokyo 169-8555, Japan~} \\
$^c${\em Waseda Institute for Astrophysics, Waseda University,
Shinjuku, Tokyo 169-8555, Japan~}  \\
$^d${\em Department of Physics, Osaka University,
Toyonaka, Osaka 560-0043, Japan}
\vs{10}

{\bf{Abstract}}
\end{center}

We present a detailed study of inflationary solutions in M-theory with higher
order quantum corrections. We first exhaust all exact and asymptotic
solutions of exponential and power-law expansions in this theory with quartic
curvature corrections, and then perform a linear perturbation analysis around
fixed points for the exact solutions  in order to see which solutions are
more generic and give interesting cosmological models. We find an interesting
solution in which the external space expands exponentially and the internal
space is static both in the original and Einstein frames.
Furthermore, we perform a numerical calculation around this solution and find
numerical solutions which give enough e-foldings.
We also briefly summarize similar solutions in type II
superstrings.
\end{titlepage}
\newpage
\renewcommand{\thefootnote}{\arabic{footnote}}
\setcounter{footnote}{0}
\setcounter{page}{2}

\section{Introduction}

The recent cosmological observations have confirmed
the existence of the early inflationary epoch and
the accelerated expansion of the present universe~\cite{WMAP}.
Though it is not difficult to construct cosmological models with these
features if one introduces scalar fields with suitable potentials,
it is desirable to derive such a model from  fundamental
theories of particle physics   without making
special assumptions. The most promising candidates for
such theories are the ten-dimensional superstrings or eleven-dimensional
M-theory, which are hoped to give models of accelerated
expansion of the universe upon compactification to four dimensions.

There is a no-go theorem which forbids such solutions if six- or
seven-dimensional internal space is
a time-independent nonsingular compact manifold without boundary~\cite{NG}.
This theorem also assumes that the gravity action does not contain higher
curvature corrections.
So we can evade this theorem by violating some of the assumptions.
In fact, it has been shown that a model with certain period of accelerated
expansion can be obtained from the higher-dimensional vacuum Einstein
equation if one assume a time-dependent hyperbolic  internal space
~\cite{TW}. It has been shown~\cite{NO} that this class of models
is obtained from what are known as S-branes
\cite{Wohlfarth:2003ni,Sbrane1,Sbrane2}
in the limit of vanishing flux of three-form fields (see also~\cite{Sbrane3}).
For  other attempts for inflation in the context of string theories,
see, for instance, Refs.~\cite{other,cosm2,cosm3}.

The scale when the acceleration occurs in this type
of models is basically governed by the Planck scale in the higher ten
or eleven dimensions. With phenomena at such high energy, it is expected
that we cannot ignore quantum corrections such as higher derivative
terms in the theories at least in the early universe.
It is known that there are terms of higher orders in the curvature to
the lowest effective supergravity action coming from superstrings or
M-theory~\cite{Be,hetero0,hetero,Mth,TBB}.
The no-go theorem does not apply to theories with higher derivatives because
they violate the assumptions.
With such corrections, they will significantly affect the inflation
at the early stage of the evolution of our universe.

The cosmological models in higher dimensions were studied intensively
in the 80's by many authors~\cite{KK_cosmology,maeda,ISHI,old1,HAL}.
It was shown that inflation is indeed possible with higher-order
curvature corrections~\cite{ISHI,old1}.
However, most of the work considered theories with higher orders
of scalar curvature, which are not typical correction terms known to
arise in the superstring theories or M-theory. It is thus important to
examine if the above result of small e-folding is modified with higher-order
corrections expected in these fundamental theories.

In particular, the leading quadratic correction for heterotic string theories
is proportional to the Gauss-Bonnet (GB) combination~\cite{Be,hetero0,hetero}.
This model has been studied in some detail in Ref.~\cite{ISHI} and it was shown
that there are two exponentially expanding solutions, which may be called
generalized de Sitter solutions since the size of the internal space also
depends on time. Despite the exponential behavior in the original frame,
those solutions give non-inflationary power-law expansion in the Einstein
frame as indicated in our previous papers~\cite{MO1,MO2}.
On the other hand, there is no quadratic and cubic curvature corrections in
type II superstring theory or M-theory, and thus the first higher order
corrections start with quartic curvature terms~\cite{Mth,TBB}.

In our previous papers~\cite{MO1,MO2}, we have reported our results on
this problem for M-theory with forth-order terms in terms of the Riemann
curvature tensors.
However, it turns out that the coefficients we took had opposite signs
to the M-theory case. Since there is ambiguity in the coefficients coming from
the field redefinition, this does not immediately invalidate the results,
but it is necessary to study how the results change.
Here we give the details of our results in M-theory as well as type II
superstrings with the correct coefficients.
We consider these corrections given in terms of the Weyl tensors which
are favorable because only corrections in this scheme do not affect the highly
symmetric tree-level solutions such as AdS$_7 \times S^4$ \cite{TBB}.
We also include an additional quartic term in scalar curvature with an arbitrary
coefficient $\d$ in order to take into account the ambiguity mentioned above.
With this action, we exhaust exact solutions as well as past and future
asymptotic solutions and then discuss inflationary solutions among them.
The past and future asymptotic solutions are useful in describing
the inflation at the early universe and the present accelerating cosmology,
respectively.  Furthermore, we perform a linear perturbation analysis in
order to see which solutions are more generic and to make interesting
cosmological models.
We find an interesting solution in which the external space expands
exponentially and the internal space is static both in the original and
Einstein frames. This may be regarded as ``moduli stabilization'' by higher
order corrections, but this is not the usual moduli stabilization
in the sense that the moduli are fixed and stable. Rather we are interested
in such solutions in which the sizes of the internal spaces do not grow too
much while they exhibit inflation. What we find is that this is possible
with higher order corrections.
We also perform a numerical calculation around this
solution and find that some spacetimes give enough e-foldings. Finally we
briefly summarize similar solutions in type II superstrings.
Necessity of the higher order corrections for inflation in M-theory is
also discussed in Ref.~\cite{Chen}.

In the next section, we present our actions and field equations to be solved.
We write down these for $D=(1+p+q)$ dimensions with $p$ external and $q$ internal
space dimensions. Though we are mainly interested in $p=3$ in this paper,
there may be interesting applications if we keep the dimension $p$ arbitrary.
We give the equations for maximally symmetric spaces with non-vanishing
curvatures. The explicit forms of the actions are given in Appendix A,
and the field equations in Appendices~B and C. Although similar equations are
given in our previous papers~\cite{MO1,MO2}, the present theory is different
in that the quartic terms are written in the Weyl tensor instead of Riemann,
and also that we have an additional term $R^4$. So we present these
equations for our new system.

We give exact solutions as well as past and future asymptotic solutions
with exponential and power-law expansions in \S~3 for $\d=0$ and in \S~4
for $\d\neq 0$. Also we present the solutions for maximally symmetric spaces
with non-vanishing curvatures.

In \S~5, we perform a linear perturbation analysis around fixed points
for the exact generalized de Sitter solutions that are given in \S~3 and \S~4
in order to see which solutions are more generic and to make interesting
cosmological scenario.

In \S~6, we also summarize exact solutions for type IIB
superstrings for constant dilaton.

Using the obtained solutions, we discuss an inflationary scenario
in \S 7. Many of our exact solutions do not seem to give successful
inflation in the sense that they do not give big enough e-foldings.
However the simple analysis of exact solutions does not tell us what
happens after the inflationary solutions decay. Actually it turns
out by numerical analysis that there are interesting solutions, which first
approach to the exact solution for which the external space expands
exponentially and the internal space is static both in the original and
Einstein frames, and then eventually go to a stable solution.
For such solutions, we show that it is possible to obtain enough e-foldings
for successful inflation.
This is a very interesting possibility of achieving inflationary solutions.

The contents of \S \S~3 -- 6 summarize our analysis of solutions
and are rather technical. The reader may find it useful to skip this part
for the first reading, and then come back to check when they study physically
interesting solutions described in \S~7.

\section{Field equations}

We consider the low-energy effective action for
M-theory ($D=11$) with higher order corrections keeping dimension
$D$ arbitrary:
\bea
S &=& \sum_{n=1}^4 S_n+S_W +S_{R^4},
\label{totaction}
\ena
with
\bea
\label{eh}
S_1 &=&S_{\rm EH} ~\equiv~\frac{\a_1}{2\kappa_{D}^2} \int d^{D} x
\sqrt{-g} R,\\
\label{gb}
S_2 &=&S_{\rm GB} ~\equiv~ \frac{\a_2}{2\kappa_{D}^2} \int d^{D} x
\sqrt{-g}\;  \left[R_{\mu\nu\rho\sigma}^2 - 4 R_{\mu\nu}^2 +R^2\right],\\
S_3 &=&
\frac{\a_3}{2\kappa_{D}^2} \int d^{D} x
\sqrt{-g}\;  \tilde{E}_{6}\,, \\
S_4 \! &=&  \frac{\a_4}{2\kappa_{D}^2}\int d^{D} x
\sqrt{-g}  ~\tilde{E}_8  \,,
\label{S4} \\
S_W \! &=&  \frac{\c}{2\kappa_{D}^2}\int d^{D} x
\sqrt{-g} ~ L_W  \,,
\label{4th1} \\
S_{R^4} \! &=&  \frac{\d}{2\kappa_{D}^2}\int d^{D} x
\sqrt{-g} ~R^4  \,,
\label{R4}
\ena
where
\bea
\tilde{E}_{2n} &=&\!\!\! -{1\over 2^n  (D-2n)!}
\e^{\a_1 \cdots \a_{D-2n} \mu_1 \nu_1 \ldots \mu_n \nu_n}
\e_{\a_1 \cdots \a_{D-2n} \rho_1 \sigma_1 \ldots \rho_n \sigma_n}
R^{\rho_1\sigma_1}{}_{\mu_1 \nu_1}
\cdots R^{\rho_n \sigma_n}{}_{\mu_n \nu_n}
\,, \\
L_W&=& C^{\lambda\mu\nu\kappa} C_{\a\mu\nu\b}
C_{\lambda}{}^{\rho\sigma\a} C^\b{}_{\rho\sigma\kappa}
+\frac12 C^{\lambda\kappa\mu\nu}C_{\a\b\mu\nu}C_{\lambda}{}^{\rho\sigma\a}
C^\b{}_{\rho\sigma\kappa},
\label{j0}
\ena
and $R^4$ is a quartic term of scalar curvature.
Here we have dropped contributions from form fields,
$\kappa_{D}^2$ is a $D$-dimensional gravitational constant,
and $\a_1,\ldots, \a_4,\c$ and $\d$ are numerical coefficients.
The coefficient $\a_1$ of the Einstein-Hilbert (EH) term is $1$ by
definition, and though $\a_3$ is zero for all superstrings and M-theory,
we have included it since it will be useful for examining other cases.
The Weyl tensors in $L_W$ are defined by
\bea
C_{\lambda\mu\nu\kappa}
&=& R_{\lambda\mu\nu\kappa}
- \frac{1}{D-2}(g_{\lambda\nu} R_{\mu\kappa}-g_{\lambda\kappa} R_{\mu\nu}
-g_{\mu\nu} R_{\lambda\kappa} + g_{\mu\kappa} R_{\lambda\nu}) \nn
&&
+\; \frac{R}{(D-1)(D-2)}
(g_{\lambda\nu} g_{\mu\kappa}-g_{\lambda\kappa} g_{\mu\nu}).
\label{weyl}
\ena

In our previous papers~\cite{MO1,MO2}, we gave field equations for this system
with arbitrary couplings, but we focused on the case of nonzero $\a_4$ and $\c$.
We discussed the flat external and internal spaces in Ref.~\cite{MO1}
with various coefficients of the higher order terms, whereas all combinations
of curved spaces are examined in \cite{MO2}.
We considered the case in which the quartic term $L_W$ are
written in terms of Riemann tensors. However, it turned out that the
coefficients we took in \cite{MO2} were opposite in sign to the M-theory
case,\footnote{See the recently replaced version of Ref.~\cite{TBB}.}
and we should set
\bea
\a_2 = \a_3 = 0, \quad
\a_4 = {\kappa_{11}^2 ~T_2\over 3^2\times 2^{9} \times (2\pi)^4} ,\qquad
\gamma = {\kappa_{11}^2 ~ T_2\over 3 \times 2^{4}\times (2\pi)^4} \,,
\label{m}
\ena
where $T_2=({2\pi^2 /\kappa_{11}^2})^{1/3}$ is the membrane tension.
Though the field equations remain valid, the numerical results on the
generalized de Sitter solutions are significantly affected by this sign change.
We find that many of the solutions found in our previous paper go away
if we simply reverse the signs of these coefficients.
However, we should also note that contributions of the Ricci tensor
$R_{\mu\nu}$
and scalar curvature $R$ are not included in the fourth-order
corrections~\p{j0} because these terms are not uniquely fixed.
This means that there is significant ambiguity in the additional terms
involving these tensors, and in particular this allows us to put the
forth-order terms in terms of the Weyl tensors as given above.
This form appears
particularly favorable because only corrections in this scheme do not
affect the highly symmetric tree-level solutions such as AdS$_7 \times S^4$
(AdS$_5 \times S^5$ for type IIB superstring theory)~\cite{TBB}.
In view of this situation, it appears more appropriate to consider the quartic
correction terms given by Weyl tensors.

Furthermore it is interesting to examine how the ambiguity
may affect the results. For this purpose, we also include additional
quartic Ricci scalar term (\ref{R4}).
Here we discuss what value of $\d$
is plausible. Writing down Eq.~(\ref{j0}) in terms of Riemann curvature tensor,
we have the following equation:
\begin{align}
L_W(R_{\mu\nu\rho\sigma}) &=
L_W(C_{\mu\nu\rho\sigma}) + {12(D-4)\over (D-1)^2(D-3)^3}R^2
R_{\alpha\beta}R^{\alpha\beta}
+ {60\over (D-1)^2(D-3)^3}R^4\nonumber \\
&\quad +({\rm terms~containing~both} ~C_{\mu\nu\rho\sigma} ~{\rm and}~
R_{\alpha\beta}, R )\ ,
\end{align}
where we omit terms which contain both $C_{\mu\nu\rho\sigma}$ and
$R_{\alpha\beta}, R$, and $L_W(R_{\mu\nu\rho\sigma})$ is defined by
\begin{align}
L_W(R_{\mu\nu\rho\sigma})
= R^{\lambda\mu\nu\kappa}R_{\a\mu\nu\b}
R_{\lambda}{}^{\rho\sigma\a} R^\b{}_{\rho\sigma\kappa}
+\frac12 R^{\lambda\kappa\mu\nu}R_{\a\b\mu\nu}R_{\lambda}{}^{\rho\sigma\a}
R^\b{}_{\rho\sigma\kappa}\ .
\end{align}
We thus find that the difference between $L_W$ in terms of the Riemann and
Weyl tensors depends on the Ricci and scalar curvatures with very suppressed
coefficient due to large $D=11$. It is then natural to consider the $R^4$
correction terms with the same order of magnitude as above. Otherwise
the scalar curvature terms will be dominant and the stringy effects may
not be seen. Thus, the appropriate value of $\delta$ appears to have the order
\begin{align}
|\delta | \sim \frac{60}{ (D-1)^2(D-3)^3} \gamma \sim 10^{-3}  \gamma\ .
\end{align}

However, since we do not know the exact contribution
from Ricci and scalar curvatures, we leave $\delta$ to be free.

This is the system that we are going to examine.

\subsection{Basic equations for cosmology}
\label{sec2.1}

The metric of our $D$-dimensional space is
\begin{align}
 ds_D^2 &=  -N^2(t)dt^2 + a^2(t) ds_{p}^2 + b^2(t) ds_{q}^2\ ,
\label{met1}
\end{align}
with
\begin{align}
 N(t) = e^{u_0(t)}\ ,\qquad  a(t) = e^{u_1(t)}\ ,\qquad b(t) = e^{u_2(t)}\ ,
\end{align}
where $D=1+p+q$.
The external $p$- and internal $q$-dimensional spaces
($ds_{p}^2$ and $ds_{q}^2$)
are chosen to be maximally symmetric.
The curvature constants of those spaces are defined by
$\sigma_p$ and $\sigma_q$. The sign of $\sigma_p$ ($\sigma_q$)
determines the type of maximally symmetric spaces, i.e.  $\sigma_p$
(or $\sigma_q$) = $-1$, 0 and 1 denote a hyperbolic space, a flat
Euclidean space, and a sphere, respectively. The hyperbolic
and flat spaces are supposed to be compactified by identifying boundaries
of those finite part.

{}From the variation of the total action~\p{totaction},
whose explicit forms with the metric (\ref{met1}) are given in Appendix
\ref{appendix_A},
 with respect to
$u_0, u_1$ and $u_2$, we find three basic field equations:
\begin{align}
 \label{basic0}
F &\equiv \sum_{n=1}^4 F_n+F_W +F_{R^4}=0\ ,
\\
\label{basic1}
F^{(p)} &\equiv \sum_{n=1}^4
f_n^{(p)}+X \sum_{n=1}^4
g_n^{(p)}+Y \sum_{n=1}^4
h_n^{(p)} +F_W^{(p)} +F_{R^4}^{(p)}=0\ ,
\\
\label{basic2}
F^{(q)}&\equiv \sum_{n=1}^4
f_n^{(q)}+Y \sum_{n=1}^4
g_n^{(q)}+X\sum_{n=1}^4
h_n^{(q)} +F_W^{(q)} +F_{R^4}^{(q)}=0\ ,
\end{align}
where $X=\ddot{u}_1-\dot{u}_0\dot{u}_1+\dot{u}_1^2$,
$Y=\ddot{u}_2-\dot{u}_0\dot{u}_2+\dot{u}_2^2$, and
\begin{align}
 F_n&=F_n(u_0,\dot{u}_1,\dot{u}_2, A_p,A_q)\ ,
\nn
F_W&=F_W(u_0,u_1,u_2,\dot{u}_0,\dot{u}_1,\dot{u}_2,\ddot{u}_1,
\ddot{u}_2,\dddot{u}_1,\dddot{u}_2,X,Y,\dot{X},\dot{Y})\ ,
\nn
 F_{R^4}&=F_{R^4}(u_0,u_1,u_2,\dot{u}_0,\dot{u}_1,\dot{u}_2,\ddot{u}_1,
\ddot{u}_2,\dddot{u}_1,\dddot{u}_2,X,Y,\dot{X},\dot{Y})\ ,
\nn
f_n^{(p)}&=f_n^{(p)}(u_0,\dot{u}_1,\dot{u}_2, A_p,A_q)\ ,
\qquad
g_n^{(p)}=g_n^{(p)}(u_0,\dot{u}_1,\dot{u}_2, A_p,A_q)\ ,
\nn
h_n^{(p)}&= h_n^{(p)}(u_0,\dot{u}_1,\dot{u}_2, A_p,A_q)\ ,
\nn
F_W^{(p)}&= F_W^{(p)}(u_0,u_1,u_2,\dot{u}_0,\dot{u}_1,
\dot{u}_2,\ddot{u}_1,\ddot{u}_2,\dddot{u}_1,\dddot{u}_2,
X,Y,\dot{X},\dot{Y},\ddot{X},\ddot{Y})\ ,
\nn
 F_{R^4}^{(p)}&= F_{R^4}^{(p)}(u_0,u_1,u_2,\dot{u}_0,\dot{u}_1,
\dot{u}_2,\ddot{u}_1,\ddot{u}_2,\dddot{u}_1,\dddot{u}_2,
X,Y,\dot{X},\dot{Y},\ddot{X},\ddot{Y})\ ,
\nn
f_n^{(q)}&= f_n^{(q)}(u_0,\dot{u}_1,\dot{u}_2, A_p,A_q)\ ,
\qquad
g_n^{(q)}=g_n^{(q)}(u_0,\dot{u}_1,\dot{u}_2, A_p,A_q)\ ,
\nn
h_n^{(q)}&= h_n^{(q)}(u_0,\dot{u}_1,\dot{u}_2, A_p,A_q)\ ,
\nn
F_W^{(q)}&=F_W^{(q)}(u_0,u_1,u_2,\dot{u}_0,\dot{u}_1,
\dot{u}_2,\ddot{u}_1,\ddot{u}_2,\dddot{u}_1,\dddot{u}_2,
X,Y,\dot{X},\dot{Y},\ddot{X},\ddot{Y})\ ,
 \nn
F_{R^4}^{(q)}&=F_{R^4}^{(q)}(u_0,u_1,u_2,\dot{u}_0,\dot{u}_1,
\dot{u}_2,\ddot{u}_1,\ddot{u}_2,\dddot{u}_1,\dddot{u}_2,
X,Y,\dot{X},\dot{Y},\ddot{X},\ddot{Y})\ ,
\end{align}
are explicitly given in Appendix B.
Here $A_p$ and $A_q$ are defined by
\bea
&&
A_p=\dot{u}_1^2+\sigma_p \exp [2(u_0-u_1)],
\nn
&&
A_q=\dot{u}_2^2+\sigma_q \exp [2(u_0-u_2)].
\ena

Since $u_0$ is a gauge freedom of time coordinate, we have three equations
for two variables $u_1$ and $u_2$. It looks like an over-determinant system.
However, these three equations are not independent.
In fact, we can derive the following equation after bothersome calculation:
\bea
\dot{F}+(p\dot{u}_1+q\dot{u}_2-\dot{u}_0) F=
p\dot{u}_1  F^{(p)}+q\dot{u}_2  F^{(q)}\,.
\label{relation:FFpFq}
\ena
The constraint equation $F=0$ is satisfied if other dynamical equations are
solved {\it and} it is initially satisfied. As argued in Ref.~\cite{MO2},
it is in general enough to solve the two equations $F=0$ and $F^{(p)}=0$
(or $F^{(q)}=0$) instead of trying to solve all three equations.

\subsection{Ansatz for solutions}
We now analyze our basic Eqs.~\p{basic0} -- \p{basic2} and
look for inflationary solutions. Since we are interested in analytic
solutions, we study the following two cases:
\begin{itemize}
\item[\textbf{(1)}] \textbf{Generalized de Sitter solutions}:\\
Using a cosmic time, i.e. $u_0=0$, an exponential expansion of each scale
factor is given by $u_1=\mu t+\ln a_0$, and
$u_2=\nu t+\ln b_0$, where $\mu, \nu, a_0$ and $b_0$ are constants.
\item[\textbf{(2)}] \textbf{Power-law solutions}:\\
To find a power-law solution, although we can discuss it with the above
cosmic time, we use a different time gauge, which is defined by $u_0=t$.
Using this time coordinate, a power-law solution is given by
$u_1=\mu t+\ln a_0$, and $u_2=\nu t+\ln b_0$, where $\mu$ and $\nu$ are
constants.
\end{itemize}

If we choose the following time coordinate $u_0$,
\bea
u_0=\epsilon t,\quad
u_1=\mu t+\ln a_0,\quad
\mbox{ and } \quad
u_2=\nu t+\ln b_0,
\label{ansatz}
\ena
 we can discuss both solutions in the same set up, that is,
$\epsilon = 0$ for case (1), while $\epsilon =1$ for case (2).
In the latter case, in terms of a cosmic time $\tau=e^t$, we see
that the solution gives the power-law behavior:
\bea
a=e^{u_1}=a_0\tau^\mu, ~~~{\rm and}~~~b=e^{u_2}=b_0\tau^\nu \,.
\ena
Note that when the curvature constant $\sigma_p$ (or $\sigma_q$) vanishes,
$a_0$ and $b_0$ are arbitrary but we can set $a_0=1$ (or $b_0=1$) because
such a numerical constant can be absorbed by rescaling of the spatial
coordinates.

Before giving the solution, we note on the unit used in our solutions.
Rescaling $\a_4$, $\c$, $\mu$ and $\nu$ as
\begin{align}
 \tilde\a_4 =\a_4 /|\c|\ ,\qquad
\tilde{\c}=\c /|\c| ~(=1)  \ ,\qquad
\tilde{\mu}= \mu |\c|^{1/6} \ ,\
\mbox{ and }\
\tilde{\nu}= \nu |\c|^{1/6}\ ,
\label{nonzeroc}
\end{align}
we can always set $\c$ to $+1$.
We also have to rescale time coordinate as $\tilde{t}=|\c|^{-1/6} t$.
The typical dynamical time scale is then given by
$|\c|^{1/6}\sim O(m_{D}^{-1})$, where $m_{D}=\kappa_{D}^{-2/(D-2)}$ is
the fundamental Planck scale. In particular, for M-theory, we find
$|\c|^{1/6}=6^{-1/6}(4\pi)^{-5/9} m_{11}^{-1}\sim 0.1818176 m_{11}^{-1}$ from
Eq.~\p{m}. After this scaling for the M-theory, we have
\bea
\tilde\a_4 = \frac{1}{3 \times 2^5}\ , \qquad
\tilde\c = 1\ ,
\label{m1}
\ena
and a free parameter $\tilde\d$.

We use the above unit throughout this paper and
omit tilde for simplicity.
We now present solutions for $\d=0$ and $\d \neq 0$ in order.

\section{Solutions in M theory for $\d=0$}
\label{sec3}
From Eq. (\ref{A_pA_q}), we expect there may exist no exact solutions expect
for the case $\sigma_p=\sigma_q=0$. However,
even for the case of $\sigma_p\neq 0$ or $\sigma_q \neq 0$, we may have some
asymptotic solutions either in the future
direction $(t\rightarrow\infty)$ or in the past direction
$(t\rightarrow -\infty)$,
which describe the universes in these
stages. We classify solutions for Eqs.~(\ref{basic0})
-- (\ref{basic2})
by the signatures of $\sigma_p$ and $\sigma_q$ in
the following subsections.

\subsection{$\s_p=\s_q=0$}
\label{sec3.1}
In this case,
$A_p=\mu^2, A_q=\nu^2$ are constants.
We will discuss the cases of $\epsilon=0$ and  $\epsilon=1$ in order.

\subsubsection{Generalized de Sitter solutions ($\epsilon=0$)}
\label{sec3.1.1}

In this case, we can take the independent equations as $F=0$ and
$F^{(p)}+F^{(q)}=0$ for $\mu,\nu\neq 0$.
{}From the explicit forms of field equations given in Appendix~C, we have
two algebraic equations for $p=3$:
\begin{align}
&  \alpha_1\bigl[6 \mu^2 + 6 q \mu \nu  + q_1\nu ^2 \bigr]
+\alpha_4\nu^5 \bigl[ 336q_4 \mu ^3 + 168q_5\mu^2\nu
+ 24q_6\mu\nu^2+ q_7\nu\bigr]
\nn
&\quad  +\frac{6\gamma q (\mu -\nu)^4}{(q+2)^4 (q+3)^3}
  \bigl[-2( q-1 )N_1(q,3)\mu^4   +   12( q-1 )( 2 q^6 + 7q^5 -31q^4
+ 39q^3 + 565q^2 -6q-1512) \mu^3\nu\nn
&\qquad   -2 ( q-1 )(14 q^6 + 13q^5 -242q^4 + 1079q^3 + 4014q^2 -2970q  -7236 )
\mu^2\nu^2\nn
&\qquad  +4 (2 q^7 -3q^6 -11q^5 + 413q^4 + 337q^3 -1494q^2
+ 900q + 216)\mu \nu^3
+(q-7)N_1(3,q)\nu^4  \bigr]\nn
&\quad +\delta(-12 \mu^2 + 6q\mu\nu  + q(q-7){\nu }^2)
 \bigl[12\mu^2 + 6q\mu\nu  + (q+1)_0 \nu^2 \bigr]^3 = 0\ ,\label{3.1.1-1}\\
  &(\mu-\nu) \bigl\{\alpha_1
  + 4\alpha_4 \bigl[30( q-1)_4 \mu^2 \nu^4  + 12  ( q-1)_5\mu \nu^5
  + ( q-1)_6\nu ^6 \bigr]\nn
 &\quad+ \frac{4\gamma  (\mu -\nu)^2}{(q+2)^4 (q+3)^3}\bigl[ q_1N_1(q,3)\mu^4
 \nn
&\quad  - 3(q-1)(2 q^7+ 17q^6 -11q^5 -142q^4 + 1109q^3 + 1995q^2 -3978q-2376)
\mu^3
\nu \nn
&\quad +(q-1)(7 q^7 + 40q^6 - 104q^5 -114q^4 + 4887q^3 + 6372q^2 -14580q - 7776)
\mu^2\nu^2 \nn
&\quad -2q(q^7+ 4q^6-11q^5+ 135q^4 + 1104q^3+ 9q^2  -3366q  + 3024)\mu \nu^3
+ 6 N_1(3,q)\nu ^4 \bigr]\nn
&\quad+4\delta \bigl[ 12 \mu^2 + 6 q \mu  \nu  + (q+1)_0 \nu^2 \bigr]^3 \bigr\}
= 0\ ,  \label{3.1.1-2}
\end{align}
where $N_1(q,p)$ is an integer constant defined by
\begin{align}
N_1(q,p)&= p^3( 3p^2( q-2)  + p( q^2+7q-14) - 2q( 2q^2 - 5q + 1)-7) \nn
&\quad + (q-1)^3(p^2( q-3)+ 3pq( q+1) - 3q^2 ) \label{n1} \ .
\end{align}
Using the values for $\tilde\alpha_4$ and $\tilde\gamma$ in Eq.~(\ref{m1})
and setting $q=7$, we can solve these equations numerically and
found the following solution for $\delta=0$:
\begin{align}
  (\mu,\nu) = \mathrm{ME}1_\pm (\pm 0.104\,65,\, \mp 0.936\,66)\ .\label{gds00}
\end{align}
Here ME1$_\pm$ (the \underline{first} \underline{e}xact solutions in
 \underline{M} theory)
are the names given to the solutions. We will use similar
names for solutions in what follows.

\subsubsection{Power-law solutions ($\epsilon=1$)}
\label{sec3.1.2}

Setting $\epsilon=1$ in Eqs.~(\ref{emn-EH}) -- (\ref{emn-S_W}), there is
no exact solutions. We find that the EH action is dominant as
$t\rightarrow\infty$, while the actions $S_4$, $S_W$ and $S_{R^4}$ become
dominant as $t\rightarrow -\infty$.
Here we present asymptotic power-law solutions for each case.

As $t \to \infty$, EH term dominates and our basic equations reduce to
\bea
&& p_1\mu^2+q_1\nu^2+2pq \mu\nu =0,
\\
&&
q\nu(\nu-\mu-1)-(p-1)\mu=0, \\
&&
p\mu(\mu-\nu-1)-(q-1)\nu=0 .
\ena
We can easily show that these three equations are equivalent to
the following two equations, if it is not Minkowski space ($\mu=\nu=0$):
\bea
p\mu^2+q\nu^2=1, ~~~
p\mu+q\nu=1 ,
\label{Kasner}
\ena
which is a special case of Kasner solutions. We have a solution
\bea
&&\mu={p\pm\sqrt{pq(p+q-1)}\over p(p+q)},
\nn
&&\nu={q\mp\sqrt{pq(p+q-1)}\over q(p+q)}\,.
\label{Kasner1}
\ena
For $p=3, q=7$, we get two future asymptotic solutions:
\begin{align}
(\mu,\nu)= \mathrm{MF}6 \left(\frac{1+ \sqrt{21}}{10},\,
\frac{7- 3\sqrt{21}}{70}
\right) \ ,
\qquad \mathrm{MF}7 \left(\frac{1- \sqrt{21}}{10},
\,\frac{7+ 3\sqrt{21}}{70}
\right)\ .
\label{Kasner_sol}
\end{align}

As $t \to -\infty$, the fourth-order terms dominate. So let us present
asymptotic power-law solutions only with quartic terms. Assuming the metric
(\ref{2.20}) with $\epsilon=1$, our basic Eqs.~(\ref{basic0}) and
(\ref{basic2})
give three algebraic equations. For $p=3,\ q=7$, we have
\begin{align}
&120960\alpha_4\mu {\nu }^5\bigl[ 7{\mu }^2 + 7\mu \nu  + {\nu }^2 \bigr]\nn
& \quad -\frac{28\gamma( \mu  - \nu  )^4}{30375}\bigl[
5913(\mu -7) (\mu-1  )^3 - 252(\mu-1  )^2(\nu-1 ) (631\mu-657)  \nn
&\qquad +3(\mu-1  )  {(\nu-1 ) }^2(50647\mu -43939 ) -
  2{(\nu-1 ) }^3(29858\mu -23191) \bigr]\nn
&  -48\delta( \mu (2\nu+ 21)  - 7\nu(\mu -7) )\bigl[ 6{\mu }^2
+ 3\mu (7\nu -1)+ 7\nu (4\nu -1) \bigr]^3=0\ ,
 \label{4thp-1}
\end{align}
\begin{align}
&40320\alpha_4{\nu }^5\bigl[6{\mu }^3 + 3{\mu }^2( 8\nu-7) + 14\mu\nu (\nu -1)
+ (\nu -1) {\nu }^2  \bigr]\nn
& \quad - \frac{28\gamma( \mu  - \nu  )^3}{91125}\bigl[
17739(\mu-1  )^5 - 27(\mu-1  )^4( 15259\nu-12631)  + 13334(\nu-1)^3({\nu }^2
- 9\nu + 5)\nn
&\qquad  - 9(\mu-1  )^3(94639{\nu }^2 - 102980\nu  + 20167   )
+  3(\mu-1  )^2(\nu-1  )  (362107{\nu }^2 - 518030\nu  + 162475) \nn
&\qquad
- 2(\mu-1  ) (\nu-1  )^2 ( 258865{\nu }^2- 468650\nu  + 179599 )  \bigr]\nn
& \quad -16\delta (6{\mu }^2 + 47\mu + 7(\mu+ 25) \nu  - 28{\nu }^2
-168)\bigl[
 6{\mu }^2 + 3\mu (7\nu -1) + 7\nu (4\nu -1)   \bigr]^3=0\ ,
\end{align}
\begin{align}
&17280\alpha_4\mu {\nu }^4 \bigl[ 5{\mu }^2( 3\mu-7 ) + 2\mu\nu (23\mu -21)
+ (38\mu -7) {\nu }^2 + 6{\nu }^3 \bigr]\nn
&\quad+\frac{4\gamma( \mu  - \nu  )^3}{30375}\bigl[
41391(\mu-1  )^5 -  9(\mu-1  )^4(70015\nu -58189)\nn
&\qquad
- 3(\mu-1  )^3(151793{\nu }^2 - 58702\nu-57613 )
 + (\mu-1  )^2(\nu-1  )  (784721{\nu }^2- 937318\nu  +  172253  )\nn
&\qquad
- 2(\mu-1  ) (\nu-1  )^2  (202339{\nu }^2 - 338869\nu  + 106344  )
- 40002 (\nu-1  )^3(2\nu -1)
 \bigr]\nn
&\quad+48\delta (\mu (2\mu  + 3\nu -25) - 49\nu  +56 )\bigl[
6{\mu }^2 + 3\mu (7\nu -1 ) + 7\nu ( 4\nu-1 )   \bigr]^3=0\ .
\label{4thp-3}
\end{align}
Two of them are independent as we have shown.
Using the values for $\tilde\alpha_4$ and $\tilde\gamma$ in Eq.~(\ref{m1}),
we have solved these equations numerically and
found the following ten solutions for $\delta=0$:
\begin{align}
(\mu,\nu) &= \mathrm{MP}1(1.588\,41,\, 0.319\,25) \ ,
&&\mathrm{MP}2( 0.733\,61,\, 0.082\,88) \ ,
&&\mathrm{MP}3( 0.722\,46,\, -0.166\,74) \ ,\nn
&\, \quad    \mathrm{MP}4( 0.622\,07,\,-0.400\,40) \ ,
&&  \mathrm{MP}5( 0.100\,32,\, -1.700\,96) \ ,
&& \mathrm{MP}6 ( 0.022\,04,\, 0.990\,55)\ ,\nn
&\, \quad \mathrm{MP}7( -0.030\,14,\, 0.620\,90) \ ,
&&\mathrm{MP}8 ( -0.335\,30,\, 0.850\,24) \ ,
&&\mathrm{MP}9( -0.668\,48,\, 0.634\,27) \ ,\nn
&\quad\, \mathrm{MP}10( -0.938\,01,\, 2.572\,50) \ .
\label{mps0}
\end{align}

\subsection{$\sigma_p =0, \sigma_q\neq 0$ (or $\sigma_p\neq 0, \sigma_q = 0$)}

\subsubsection{Generalized de Sitter solutions ($\epsilon=0$)}
\label{sec3.2.1}

Here we have $A_p=\mu^2, A_q=\nu^2 +\tilde \s_q e^{-2\nu t}, X=\mu^2$
and $Y=\nu^2$, where $\tilde \s_q\equiv \s_q/b_0^2$.
It is easy to see that there is no exact solution unless
$\nu=0$, in which case we have constant $A_p=X=\mu^2,\ A_q=\tilde \s_q$
and $Y=0$. Our basic Eqs.~(\ref{basic0}) and (\ref{basic2}) now read
\begin{align}
& \alpha_1\bigl[p_1 \mu^2 + q_1\tilde\sigma_q\bigr]
 +\alpha_4\bigl[p_7\mu^8 + 4 p_5 q_1 \mu^6\tilde\sigma_q
+ 6{p_3}{q_3}{\mu }^4\tilde\sigma_q +
  4{p_1}{q_5}{\mu }^2 \tilde\sigma_q {^3} + {q_7}\tilde\sigma_q {^4}\bigr]\nn
& \quad+\frac{\gamma pq_1N_1(q,p)}{(D-1)^3(D-2)^4}\bigl[(p-7)\mu^2
+ (p+1)\tilde\sigma_q\bigr]\bigl[
  \mu^2 +\tilde\sigma_q \bigr]^3\nn
& \quad+ \delta\bigl[p(p-7)\mu^2 + q_1 \tilde\sigma_q\bigr]
\bigl[p_1\mu^2 + q_1\tilde\sigma_q\bigr]^3 = 0\ ,
 \label{3.2.1-1}\\
&\alpha_1\bigl[(p+1)_0 \mu^2 + (q-1)_2 \tilde\sigma_q \bigr] \nn
&  \quad + \alpha_4\bigl[( p+1 )_6 \mu^8 + 4(p+1)_4 q_2 \mu^6\tilde\sigma_q +
  6(p+1)_2 q_4\mu^4\tilde\sigma_q^2 + 4(p+1)_0q_6\mu^2\tilde\sigma_q^3
+ q_8\tilde\sigma_q^4\bigr]\nn
& \quad+\frac{\gamma (q-1)(p+1)_0N_1(q,p)}{(D-1)^3(D-2)^4}\bigl[q\mu^2
+ ( q-8)\tilde\sigma_q\bigr]\bigl[
 \mu^2 +\tilde\sigma_q\bigr]^3\nn
&\quad  +\delta\bigl[ (p+1)_0 \mu^2 + (q-1)(q-8)\tilde\sigma_q\bigr]
  \bigl[(p+1)_0 \mu^2 + q_1\tilde\sigma_q\bigr]^3 = 0\ .
  \label{3.2.1-2}
\end{align}
We note that Eq.~(\ref{basic1}) gives the same equation as Eq.~(\ref{basic2})
for $\nu=0$ and need not be taken into account.
Setting $\delta=0$, it turns out that there is no solution for
$\sigma_p= 0, \sigma_q \neq 0$.
For the case of $\sigma_p\neq 0$ and $\sigma_q=0$,
exchanging $\mu, p$ and $\nu, q$, we find that there is also no solution.
Although there is no this kind of solution for $\delta=0$, we find solutions
for $\delta\neq 0$ as discussed in \S~\ref{sec4.2.1}.

\subsubsection{Power-law solutions ($\epsilon=1$)}
\label{sec3.2.2}

Here we have $A_p=\mu^2, A_q=\nu^2 +\tilde \s_q e^{2(1-\nu) t},
 X=\mu(\mu-1)$
and $Y=\nu(\nu-1)$. We have only asymptotic solutions in most cases.

\begin{description}
\item[(1)] $\nu>1:$

For $t \to \infty$, the EH term dominates and $A_q \to \nu^2$.
The solutions are the same as $\s_p=\s_q=0$ case in section~\ref{sec3.1.2}.
However, there is no solution with $\nu>1$.

For $t \to -\infty$, $A_q \to \tilde \s_q e^{2(1-\nu)t}$ and
there is no solution.
\item[(2)] $\nu<1:$

For $t \to \infty$, $A_q \to  \tilde \s_q e^{2(1-\nu)t}$ and there is
no solution.

For $t \to -\infty$, $A_q \to \nu^2$ and the solutions are the same as
$\s_p=\s_q=0$ case in section~\ref{sec3.1.2}.
Choosing solutions with $\nu<1$ from Eq.~\p{mps0}, we get nine solutions
$\mathrm{MP}1$ -- $\mathrm{MP}9$.

\item[(3)] $\nu=1:$

We have $A_p=\mu^2, A_q =1+\tilde\s_q, X=\mu(\mu-1)$ and $Y=0$.

\end{description}

For $t \to \infty$, the EH term dominates and the solution is
\begin{align}
(\mu,\nu,\tilde\sigma_q) &= \mathrm{ME}12 (0,\,1,\,-1)\ .
\label{sp1}
\end{align}
Actually this is an exact solution.

For $t \to -\infty$, fourth-order terms dominate. Our basic independent
Eqs.~(\ref{basic0}) and (\ref{basic2}) give
\begin{align}
&\alpha_4\bigl[  {p_7}{\mu }^8 + 8{p_6}q{\mu }^7 + 4 {p_5}{q_1}{\mu }^6
( 6 + {A_q} ) + 8 {p_4}{q_2} {\mu }^5( 4 + 3{A_q} )+ 2 {p_3}{q_3}{\mu }^4
( 8 + 24{A_q} + 3{{A_q}}^2 ) \nn
&\qquad+  8{p_2}{q_4} {\mu }^3{A_q}( 4 + 3{A_q} ) +
4 {p_1}{q_5}{\mu }^2{{A_q}}^2( 6 + {A_q} ) + 8p{q_6}\mu {{A_q}}^3
+ {q_7}{{A_q}}^4\bigr]\nn
&\quad+\frac{\gamma pq_1N_1(q,p)}{(D-1)^3(D-2)^4}\bigl[(p-7) {\mu }^2
+ 2(3p+4q -21) \mu - 8(p+q-6) + (p+1) {A_q}\bigr]
\bigl[\mu(\mu-2) + {A_q} \bigr]^3\nn
&\quad +\delta\bigl[p(p-7){\mu }^2 + 2p(q-21)\mu - 48q+ {q_1}{A_q}\bigr]
 \bigl[p(p+1) {\mu }^2 + 2p( q-1 ) \mu  + {q_1}{A_q} \bigr]^3=0\ ,
\label{3.2.2-1}
\end{align}
\begin{align}
 &  \alpha_4\bigl[{(p+1)_6}{\mu }^8 + 8p_5(pq-2p+6) {\mu }^7
 + 4p_4 (q-1){\mu }^6 ( 6(pq-4p-q+12) + (p+1) ( q-2 ) {A_q} )\nn
&\qquad+ 8p_3(q-1)_2 {\mu }^5 ( 4p(q-6) -8(q-9) +3(pq-4p+4) {A_q})\nn
&\qquad+ 2p_2 (q-1)_3 {\mu }^4 ( 8(p-3) (q-8) + 24(pq-6p-q+10){A_q}
+ 3(p+1) (q-4) {A_q}{^2})\nn
&\qquad+ 8p_1 (q-1)_4 {\mu }^3 A_q ( 4(p-2) (q-8) + 3( pq - 6p + 2) {A_q})\nn
&\qquad+ 4p(q-1)_5{\mu }^2 A_q{^2}( 6(p-1) (q-8) +(p+1) (q-6) {A_q})
+ 8p(q-8)(q-1)_6 \mu A_q{^3} + (q-1)_8 A_q{^4}
 \bigr]\nn
&\quad + \frac{\gamma p(q-1) N_1(q,p)}{(D-1)^3(D-2)^4}\bigl[\mu(\mu-2)
+ A_q\bigr]^3
 \bigl[(p+1)q \mu^2 - 2(4p^2 + 5pq-28p+q)\mu \nn
&\qquad - 8(q-8)(p+q-6)+ (p+1) (q-8) {A_q}) \bigr]\nn
&\quad +\delta\bigl[(p+1)_0 {\mu }^2 + 2p(q-29)\mu + (q-8)((q-1) {A_q}-48 )
\bigr]
  \bigl[(p+1)_0 \mu^2 + 2p(q-1)\mu + q_1{A_q}\bigr]^3=0\ .\label{3.2.2-2}
\end{align}
For $p=3$ and $q=7$, we find the solutions $\mathrm{ME}10$ in Eq.~(\ref{sp1})
and
\begin{align}
(\mu,\nu,\tilde\sigma_q) &=\mathrm{MP}11(6.679\,58,\,1\, -2.106\,62)\ ,
&&\mathrm{MP}12(6.085\,83,\,1,\, -1.357\,93) \ ,
&&\mathrm{MP}13(2,\,1,\, -1)\ ,\nn
&\quad\, \mathrm{MP}14 (0.481\,80,\, 1,\, -1.238\,02)\ ,
&&\mathrm{MP}15(0.072\,89,\, 1,\, -2.208\,58)\ .
\label{sp2}
\end{align}

For the case of $\sigma_p\neq 0$ and $\sigma_q=0$,
exchanging $\mu, p$ and $\nu, q$, we obtain one exact solution ME13
and thirteen asymptotic solutions MP2 -- MP10 in Eq.~(\ref{mps0}) and
the following solutions MP16 -- MP19:
\bea
(\mu,\nu,\tilde\s_p)\! &=&\!
 \mathrm{ME}13(1,\,0,\,-1)\ ,\label{3.2.2-4} \\
(\mu,\nu,\tilde\sigma_p)\! &=&\!
\mathrm{MP}16 (1,\, 0.268\,18,\, 1.438\,12)\ ,~
\mathrm{MP}17 (1,\, -9.177\,79,\, 4.633\,79)\ ,\nn
&& \mathrm{MP}18 (1,\, 0.931\,76,\, -3.965\,11)\ ,~
\mathrm{MP}19 (1,\, -0.122\,50,\, -1.414\,71) \ ,
  \label{3.2.2-6}
\ena
where $\tilde\s_p\equiv \s_p/a_0^2$.
\subsection{$\sigma_p\sigma_q\neq 0$}

\subsubsection{Generalized de Sitter solutions ($\epsilon=0$)}
\label{sec3.3.1}
If $\mu=\nu=0$, our basic Eqs.~(\ref{basic0}) and (\ref{basic1}) reduce to
\begin{align}
& \alpha_1\bigl[p_1 \tilde\sigma_p + q_1 \tilde\sigma_q\bigr]
 +\alpha_4\bigl[p_7 \tilde\sigma_p{^4} + 4 p_5 q_1 \tilde\sigma_p{^3}
 \tilde\sigma_q +
  6 p_3 q_3 \tilde\sigma_p{^2}\tilde\sigma_q{^2}
+ 4p_1q_5\tilde\sigma_p\tilde\sigma_q{^3} +
  q_7\tilde\sigma_q{^4}\bigr]\nn
 &\quad + \frac{\gamma}{(D-1)^3(D-2)^4}\bigl[p_1 (q+1)_0 N_1(p,q)
\tilde\sigma_p{^4}+
  4p_1 q_1 N_1(p,q) \tilde\sigma_p{^3}\tilde\sigma_q  \nn
 & \quad + 2p_1 q_1  N_2(p,q)\tilde\sigma_p{^2} \tilde\sigma_q{^2}
 + 4p_1q_1 N_1(q,p) \tilde\sigma_p \tilde\sigma_q{^3}
 + (p+1)_0 q_1 N_1(q,p)\tilde\sigma_q{^4}  \bigr]\nn
 &\quad +\delta\bigl[p_1 \tilde\sigma_p + q_1 \tilde\sigma_q\bigr]^4 = 0\ ,
\label{3.3.1-1}\\
 &\alpha_1\bigl[(p-1)_2\tilde\sigma_p + q_1\tilde\sigma_q\bigr]\nn
 &\quad +\alpha_4\bigl[(p-1)_8\tilde\sigma_p{^4} +  4(p-1)_6 q_1
 \tilde\sigma_p{^3} \tilde\sigma_q
 + 6(p-1)_4 q_3 \tilde\sigma_p{^2} \tilde\sigma_q{^2} + 4(p-1)_2 q_5
\tilde\sigma_p\tilde\sigma_q{^3}
 + q_7 \tilde\sigma_q{^4}\bigr]\nn
  &\quad + \frac{\gamma }{(D-1)^3(D-2)^4}\bigl[
( p-8) (p-1) (q+1)_0 {N_1}(p,q)  \tilde\sigma_p{^4} + 4(p-6) (p-1)q_1
 {N_1}(p,q) \tilde\sigma_p{^3}\tilde\sigma_q\nn
& \qquad + 2( p-4 ) (p-1) q_1{N_2}(p,q)
\tilde\sigma_p{^2}\tilde\sigma_q{^2}
 +  4(p-2) (p-1) q_1 {N_1}(q,p) \tilde\sigma_p \tilde\sigma_q{^3}
    + ( p+1)_0 q_1 {N_1}(q,p) \tilde\sigma_q{^4}
 \bigr]\nn
 &\quad +\delta\bigl[( ( p-8) (p-1) \tilde\sigma_p + {q_1}\tilde\sigma_q )
  {( {p_1}\tilde\sigma_p + {q_1}\tilde\sigma_q) }^3\bigr] = 0\ ,\\
 &\alpha_1\bigl[p_1 \tilde\sigma_p + (q-1)_2\tilde\sigma_q\bigr]\nn
 &\quad  +\alpha_4\bigl[p_7\tilde\sigma_p{^4}
+ 4p_5(q-1)_2\tilde\sigma_p{^3}\tilde\sigma_q +
  6p_3(q-1)_4\tilde\sigma_p{^2}\tilde\sigma_q{^2} +
  4p_1(q-1)_6\tilde\sigma_p\tilde\sigma_q{^3}
+ (q-1)_8\tilde\sigma_q{^4}\bigr]\nn
 &\quad +\frac{\gamma }{(D-1)^3(D-2)^4}\bigl[p_1 (q+1)_0 {N_1}(p,q)
\tilde\sigma_p{^4} +
 4p_1(q-1)_2{N_1}(p,q) \tilde\sigma_p{^3}\tilde\sigma_q\nn
 &\qquad  + 2p_1(q-4) (q-1){N_2}(p,q)\tilde\sigma_p{^2}\tilde\sigma_q{^2}
 + 4p_1(q-6) (q-1){N_1}(q,p) \tilde\sigma_p\tilde\sigma_q{^3}
 + (p+1)_0 ( q-8) (q-1) {N_1}(q,p) \tilde\sigma_q{^4}\bigr]\nn
 &\quad +\delta\bigl[( {p_1}\tilde\sigma_p + (q -8) (q-1) \tilde\sigma_q )
  {( {p_1}\tilde\sigma_p + {q_1}\tilde\sigma_q) }^3\bigr] = 0\ ,\label{3.3.1-3}
\end{align}
where $N_2(p,q)$ is defined by
 \begin{align}
N_2(p,q) &= p^5( 9q-16) + p^4(3q^2 -1) + p^3(14q^2-21q+ 26) +  3p^2(q-3)
- 12p^3q^3 \nn
  &\quad - 13p^2q^2 + 9pq + 3(p-3) q^2 + (14p-21p+ 26) q^3 +
    (3p^2-1) q^4 + (9p-16) q^5\ .
 \end{align}
For $p=3,\ q=7$ and $\delta=0$, we find that there is no solution.

If $\mu\neq 0$ and $\nu=0$, it is clear that
there is no exact solution. For asymptotic solutions, we can search for
them by setting $A_p=\mu^2$, $A_q=\tilde \sigma_q$, $X=\mu^2$ and $Y=0$.
 This is the same condition in \S~\ref{sec3.2.1},
and we have no solution. The  case
with $\mu=0$ and $\nu\neq 0$ is obtained by exchanging
$p,\ \mu$ and $q, \ \nu$. We find again that there is no solution.

For $\mu\nu\neq 0$, if our ansatz for solution is imposed, it is easy to
see that there is no solution if $\mu$ and $\nu$
are of the opposite signs. If they are of the same sign, either
$t\rightarrow +\infty$ or $t\rightarrow -\infty$ gives
$A_p\rightarrow\mu^2$, $A_q \rightarrow\nu^2$ and there may be solutions.
For $\delta=0$, however, we see that Eq.~(\ref{gds00}) gives no solution
of the same sign.

\subsubsection{Power-law solutions ($\epsilon=1$)}
\label{sec3.3.2}

In this case, we first consider the cases when both $\mu$ and $\nu$ are
not equal to 1.
\begin{description}
\item[(1)] $\mu>1$ and $\nu>1$:

For $t\to \infty$, the EH term dominates and we obtain
the asymptotic solutions in \S~\ref{sec3.1.2}. Again no solutions
satisfy the condition of $\mu>1$ and $\nu>1$ (see Eq. (\ref{Kasner}))
and hence there is no asymptotic solution of our form.

As $t\to -\infty$, the fourth-order terms become dominant and
we find no consistent solution from the forth-order terms.

\item[(2)] $\mu<1$ and $\nu<1$:

As $t\to \infty$ with EH dominance, we again find no consistent solution.
As $t\to -\infty$ with fourth-order-term dominance, we obtain eight asymptotic
solutions $\mathrm{MP}2$ -- $\mathrm{MP}9$ from Eq.~\p{mps0} in
\S~\ref{sec3.1.2}:

\item[(3)] $\mu>1$ and $\nu<1$:

As $t\to \infty$, $A_p\to \mu^2$ and $A_q\to \tilde{\sigma}_q e^{2(1-\nu)t}$.
           This is similar to the case (2) in \S~\ref{sec3.2.2}
and there is no solution of our form.

As $t\to -\infty$, $A_p\to \tilde{\sigma}_p e^{2(1-\mu)t}$ and
$A_q\to \nu^2$. We find no solution.

\item[(4)] $\mu<1$ and $\nu>1$:

Here we reach the same result by exchanging $p,\mu$ and $q,\nu$.
No asymptotic solution of our form is obtained for both $t\to \pm\infty$.
\end{description}

Next, we discuss the cases in which one of $\mu$ or $\nu$ is equal to
1 and the other is not:

\begin{description}
\item[(5)] $\mu>1$ and $\nu=1$:

As $t\to \infty$ with EH dominance, $A_p\to \mu^2$,
and we recover the case of $\sigma_p=0, \sigma_q\neq 0$.
However, there is no solution with $\mu>1$.
We do not have any asymptotic solution of our form.
As $t\to -\infty$ with fourth-order-term dominance,
$A_p\to\tilde\s_p e^{2(1-\mu)t}$.
We again do not have any asymptotic solution of our form.

\item[(6)] $\mu<1$ and $\nu=1$:

As $t\to \infty$ with EH dominance, $A_p$ diverges as $\tilde{\sigma}_p
e^{2(1-\mu)t}$. There is no asymptotic solution of our form.
As $t\to -\infty$, we again recover the case of $\sigma_p=0,
\sigma_q\neq 0$ with the fourth-order-term dominance and
solutions in~\p{sp1} and~\p{sp2}. (Note that \p{sp1} was
an exact solution for $\s_p=0$.) Choosing those with $\mu<1$,
we get asymptotic power-law solutions
\begin{align}
 (\mu,\nu,\tilde\sigma_q) &= \mathrm{MP}23 (0,\, 1,\,-1)\ .
\end{align}
{}form Eq.~\p{sp1}, and $\mathrm{MP}14$ and $\mathrm{MP}15$ from
Eq.~(\ref{sp2}).

\item[(7)] $\mu=1$ and $\nu>1$:

The analysis is almost the same as the case (5).
There is no asymptotic solution.

\item[(8)] $\mu=1$ and $\nu<1$:

The analysis is almost the same as the case (6),
and we find the asymptotic solutions
\begin{align}
(\mu,\nu,\tilde\sigma_q) &= \mathrm{MP}24 (1,\, 0,\,-1)\ .
\end{align}
{}form Eq.~\p{3.2.2-4}, and $\mathrm{MP}16$ -- $\mathrm{MP}19$ for
$t\to -\infty$,
which are the same as the case of $\sigma_p\neq 0, \sigma_q= 0$ given
in Eq.~\p{3.2.2-6}.
\end{description}

Finally, we consider the remaining case.
\begin{description}
\item[(9)] $\mu=1$ and $\nu=1$:

Here we have constant $A_p=1+\tilde{\sigma}_p$ and $A_q=1+\tilde{\sigma}_q$.
As $t \to +\infty$, the EH term is dominant, and we have
\bea
&& p_1 A_p +q_1 A_q +2pq=0, \nn
&& (p-1)_2 A_p +q_1 A_q +2(p-1)q=0, \nn
&& p_1 A_p +(q-1)_2 A_q +2p(q-1)=0.
\ena
The solution is given by
\bea
A_p = -\frac{q}{p-1}, \quad
A_q = -\frac{p}{q-1}.
\label{mn1}
\ena
This is the solution found in Ref.~\cite{cosm3} which exhibits eternal
accelerating expansion when higher order effects are taken into account.
For $p=3,q=7$, we get the following future asymptotic solution:
\begin{align}
(\mu,\nu,\tilde\s_p,\tilde\s_q) &=\mathrm{MF}8 \left(1,\,1,\, -\frac{9}{2},
-\frac{3}{2}\right)\ .
\end{align}

For $t\to -\infty$ with fourth-order-term dominance, we get
two independent equations for $\tilde\s_p$ and $\tilde\s_q$ from
Eqs.~(\ref{basic0}) and (\ref{basic1}).
\begin{align}
&  \alpha_4\bigl[ {p_7}{( 1 + \tilde\sigma_p ) }^4
+ 4( 1+\tilde\sigma_p )^3( 2q{p_6} + {p_5}{q_1}( 1 + \tilde\sigma_q ))
+ 24{( 1 + \tilde\sigma_p ) }^2( {p_5}{q_1} + {p_4}{q_2}( 1
+ \tilde\sigma_q ))
\nn
 & \qquad
+ 32{p_4}{q_2}( 1 + \tilde\sigma_p )
+ 32{p_2}{q_4}( 1 + \tilde\sigma_q )
+ 24( {p_1}{q_5} + {p_2}{q_4}( 1 + \tilde\sigma_p ))
( 1 + \tilde\sigma_q )^2
\nn
& \qquad
+ 4 ( 2p{q_6} + {p_1}{q_5}( 1 + \tilde\sigma_p )  ) ( 1
+ \tilde\sigma_q )^3
+ {q_7}{( 1 + \tilde\sigma_q ) }^4
+ 6{p_3}{q_3}{( 1 + \tilde\sigma_p ) }^2{( 1 + \tilde\sigma_q ) }^2
\nn
& \qquad
+48{p_3}{q_3}( 1 + \tilde\sigma_p ) ( 1 + \tilde\sigma_q )
 +16{p_3}{q_3}  \bigr]
\nn
& \quad
 +\frac{\gamma }{(D-1)^3(D-2)^4}\left[
 p_1 (q+1)_0 {N_1}(p,q) \tilde\sigma_p{^4}
+ 4p_1q_1{N_1}(p,q) \tilde\sigma_p{^3} \tilde\sigma_q
+ 2p_1q_1{N_2}(p,q) \tilde\sigma_p{^2} \tilde\sigma_q{^2}
\right.
\nn
& \qquad
\left.
+ 4p_1q_1{N_1}(q,p)  \tilde\sigma_p \tilde\sigma_q{^3}
+ (p+1)_0q_1{N_1}(q,p) \tilde\sigma_q{^4} \right]
\nn
&\quad
+\delta\bigl[
( (p + q-49) (p + q) + {p_1}\tilde\sigma_p + {q_1}\tilde\sigma_q )
(( p + q-1 ) (p + q)+ {p_1}\tilde\sigma_p + {q_1}\tilde\sigma_q )^3 \bigr]
=0\ ,
\label{3.3.2-3}
\end{align}
\begin{align}
& \alpha_4\bigl[
{{(p-1 ) }_8}{( 1 + \tilde\sigma_p ) }^4
+ 4{( 1 + \tilde\sigma_p ) }^3( 2q{{(p-1 ) }_7} +  {{(p-1 ) }_6}{q_1}
( 1 + \tilde\sigma_q ))\nn
& \qquad
+ 32{{(p-1 ) }_5}{q_2} ( 1 + \tilde\sigma_p )
 + 32{{(p-1 ) }_3}{q_4}( 1 + \tilde\sigma_q )
+ 24 ( {{(p-1 ) }_2}{q_5} + {{(p-1 ) }_3}{q_4}( 1 + \tilde\sigma_p ))
{( 1 + \tilde\sigma_q ) }^2\nn
& \qquad
+ 4 ( 2(p-1 ) {q_6} + {{(p-1 ) }_2}{q_5} ( 1 + \tilde\sigma_p ) )
{( 1 + \tilde\sigma_q ) }^3
+ {q_7}{( 1 + \tilde\sigma_q ) }^4
+ 6{{(p-1 ) }_4}{q_3}{( 1 + \tilde\sigma_p ) }^2 {( 1 + \tilde\sigma_q ) }^2
\nn
& \qquad
+ 48{{(p-1 ) }_4}{q_3}( 1 + \tilde\sigma_p ) ( 1 + \tilde\sigma_q )
+ 16{{(p-1 ) }_4}{q_3} \bigr]\nn
& \quad
+\frac{\gamma}{(D-1)^3(D-2)^4} \bigl[
-(p-1)N_1(p,q)\tilde\sigma_p{^3}\bigl\{(p-8)(q+1)_0\tilde\sigma_p
+ 4(p-6)q_1\tilde\sigma_q-8(p+q-7)^2\bigr\}\nn
&\qquad -q_1N_1(q,p)\tilde\sigma_q{^3}\bigl\{4(p-1)_2\tilde\sigma_p
+ (p+1)_0\tilde\sigma_q +8 (p+q-7)^2\bigr\}\nn
&\qquad +4(p-1)q_1(p+q-7)^2\tilde\sigma_p\tilde\sigma_q(N_3(p,q)\tilde\sigma_p
- N_3(q,p)\tilde\sigma_q)
-2(p-1)(p-4)N_2(p,q)\tilde\sigma_p{^2}\tilde\sigma_q{^2} \bigr]\nn
&\quad +\delta\bigl[( (  p + q-49  ) (p + q-8 )  +
 ( p-8 ) (p-1) \tilde\sigma_p + {q_1}\tilde\sigma_q )
{( (p+ q-1 ) ( p + q )  + {p_1}\tilde\sigma_p + {q_1}\tilde\sigma_q )}^3\bigr]
=0\ ,\label{3.3.2-4}
\end{align}
where we define $N_3(p,q)$ as
\begin{align}
N_3(p,q) &= pp_1(7p^2+35p-18) - 3p_1(6p^2-p -6) q  - (32p^3-49p^2+33p-18) q^2
 \nn
& \quad+ (16p^2+15p -47) q^3 + (21p -41) q^4 - 2q^5\ .
\end{align}
For $p=3,\ q=7$ and $\delta=0$, we find four solutions
\begin{align}
(\mu,\nu, \tilde \sigma_p, \tilde \sigma_q) &=  \mathrm{MP}25(1,\,1,\, 2.370\,
16, -1.262\,93)\ ,
&& \mathrm{MP}26(1,\,1,\,-1.412\,83, -120.598\,08) \ ,\nn
&\, \quad \mathrm{MP}27(1,\,1,\, -4.376\,02, -0.120\,21)\ ,
&& \mathrm{MP}28(1,\,1,\,-5.762\,67, -2.588\,03)\ .
\end{align}
\end{description}

We summarize our results obtained here in Appendix \ref{summary}.
The exact solutions for $\delta=0$ are listed in Table~\ref{sum1},
future asymptotic solutions in Table~\ref{sum2}
and past asymptotic solutions in Table~\ref{sum3}.
The numbering of solutions are given for those with
$\delta\neq0$ (see next section).
For example, in Table \ref{sum1}, we do not find solutions
ME$2_\pm$ -- ME$11_\pm$.

\section{Solutions  in M theory for $\delta\neq 0$}

In this section, we search for exact generalized de Sitter solutions for
$-10\leq\delta\leq 10$ whereas we give a few examples for future and past
asymptotic solutions for particular $\delta$ to avoid vexatious complications.
As typical examples, we mainly focus on the case of $\delta= -0.001$
and $\delta = -0.1$ as to future and past asymptotic solutions
in \S~\ref{sec4.1.2}, \ref{sec4.2.2} and \ref{sec4.3.1} -- \ref{sec4.3.2},
but we also study how the results change depending on $\d$.

  \subsection{$\sigma_p = \sigma_q=0$}

\subsubsection{Generalized de Sitter Solutions ($\epsilon=0$)}
\label{sec4.1.1}
\begin{figure}[htb]
\begin{minipage}{.475\textwidth}
\includegraphics[width=\linewidth, height=\linewidth]{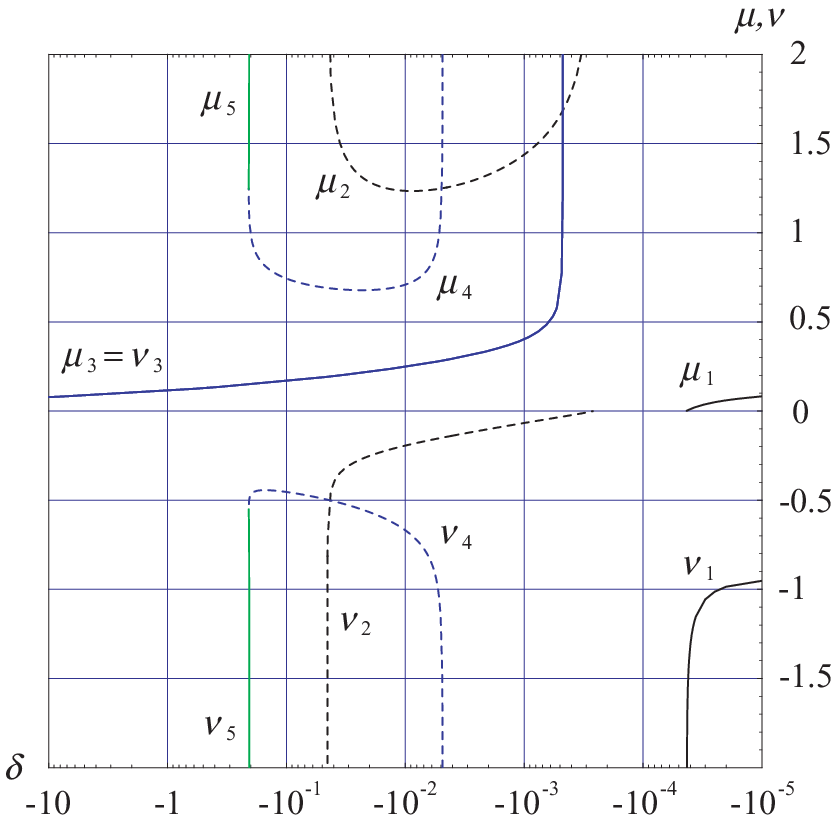}
\caption{\small{Five Generalized de Sitter solutions with $\sigma_p =\sigma_q=0$
with respect to $\delta<0$.}}\label{delta_log}
\end{minipage}\hfill
\begin{minipage}{.475\textwidth}
\includegraphics[width=\linewidth,height=\linewidth]{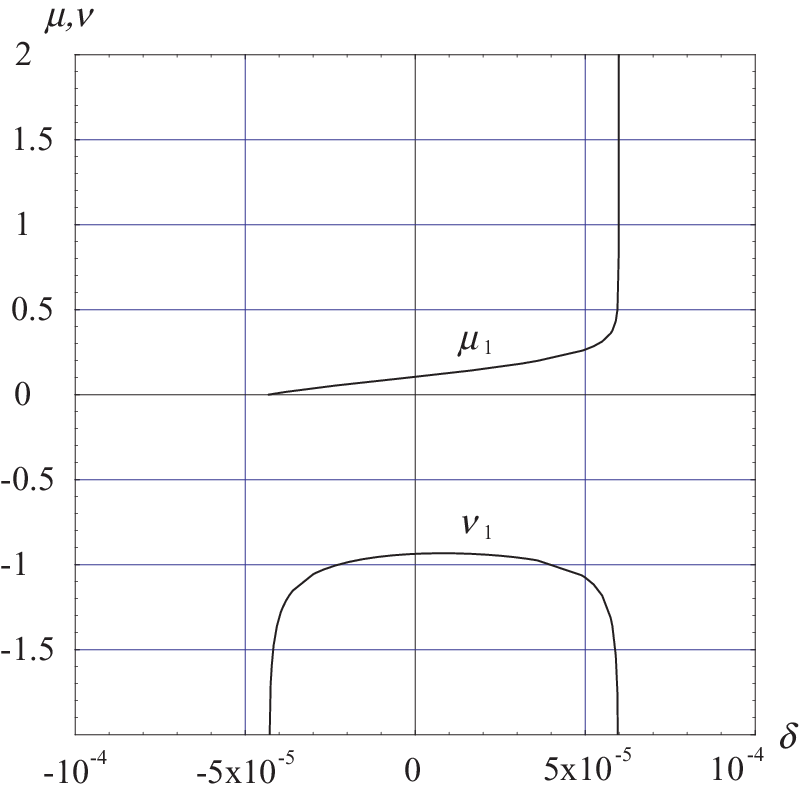}
\caption{\small{One Generalized de Sitter solution with $\sigma_p =\sigma_q=0$
near the origin of $\delta$.}}\label{delta_linear2}
\end{minipage}\\
\\
\small{Each pair of points $(\delta,\mu_i,\nu_i)$ with the same value of
$\delta$ gives one solution. Number of solutions changes with the value of
$\delta$. There is another set of time-reversed solutions
$\mathrm{ME}i_{-}$ with $\mu_i<0$. No solution exists for $10^{-4}<\delta$.}
\end{figure}

In this case, we have the same Eqs.~(\ref{3.1.1-1}) and (\ref{3.1.1-2})
with a free parameter $\delta\neq 0$. For the given
value of $\delta$, we may solve these equations numerically.
In Figs.~\ref{delta_log} and \ref{delta_linear2}, we depict numerical solutions
$\mathrm{ME}i_{+}(\delta,\mu_i,\nu_i)\ (i=1,\cdots,5)$ with $\mu_i\geq 0$
when $\delta$ is varied. We note that there are always time-reversed
solutions $\mathrm{ME}i_{-}(\delta,\mu'_i,\nu'_i)\ (i=1,\cdots,5)$  obtained by
$(\mu'_i,\nu'_i) =(-\mu_i,-\nu_i)$ which are not shown explicitly.
We find five solutions $\mathrm{ME}1_{+}$ -- $\mathrm{ME}5_{+}$ for $\delta<0$,
while just one $\mathrm{ME}1_{+}$ for $\delta>0$.
For the case of $\delta = 0$, we have a solution $\mathrm{ME}1_{+}$ which
is consistent with the result of \S~\ref{sec3.1.1}.
In Table~\ref{delta_table1}, we summarize their properties.

\begin{table}[htb]
\begin{center}
\small{
\caption{\small{Generalized de Sitter Solutions ME$i_{+}\ (i=1,\cdots,5)$ with
$\mu_i\geq 0$ for various values of $\delta$. Five eigenmodes for linear
perturbations are also shown. ($m$s,$n$u) means that there are $m$ stable
modes and $n$ unstable modes. The solution has many stable modes if its
10-volume expansion rate $3\mu+7\nu$ is positive.}}
\vspace{2mm}\label{delta_table1}
\begin{tabular}{lcccc}
\noalign{\global\arrayrulewidth1pt}
\hline
\noalign{\global\arrayrulewidth.4pt}
Solution & Property &Range & Stability &$3\mu_i +7\nu_i$\\ \hline
ME$1_{+}$ & $\nu_1<0<\mu_1$ & $-0.000\,043\,11<\delta<0$ &(0s,5u) &$-$\\
&& $\delta=0$&(1s,2u)&$-$\\
&& $0<\delta<0.000\,059\,88$&(1s,4u)&$-$\\
ME$2_{+}$ &$\nu_2<0<\mu_2$ &$-0.045\,20<\delta<-0.002\,649$ &(4s,1u)&$+$\\
ME$3_{+}$ &$0<\mu_3=\nu_3$ &$\delta<-0.000\,4732$ &(3s,0u)&$+$\\
ME$4_{+}$ & $\nu_4<0<\mu_4$&$-0.2073<\delta<-0.004\,852$ &(1s,4u)&$-$\\
ME$5_{+}$ & $\nu_5<0<\mu_5$&$-0.2073<\delta<-0.2056$ &(2s,3u)&$-$\\
\noalign{\global\arrayrulewidth1pt}
\hline
\noalign{\global\arrayrulewidth.4pt}
\end{tabular}}
\end{center}
\end{table}

We find that there are two solutions around the value $\delta \sim -0.001$
in Fig.~\ref{delta_log}.
Especially, we have the following solutions for $\delta= -0.001$:
\begin{align}
(\delta,\mu,\nu) &= \mathrm{ME}2_{+} (-0.001,\,1.437\,87,\,-0.067\,0662)\ ,
&& \mathrm{ME}3_{+} (-0.001,\,0.402\,934,\,0.402\,934 ) \ .\label{4.1.1-1}
\end{align}
It is interesting to see how the solutions change for other value of $\d$.
For instance, we have the following two solutions for $\delta= -0.1$:
\begin{align}
(\delta,\mu,\nu) & =\mathrm{ME}3_{+} (-0.1,\, 0.168\,203,\, 0.168\,203)\ ,
&& \mathrm{ME}4_{+} (-0.1 ,\,  0.742\,918,\, -0.453\,997)\ .
  \label{4.1.1-2}
\end{align}
where we use the same names for the solutions connected when $\d$ is changed.
The value of $\mu_i$ and $\nu_i$ of the solution with the same sign change
with the value of $\delta$ as in Eqs.~(\ref{4.1.1-1}) and (\ref{4.1.1-2}).
We will also study a linear perturbation around these solutions
in \S~\ref{stability1}.

\subsubsection{Power-law solutions ($\epsilon=1$)}
\label{sec4.1.2}

There is no exact solutions, but we have asymptotic solutions for various
values of $\delta$. We will give them explicitly for $\delta= -0.001$
and $\delta = -0.1$.

As $t\rightarrow\infty$ with EH dominance, we get the same solutions
$\mathrm{MF}6$ and $\mathrm{MF}7$ given by Eq.~(\ref{Kasner_sol}) in
\S~\ref{sec3.1.2}

As $t\rightarrow -\infty$, the forth order terms dominate.
We have the same Eqs.~(\ref{4thp-1}) -- (\ref{4thp-3}) in \S~\ref{sec3.1.2}
with $\delta\neq 0$.
For $\delta=-0.001 $, we find the following twelve solutions
\begin{align}
 (\delta,\mu,\nu) =\
  &\mathrm{MP}6 (-0.001, 121.218, -5.487\,83)\ ,
 &&\mathrm{MP}7 (-0.001, 27.0789, 27.0789)\ ,\nn
& \mathrm{MP}8(-0.001, 26.6578, -37.1453)\ ,
 &&\mathrm{MP}9(-0.001, 2.610\,38, -0.118\,736)\ ,\nn
 &\mathrm{MP}10(-0.001, 0.737\,553, -0.086\,3059)\ ,
 &&\mathrm{MP}11 (-0.001, 0.726\,753, -0.151\,41)\ ,\nn
 &\mathrm{MP}12(-0.001, 0.190\,928, 0.139\,524)\ ,
 &&\mathrm{MP}13 (-0.001, 0.154\,834, 0.154\,834)\ ,\nn
 &\mathrm{MP}14 (-0.001, 0.120\,104, 0.169\,926)\ ,
 &&\mathrm{MP}15(-0.001, -0.757\,551, 0.625\,032)\ ,\nn
 &\mathrm{MP}16 (-0.001, -1.161\,61, 1.497\,83)\ ,
 &&\mathrm{MP}17(-0.001, -2.407\,56, 0.598\,625)\ ,
  \label{4.1.2-1}
\end{align}
while the following eight for $\delta= -0.1$
\begin{align}
(\delta,\mu,\nu) = \
  &\mathrm{MP}6 (-0.1,\, 14.0692,\, 14.0692 )\ ,&
  &\mathrm{MP}7 (-0.1,\, 8.240\,22,\, -2.531\,52)\ ,\nn
  &\mathrm{MP}8 (-0.1,\, 0.229\,099,\, 0.151\,155)\ ,&
  &\mathrm{MP}9 (-0.1,\, 0.174\,972,\, 0.174\,972)\ ,\nn
  &\mathrm{MP}10(-0.1,\, 0.123\,489,\, 0.196\,532)\ ,&
  &\mathrm{MP}11(-0.1,\, -5.576\,78,\, 3.390\,05)\ ,\nn
  &\mathrm{MP}12(-0.1,\, -60.445,\, 36.7782)\ ,&
  &\mathrm{MP}13(-0.1,\, -225.859,\, 69.5863)\ .
  \label{4.1.2-2}
  \end{align}

\subsection{$\sigma_p =0, \sigma_q\neq 0$ (or $\sigma_p\neq 0, \sigma_q = 0$)}

\subsubsection{Generalized de Sitter solutions ($\epsilon=0$)}
\label{sec4.2.1}
\begin{figure}[htb]
\begin{minipage}{.475\textwidth}
\includegraphics[width=\linewidth, height=\linewidth]{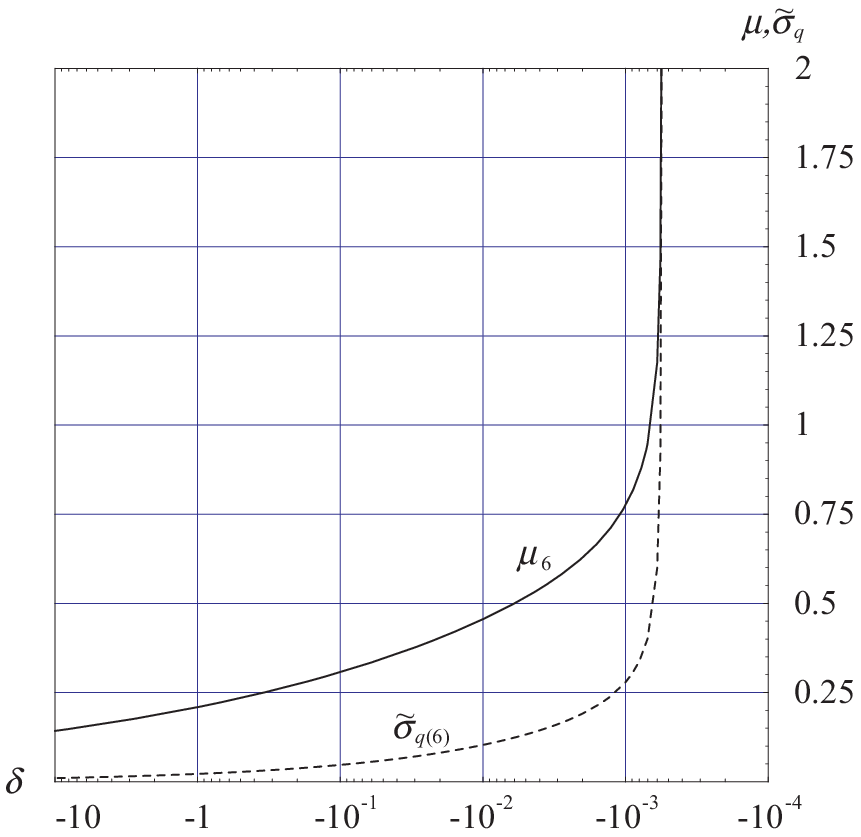}
\caption{\small{One generalized de Sitter solution with $\sigma_p=0,
\ \sigma_q\neq0$ for $\delta<0$.}}
\label{delta_mu01}
\end{minipage}\hfill
\begin{minipage}{.475\textwidth}
\includegraphics[width=\linewidth,height=\linewidth]{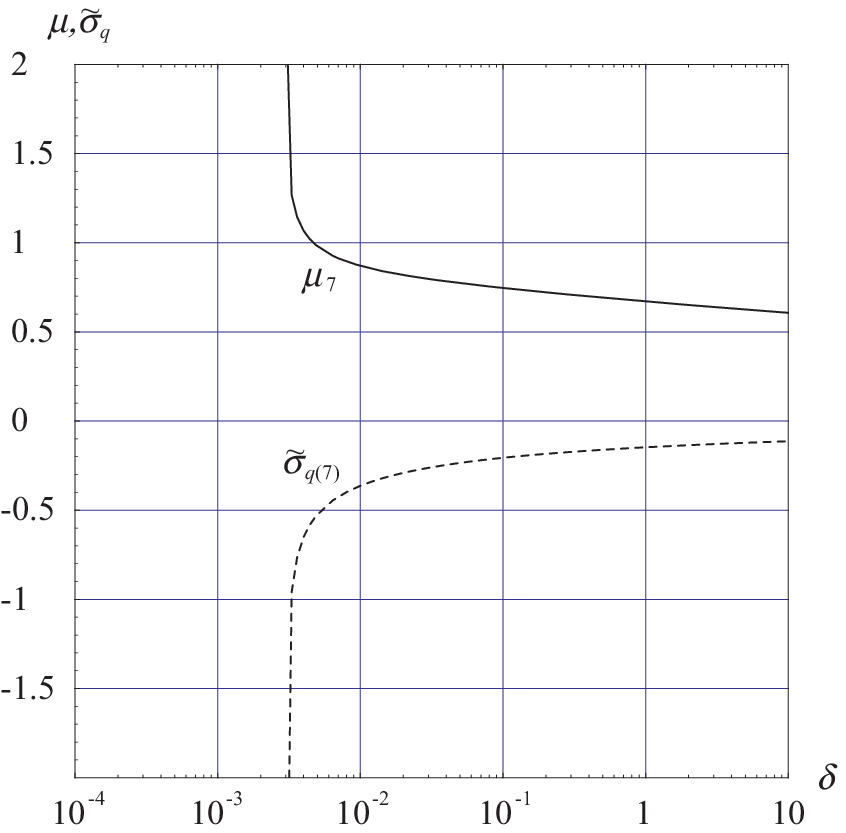}
\caption{\small{One Generalized de Sitter solution with $\sigma_p=0,
\ \sigma_q\neq0 $ for $0<\delta$.}}
\label{delta_mu02}
\end{minipage}
\end{figure}

Here we have $A_p=\mu^2, A_q=\nu^2 +\tilde \s_q e^{-2\nu t}, X=\mu^2$
and $Y=\nu^2$. It is easy to see that there is no exact solution unless
$\nu=0$, in which case we have constant $A_p=X=\mu^2, A_q=\tilde \s_q$
and $Y=0$. Our basic equations reduce to Eqs.~(\ref{3.2.1-1}) and (\ref{3.2.1-2}).
In Figs.~\ref{delta_mu01} and \ref{delta_mu02}, we depict numerical solutions
ME$i_+(\delta,\mu_i,\nu_i=0,\tilde\sigma_q)\ (i=6,7)$ with $\mu_i>0$ as a
function of $\delta$. There are also time-reversed solutions
ME$i_-(\delta,\mu'_i,\nu_i=0,\tilde\sigma_q)\ (i=6,7)$ obtained by $\mu'_i=-\mu_i$
which are not shown.
We find a solution ME$6_+$ for $\delta<0$, while ME$7_+$ for $\delta<0$.
In the vicinity of $\delta=0$, we find no
solution as discussed in \S~\ref{sec3.2.1}.
In the anterior part of Table~\ref{delta_table2}, we summarize their properties.

\begin{table}[htb]
\begin{center}
\caption{\small{Generalized de Sitter Solutions ME$i_{+}(\delta,\mu_i,\nu_i)
\ (i=6,\ldots,11)$ with $\mu_i,\nu_i\geq 0$ for various values of
$\delta$. Six eigenmodes for linear perturbations are also shown.
($m$s,$n$u) means that there are $m$ stable modes and $n$ unstable modes.}}
\label{delta_table2}
\vspace{2mm}
\small{
\begin{tabular}{lcccc}
\noalign{\global\arrayrulewidth1pt}
\hline
\noalign{\global\arrayrulewidth.4pt}
Solution & Property &Range & Stability &$3\mu_i +7\nu_i$\\ \hline
ME$6_{+}$ &$\nu_6=0,\ 0<\mu_6,\tilde\sigma_{q(6)}$ &$\delta<-0.000\,5589$ &(5s,1u)
&$+$\\
ME$7_{+}$ & $\nu_7=0, \ \tilde\sigma_{q(7)}<0<\mu_6$ & $0.002\,999<\delta$
&(4s,2u) &$+$\\ \hline
ME$8_{+}$ & $\mu_8=0,\ 0<\tilde\sigma_{p(8)},\nu_8$&$-0.003\,163<\delta<-0.000\,
5650$ &(4s,2u)&$+$\\
ME$9_{+}$ & $\mu_{9}=0,\ 0<\tilde\sigma_{p(9)},\nu_{9}$&$\delta<-0.000\,5657$
&(5s,1u)&$+$\\
ME$10_{+}$ &$\mu_{10}=0,\ \tilde\sigma_{p(10)}<0<\nu_{10}$
  &$\delta<-0.000\,043\,49$ &(5s,1u)&$+$\\
ME$11_{+}$ & $\mu_{11}=0,\ \tilde\sigma_{p(11)}<0<\nu_{11}$
  &$-0.085\,22<\delta<-0.003\,164$ &(4s,2u)&$+$\\
\noalign{\global\arrayrulewidth1pt}
\hline
\noalign{\global\arrayrulewidth.4pt}
\end{tabular}}
\end{center}
\end{table}

Especially, we find the following solution for $\delta=-0.001 $:
\begin{align}
(\delta,\mu,\nu,\tilde \sigma_q) &=
  \mathrm{ME}6_+ (-0.001,\, 0.775\,074,\,0,\, 0.278\,981)\ .\label{4.2.1-1}
\end{align}
while the following for $\delta=-0.1$
\begin{align}
(\delta,\mu,\nu,\tilde \sigma_q)&=
  \mathrm{ME}6_+(-0.1,\,\ 0.307\,198,\,0,\, 0.047\,1560)\ ,\label{4.2.1-2}
 \end{align}

\begin{figure}[htb]
\begin{center}
\includegraphics[width=.475\textwidth, height=.475\textwidth]{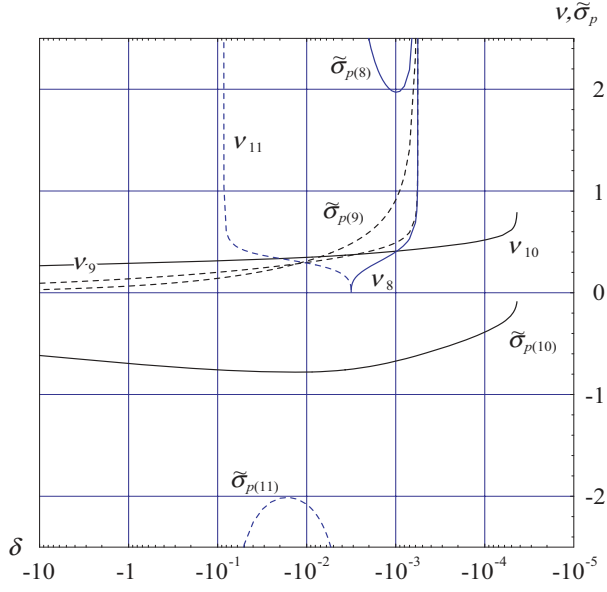}
\caption{\small{Four Generalized de Sitter solutions with $\sigma_p\neq0,
\ \sigma_q=0$ with respect to $\delta>0$.
There is no solution for $\delta \geq 0$}}\label{delta_nu0}
\end{center}
\end{figure}

For the case of $\sigma_p \neq 0$ and $\sigma_q=0$, exchanging $\mu,\ p$ and
$\nu,\ q$, we find four solutions
ME$8_+$ -- ME$11_+$. In the vicinity of $\delta=0$, we find again
no solution as discussed in \S~\ref{sec3.2.1}.
In Fig.~\ref{delta_nu0}, we depict these four numerical solutions with $\nu_i>0$
with respect to $\delta$. There are also time-reversed
solutions ME$i_-(\delta,\mu_i=0,\nu'_i,\tilde\sigma_p)\ (i=6,7)$ obtained by
$\nu'_i=-\nu_i$ which are not shown.
In the posterior part of Table~\ref{delta_table2}, we summarize their properties.

Especially, for $\delta =-0.001$ and $\delta=-0.1$, we have the following
solutions, respectively:
\begin{align}
  (\delta,\mu,\nu,\tilde\sigma_p) =\
  &\mathrm{ME}8_+ ( -0.001 ,\, 0 ,\, 0.491\,829,\, 0.911\,910)\ ,&
  &\mathrm{ME}9_+  ( -0.001 ,\, 0 ,\, 0.401\,567,\, 1.970\,26)\ , \nn
  &\mathrm{ME}10_+  ( -0.001 ,\, 0 ,\, 0.408\,387,\, -0.672\,260)\ .
  \label{4.2.1-3}\\
  &\mathrm{ME}9_+(-0.1 ,\, 0 ,\, 0.201\,057,\, 0.141\,557 )\ ,&
  &\mathrm{ME}10_+  (-0.1 ,\, 0 ,\, 0.314\,065,\, -0.757\,016)\ .
  \label{4.2.1-4}
\end{align}

\subsubsection{Power-law solutions ($\epsilon=1$)}
\label{sec4.2.2}

Here we have $A_p=\mu^2, A_q=\nu^2 +\tilde \s_q e^{2(1-\nu) t}, X=\mu(\mu-1)$
and $Y=\nu(\nu-1)$. We have only asymptotic solutions in most cases.

\begin{description}
\item[(1)] $\nu>1:$

In this case, we have same result in \S~\ref{sec3.2.2} (1), and there is
no asymptotic solution of our form.

\item[(2)] $\nu<1:$

For $t \to \infty$, $A_q \to  \tilde \s_q e^{2(1-\nu)t}$ and there is
no solution.

For $t \to -\infty$, $A_q \to \nu^2$ and the solutions are the same as
$\s_p=\s_q=0$ case in \S~\ref{sec4.1.2}. From Eqs.~\p{4.1.2-1} and
\p{4.1.2-2}, we get ten past asymptotic solutions
$\mathrm{MP}6$, $\mathrm{MP}8$ -- $\mathrm{MP}15$
and $\mathrm{MP}17$ for $\delta=-0.001$,
while the four solutions $\mathrm{MP}7$  -- $\mathrm{MP}10$ for $\delta=-0.1$.

\item[(3)] $\nu=1:$

We have $A_p=\mu^2, A_q =1+\tilde\s_q, X=\mu(\mu-1)$ and $Y=0$.

For $t \to \infty$, the EH term dominates and the solution is
\begin{align}
(\mu,\nu,\tilde\sigma_q) &= \mathrm{ME}12 (0,\, 1,\, -1)\ .
\label{4.2.2-1}
\end{align}
Actually this is an exact solution for all $\delta$.

For $t \to -\infty$, fourth-order terms dominate. Our basic Eqs.~(\ref{basic0})
and (\ref{basic2}) give the same Eqs.~(\ref{3.2.2-1}) and (\ref{3.2.2-2}) in
\S~\ref{sec3.2.2} with $\delta\neq 0$.
We get the following two solutions for $\delta=-0.001$
\begin{align}
(\delta,\mu,\nu,\tilde \sigma_q) =\
&\mathrm{MP}18 ( -0.001 ,\, 32.4979 ,\, 1 ,\, 482.327)\ ,
&& \mathrm{MP}19 ( -0.001 ,\, -47.5769 ,\, 1 ,\, -172.534)\ ,
\label{4.2.2-2}
\end{align}
while the following for $\delta=-0.1$
\begin{align}
(\delta,\mu,\nu,\tilde \sigma_q) =\
&\mathrm{MP}14(-0.1 ,\, 14.0811,\, 1 ,\, 92.0567)\ ,&
&\mathrm{MP}15 (-0.1 ,\, -31.2288,\, 1 ,\, -201.029)\ ,
\label{4.2.2-3}
\end{align}
\end{description}

For the case of $\sigma_p\neq 0$ and $\sigma_q=0$,
exchanging $\mu, p$ and $\nu, q$, we obtain one exact solution ME13 and
eleven past asymptotic solutions MP10 -- MP17 in Eq.~(\ref{4.1.2-1}) and
the following solutions MP20 -- MP 22 for $\delta=-0.001$,
while one exact solution ME13 and nine past asymptotic solutions MP8 -- MP13
in Eq.~(\ref{4.1.2-2}) and the following solutions MP16 -- MP18 for $\delta= -0.1$:
\begin{align}
  (\mu,\nu,\s_p,a_0) =\
  &\mathrm{ME}13 (1,0,-1,1)\ . \label{4.2.2-4}\\
   (\delta,\mu,\nu,\tilde \sigma_p) =\
  & \mathrm{MP}20(-0.001,\,1,\, 31.7651,\, 3563.36) \ ,&
  & \mathrm{MP}21(-0.001,\, 1,\,24.9773,\, 8412.47) \ ,\nn
  & \mathrm{MP}22(-0.001,\, 1,\,-0.841\,76,\, -1.256\,34) \ ,
  \label{4.2.2-5}\\
  & \mathrm{MP}16 (-0.1 ,\, 1 ,\,  14.0787,\, 618.521)\ ,&
  & \mathrm{MP}17(-0.1 ,\, 1 ,\, -0.634\,680,\, -1.447\,37)\ ,\nn
  & \mathrm{MP}18(-0.1 ,\, 1 ,\, -279.628,\, -1.00255 \times 10^{6} )\ .
  \label{4.2.2-6}
   \end{align}

\subsection{$\sigma_p\sigma_q\neq 0$}

\subsubsection{Generalized de Sitter solutions ($\epsilon=0$)}
\label{sec4.3.1}

If $\mu=\nu=0$, our basic equations reduce to Eqs.~(\ref{3.3.1-1}) --
(\ref{3.3.1-3}). We find again that there is no solution for $\delta\neq 0$.
If either $\mu=0$ or $\nu=0$ and the other is nonzero, it is clear that there
is no exact solution. For asymptotic solutions, we can search them by setting
$A_p=\mu^2$, $A_q=\tilde\sigma_q$, $X=\mu^2$ and $Y=0$ for the latter case.
This case is actually the same as \S~\ref{sec4.2.1}, and thus
Eqs.~(\ref{4.2.1-1}) and (\ref{4.2.1-2}) give asymptotic solutions
with $\nu=0$. For $\delta= -0.001$, we have the following future
and past asymptotic solutions:
\begin{align}
(\delta,\mu,\nu,\tilde \sigma_q) &= (-0.001,\, \pm 0.775\,074,\,0,\, 0.278\,981)
: \mathrm{MF}2(\mathrm{MP}2)
& \text{for}\  t\rightarrow \pm\infty\ .
\end{align}
For $\delta=-0.1$, we have
\begin{align}
(\delta,\mu,\nu,\tilde \sigma_q) &=
  (-0.1,\, \pm 0.307\,198 ,\, 0 ,\, 0.047\,1560) : \mathrm{MF}2 (\mathrm{MP}2)
& \text{for}\  t\rightarrow \pm\infty\ .
\end{align}

The first case is obtained by exchanging $p$, $\mu$ and $q$, $\nu$. From
Eqs.~(\ref{4.2.1-3}) and (\ref{4.2.1-4}), the solutions are
\begin{align}
  (\delta,\mu,\nu,\tilde\sigma_p) = \ & (-0.001,\,0,\, \pm 0.491\,829,\, 0.911\,910)
:\mathrm{MF}3 (\mathrm{MP}3 )\ ,\nn
& (-0.001,\, 0,\,\pm 0.401\,567,\, 1.970\,26) :\mathrm{MF}4 (\mathrm{MP}4 )\ ,\nn
& (-0.001,\,0,\, \pm 0.408\,387,\, -0.672\,260) :\mathrm{MF}5 (\mathrm{MP}5 )
& \text{for}\  t\rightarrow \pm\infty\ ,\label{futuregds1}
\end{align}
for $\delta= -0.001$ and
\begin{align}
(\delta,\mu,\nu,\tilde\sigma_p) = \ &(-0.1 ,\, 0 ,\, \pm 0.201\,057,\, 0.141\,557)
:\mathrm{MF}3 (\mathrm{MP}3 )\ ,\nn
& (-0.1 ,\, 0 ,\, \pm 0.314\,065,\, -0.757\,016) :\mathrm{MF}4 (\mathrm{MP}4 )
& \text{for}\  t\rightarrow \pm\infty\ ,\label{futuregds2}
\end{align}
for $\delta=-0.1$. Note that Eqs.~(\ref{futuregds1}) -- (\ref{futuregds2})
were exact solutions for $\sigma_p=0$ or
$\sigma_q = 0$ in \S~\ref{sec4.2.1}.

For $\mu,\nu \neq 0$, if our ansatz for solution is imposed, it is easy to
see that there is no asymptotic solution if $\mu$ and $\nu$ are of the opposite
signs. If they are of the same sign, either $t\rightarrow +\infty$ or
$t\rightarrow - \infty$ gives $A_p\rightarrow \mu^2$, $A_q\rightarrow \nu^2$
and there may be solutions which eventuate
in these of \S~\ref{sec4.1.1} with the same sign. This implies
that the inflationary solutions with positive eigenvalues are obtained for
asymptotic infinite future, so that they are not interesting from the
cosmological point of view. However, it may be useful to check if there are
any solutions of this type. In fact, Eq.~(\ref{4.1.1-1}) gives set of
asymptotic solutions for $\delta=-0.001$
\begin{align}
(\delta,\mu,\nu) &=  (-0.001,\,\pm0.402\,934,\,\pm0.402\,934 )
:\mathrm{MF}1(\mathrm{MP}1)
& \text{for}\  t\rightarrow \pm\infty\ .\label{futuregds3}
\end{align}
meanwhile, Eq.~(\ref{4.1.1-2}) gives solutions for $\delta = -0.1$ as
\begin{align}
(\delta,\mu,\nu) &= (-0.1,\, \pm 0.168\,203,\, \pm 0.168\,203): \mathrm{MF}1
(\mathrm{MP}1)
& \text{for}\  t\rightarrow \pm\infty\ ,\label{futuregds4}
\end{align}
Note that Eqs. (\ref{futuregds3}) and (\ref{futuregds4}) were exact solutions
for $\sigma_p,\sigma_q =0$ in \S~\ref{sec4.1.1}

\subsubsection{Power-law solutions ($\epsilon=1$)}
\label{sec4.3.2}

In this case, we first consider the cases when both $\mu$ and $\nu$ are
not equal to 1.
\begin{description}
\item[(1)] $\mu>1$ and $\nu>1$:

In this case, we have same result (1) in \S~\ref{sec3.3.2},
and there is no asymptotic solution of our form.

\item[(2)] $\mu<1$ and $\nu<1$:

As $t\to \infty$ with EH dominance, we again find no consistent solution.
As $t\to -\infty$ with fourth-order-term dominance, and we have same result
in \S~\ref{sec4.1.2} with $\mu<1$ and $\nu<1$.
We obtain the  seven asymptotic solutions
$\mathrm{MP}10$ -- $\mathrm{MP}15$ and $\mathrm{MP}17$
{}from Eq.~\p{4.1.2-1} for $\delta= -0.001$, while three asymptotic solutions
           $\mathrm{MP}8$ -- $\mathrm{MP}10$ from
Eq.~\p{4.1.2-2} for $\delta= -0.1$.

\item[(3)] $\mu>1$ and $\nu<1$:

The analysis is almost the same as the case (3) in \S~\ref{sec3.3.2},
and there is no asymptotic solution of our form for both $t\to \pm\infty$.

\item[(4)] $\mu<1$ and $\nu>1$:

The analysis is almost the same as the case (4) in \S~\ref{sec3.3.2},
and there is no asymptotic solution of our form for both $t\to \pm\infty$.
\end{description}

Next, we discuss the cases in which one of $\mu$ or $\nu$ is equal to
1 and the other is not:

\begin{description}
\item[(5)] $\mu>1$ and $\nu=1$:

The analysis is almost the same as the case (5) in \S~\ref{sec3.3.2},
and there is no asymptotic solution of our form for both $t\to \pm\infty$.

\item[(6)] $\mu<1$ and $\nu=1$:

As $t\to \infty$ with EH dominance, $A_p$ diverges as
$\tilde{\sigma}_p e^{2(1-\mu)t}$. There is no asymptotic solution of our form.
As $t\to -\infty$, we again recover the case of $\sigma_p=0,\sigma_q\neq 0$
with the fourth-order-term dominance and solutions in Eqs.~\p{4.2.2-1}, \p{4.2.2-2}
and~\p{4.2.2-3}. (Note that Eq.~\p{4.2.2-1} was an exact solution for $\s_p=0$.)
Choosing those with $\mu<1$, we get asymptotic power-law solutions
\begin{align}
(\mu,\nu,\tilde\sigma_q) &= \mathrm{MP}23(0,\, 1,\,-1 )\ ,
\end{align}
for all $\delta$, and $\mathrm{MP}19$ in Eq.~\p{4.2.2-2}  for $\delta = -0.001$
and $\mathrm{MP}15$ in Eq.~\p{4.2.2-3} for $\delta= -0.1$.

\item[(7)] $\mu=1$ and $\nu>1$:

The analysis is almost the same as the case (5).
There is no asymptotic solution.

\item[(8)] $\mu=1$ and $\nu<1$:

The analysis is almost the same as the case (6),
then we find the asymptotic solutions as $t\to -\infty$, which are
the same as the case of $\sigma_p\neq 0, \sigma_q= 0$ given in Eqs.~\p{4.2.2-4},
\p{4.2.2-5} and \p{4.2.2-6}. We have asymptotic power-law solutions
\begin{align}
(\mu,\nu,\tilde\sigma_p) &= \mathrm{MP}24(1,\, 0,\,-1 )\ ,
\end{align}
for all $\delta$, and $\mathrm{MP}22$ in Eq.~(\ref{4.2.2-5}) for
$\delta = -0.001$ and $\mathrm{MP}17$ and $\mathrm{MP}18$ in Eq.~(\ref{4.2.2-6}) for
$\delta= -0.1$.
\end{description}

Finally, we consider the remaining case.
\begin{description}
\item[(9)] $\mu=1$ and $\nu=1$:

Here we have constant $A_p=1+\tilde{\sigma}_p$ and $A_q=1+\tilde{\sigma}_q$.
For $t \to +\infty$, we have the same solution $\mathrm{MF}8$ in
\S~\ref{sec3.3.2}.

For $t\to -\infty$ with fourth-order-term dominance, we get same
two independent Eqs.~(\ref{3.3.2-3}) and (\ref{3.3.2-4}) in
\S~\ref{sec3.3.2}. We find solutions
           \begin{align}
(\delta,\tilde\sigma_p,\tilde\sigma_q) = \
  &\mathrm{MP}25 (-0.001,\, 116.501,\, 9.455)\ , &&\mathrm{MP}26 (-0.001,
\, 33.9247,\, 10.4469),\nn
&\mathrm{MP}27(-0.001,-0.886\,667,\, -2.590\,21)\ , &&\mathrm{MP}28(-0.001,
\, -3.020\,49,\, -1.423\,539)\ ,\\
&\mathrm{MP}25 (-0.1 ,\, 19.5872,\, 6.525\,89)\ ,
  && \mathrm{MP}26 (-0.1 ,\, -219.014 ,\, 20.5567),\nn
&\mathrm{MP}27(-0.1 ,\, -0.448\,085 ,\, -2.186\,91 )\ ,
  &&\mathrm{MP}28 (-0.1 ,\, -3.443\,20 ,\, -1.582\,47)\ .
\end{align}
\end{description}

We summarize our results obtained in this section in Appendix \ref{summary}.
Exact solutions for $\delta=-0.001$ are listed in Table~\ref{sum4},
future asymptotic solutions for in Table~\ref{sum5}
and past asymptotic solutions for in Table~\ref{sum6}.
For $\delta=-0.1$, we summarize only exact solutions in Table~\ref{sum7}.

\section{Stability Analysis of Generalized de Sitter Solutions}
\label{stability}
Since the exact generalized de Sitter solutions ME$1_\pm$ -- ME$11_\pm$
obtained in \S~\ref{sec3.1.1}, \S~\ref{sec4.1.1} and \S~\ref{sec4.2.1}
correspond to fixed points in our dynamical system, we have to analyze
their stabilities in order to see which solutions are more generic and also
to find interesting cosmological solutions. We have performed a linear
perturbation analysis around those fixed points for the solutions.
In the following subsections, we classify the result by the signature of
$\sigma_p$ and $\sigma_q$.

\subsection{$\sigma_p = \sigma_q=0$}
\label{stability1}
In this case, we have exact solutions ME$1_\pm$ -- ME$5_\pm$
obtained in \S~\ref{sec3.1.1} and \S~\ref{sec4.1.1}.
Setting
\begin{align}
\frac{d u_1^{(i)}}{d {t}} &=  {\mu}_i +A_i e^{\sigma^{(i)}  {t}}\ , \qquad
\frac{du_2^{(i)}}{d {t}}  =  {\nu}_i +B_i e^{\sigma^{(i)} {t}}\ ,
\end{align}
where $|A_i|,|B_i|\ll 1$, we write down the linear perturbation equations:
\begin{align}
F_A(\mu_i, \nu_i,\s^{(i)}) A_i + F_B(\mu_i, \nu_i,\s^{(i)}) B_i =0\ , \nn
G_A(\mu_i, \nu_i,\s^{(i)}) A_i + G_B( \mu_i, \nu_i,\s^{(i)}) B_i =0\ ,
\label{fluct}
\end{align}
obtained from Eqs.~\p{basic0} and \p{basic1}. Quantities $F_A$, $F_B$, $G_A$
and $G_B$ are functions of $\mu_i$, $\nu_i$ and $\sigma^{(i)}$ given by
\begin{align}
F_A &=
6\alpha_1\bigl[2\mu  + 7\nu\bigr]
+120960 \alpha_4 \nu^5\bigl[ 21{\mu }^2 + 14\mu \nu  + {\nu }^2 \bigr]\nn
&\quad - \frac{56\gamma(\mu-\nu)^3}{30375}\gamma\bigl[ 23652{\mu }^4
- 568368{\mu }^3\nu  + 694341{\mu }^2{\nu }^2 - 301231\mu {\nu }^3
+ 29858{\nu }^4 \nn
&\qquad  - ( 16647{\mu }^2 - 8892\mu \nu  + 3313{\nu }^2 )(3\mu  + 6\nu+\sigma )
\sigma  \bigr]\nn
&\quad -192 \delta ( 6{\mu }^2 + 21\mu \nu  + 28{\nu }^2 )^2\bigl[
24{\mu }^3 - 21{\mu }^2\nu  - 119\mu {\nu }^2 - 49{\nu }^3 \nn
&\qquad - 3( 9{\mu }^2 + 49\mu \nu  + 42{\nu }^2 ) \sigma - 3( 3\mu + 7\nu)
{\sigma}^2 \bigr]\ ,\\
F_B &=
42\alpha_1 \bigl[ \mu  + 2\nu \bigr]
+846720\alpha_4\mu {\nu }^4(\mu  + \nu  )\bigl[5\mu  + \nu \bigr]\nn
&\quad +\frac{56\gamma(\mu -\nu)^3}{30375}\gamma\bigl[
\mu(91332{\mu }^3 - 549471{\mu }^2\nu  + 545397\mu {\nu }^2 - 209006{\nu }^3 )\nn
&\qquad  - ( 16647{\mu }^2 - 8892\mu \nu  + 3313{\nu }^2 )( 2\mu  + 7\nu+\sigma)
\sigma\bigr]\nn
&\quad - 1344 \delta ( 6{\mu }^2 + 21\mu \nu  + 28{\nu }^2 )^2\bigl[
\mu ( 3{\mu }^2 - 9\mu \nu  - 49{\nu }^2 ) - (6{\mu }^2 + 45\mu \nu
+ 49{\nu }^2 ) \sigma  - (3\mu + 7\nu) {\sigma}^2 \bigr]\ ,  \\
G_A &=
4\alpha_1\bigl[3\mu  + 7\nu  + \sigma\bigr]
+80640\alpha_4 \nu^5 (3\mu +\nu)\bigl[3\mu  + 7\nu  + \sigma\bigr]\nn
&\quad - \frac{56\gamma (\mu -\nu)^2( 3\mu  + 7\nu +\sigma)}{91125}\bigl[
23652{\mu }^4 - 550629{\mu }^3\nu  + 707712{\mu }^2{\nu }^2 - 320185\mu {\nu }^3 +
39838{\nu }^4 \nn
&\qquad - ( 16647{\mu }^2 - 8892\mu \nu  + 3313{\nu }^2 )( 3\mu  + 7\nu+\sigma )
\sigma  \bigr]\nn
&\quad  -64 \delta ( 6{\mu }^2 + 21\mu \nu  + 28{\nu }^2 )^2 ( 3\mu  + 7\nu
+ \sigma  )\bigl[ 24{\mu }^2 + 21\mu \nu  - 56{\nu }^2 - 9( 3\mu  + 7\nu +\sigma )
\sigma \bigr]\ ,\\
G_B&=
14\alpha_1 \bigl[ 2\mu  + 8\nu  + \sigma \bigr]
+40320\alpha_4{\nu }^4(15{\mu }^2 + 12\mu \nu  + {\nu }^2 )\bigl[ 2\mu  + 8\nu
+ \sigma\bigr]\nn
&\quad +\frac{56\gamma (\mu -\nu)^2}{91125}\bigl[  232605{\mu }^5 + 27765{\mu }^4
\nu  - 3758859{\mu }^3{\nu }^2 + 4294423{\mu }^2{\nu }^3 - 1845390\mu {\nu }^4
+ 53336{\nu }^5 \nn
&\qquad - ( 8550{\mu }^4 + 1094319{\mu }^3\nu + 80985{\mu }^2{\nu }^2
- 156303\mu {\nu }^3 + 178861{\nu }^4 ) \sigma \nn
&\qquad - ( 16647{\mu }^2 - 8892\mu \nu  + 3313{\nu }^2 )(5\mu+15\nu +\sigma)
\sigma^2 \bigr]\nn
&\quad -448\delta ( 6{\mu }^2 + 21\mu \nu  + 28{\nu }^2 )^2 \bigl[
15{\mu }^3 + 45{\mu }^2\nu  - 56\mu {\nu }^2 - 224{\nu }^3 - (15{\mu }^2
+ 111\mu \nu  + 196{\nu }^2 ) \sigma\nn
&\qquad - 3( 5\mu  + 15\nu+\sigma ) {\sigma }^2 \bigr]\ ,
\end{align}
for $p=3,\ q=7$, where we omit the subscript of $\mu_i$, $\nu_i$ and
$\sigma^{(i)}$. The condition that Eq.~\p{fluct} has nontrivial solutions
for $A_i$ and $B_i$ is
\begin{align}
F_A( \mu_i,\nu_i,\s^{(i)}) G_B(\mu_i,\nu_i,\s^{(i)})
- F_B( \mu_i,\nu_i,\s^{(i)}) G_A( \mu_i, \nu_i,\s^{(i)}) = 0\ ,
\label{nontri}
\end{align}
which yields five modes ($\sigma=\sigma_a^{(i)},~a=1,2,\cdots, 5$) for
each solutions $i=1, \cdots, 5$ with fixed $\delta\neq 0$. This is because
the basic equations for $\dot u_1$ and $\dot u_2$ are two third-order
differential equations plus one constraint which is second order. Eqs.~\p{fluct}
are derived from Eqs.~\p{basic0} and \p{basic1},
but we have checked that the results for $\s^{(i)}_a$ remain the same
if we use any two combinations of Eqs.~\p{basic0} and \p{basic2}.
For instance, we have five modes
\begin{align}
\mathrm{ME}2_{\pm} :  \sigma^{(i)} &=
  (\mp5.364\,98 ,\, \mp3.844\,14 ,\, \mp3.773\,01
  ,\, \mp0.071\,1343 ,\,  \pm1.520\,84)\ ,\nn
\mathrm{ME}3_{\pm}:  \sigma^{(i)} &=
  (\mp 4.029\,34 ,\, \mp 3.939\,81 ,\, \mp 0.089\,5274 )
\end{align}
for $\delta = -0.001$ and
\begin{align}
\mathrm{ME}3_{\pm}: \sigma^{(i)} &=
  (\mp 1.682\,03 ,\, \mp 1.609\,88 ,\, \mp 0.072\,1502)\ ,\nn
\mathrm{ME}4_{\pm}:  \sigma^{(i)} &=
(\mp 3.82\,81 ,\, \pm 0.072\,7295 ,\, \pm 0.876\,493 ,\, \pm 0.949\,222 ,\, \pm 4.777\,32)\ ,
\end{align}
for $\delta=-0.1$. The class of solutions $\mathrm{ME}3_\pm$ has only three modes
because these solutions are special case with $\mu=\nu$.

For $\delta=0$, we have only three modes $\sigma^{(i)}{_a}\ (a=1,2,3)$ because
of the conformal invariance of the Weyl tensor.
Specifically, we have the following three modes for $\mathrm{ME}1_\pm$
\begin{align}
\mathrm{ME}1_{\pm} : \sigma^{(i)} &= (\mp 3.871\,09,\, \pm 6.242\,68,
\, \pm 10.1138)\ .
\end{align}

The numbers of stable and unstable modes for various values of $\delta$ are
summarized in Table \ref{delta_table1}.
The number of unstable modes
is important to
discuss generality of inflation.
For example, for the solution
$\mathrm{ME}4_{+}(\delta,\mu_4,\nu_4)$,
there are one stable and four unstable modes.
This implies that this solution may
not be generic because there are many unstable modes.
The probability to approach such generalized de Sitter solution
will be very low. On the other hand,
the solution$\mathrm{ME}2_+$ has four stable modes
as well as one unstable mode.
Hence, except for one direction in the phase space,
this solution is stable.
There may be a finite probability that a
generic spacetime first
approaches to this solution and eventually
evolves into other solution.
The solution $\mathrm{ME}3_+$ has three stable
modes, which means that this solution is stable against linear
perturbations. We find that preferable solutions, that is,
the solutions with many stable modes are obtained
when its 10-volume expansion rate
($3\mu_i+7\nu_i$) is positive.

\subsection{$\sigma_p =0, \sigma_q\neq 0$ (or $\sigma_p\neq 0, \sigma_q = 0$)}

We have exact solutions $\mathrm{ME}6_\pm$ and $\mathrm{ME}7_\pm$ obtained
in \S~\ref{sec4.2.1} for $\sigma_p =0$ and $\sigma_q\neq 0$.
We have also carried out the linear perturbation similar to that
in \S~\ref{stability1}. Writing
\bea
\frac{du_1^{(i)}}{dt} &=& \mu_i + A_i e^{\s^{(i)}t}, \nn
u_2^{(i)} &=& \ln b_0 + B_i e^{\s^{(i)}t},
\ena
we derive the linear perturbation equations~\p{fluct}. We find
\begin{align}
F_A &=
2 p_1 \alpha_1 \mu
+ 8 \alpha_4 \mu \bigl[p_7 \mu^6 + 3p_5q_1 \mu^4 \tilde\sigma_q +
3p_3q_3 \mu^2 \tilde\sigma_q{}^2 + {p_1}{q_5}\tilde\sigma_q{}^3 \bigr]\nn
&\quad + \frac{4pq_1\gamma \mu (\mu^2 + \tilde\sigma_q)^2}{(D-1)^3 (D-2)^4}
\bigl[2 N_1(q,p) ((p-7)\mu^2 + (p-1) \tilde\sigma_q) + ( 2N_2(p,q) - (p-3)
N_3 (q,p) )( p\mu +\sigma)\sigma \bigr]\nn
&\quad + 8p\delta \mu ((p+1)_0\mu ^2 + q_1 \tilde\sigma_q)^2\bigl[
(p-7)(p+1)_0 \mu^2  + {p_1}{q_1}\tilde\sigma_q+ 6( p^2 \mu +\sigma)\sigma
\bigr] \ , \\
F_B &= 2\alpha_1 \bigl[ pq\mu\sigma  - q_1\tilde\sigma_q\bigr]
-8 \alpha_4 \bigl[\tilde\sigma_q (p_5q_1\mu^6 +3 p_3q_3\mu^4 \tilde\sigma_q +
3p_1q_5\mu^2 \tilde\sigma_q{}^2 + q_7\tilde\sigma_q{}^3 ) \nn
&\qquad  -\sigma ( p_6q\mu^7 + 3p_4q_2\mu^5 \tilde\sigma_q + 3p_2q_4\mu^3
\tilde\sigma_q{}^2
+ pq_6 \mu \tilde\sigma_q{}^3 )\bigr]\nn
&\quad + \frac{4pq_1\gamma (\mu^2 + \tilde\sigma_q)^2} {(D-1)^3 (D-2)^4}
\bigl[ 2 N_1(q,p) \{ \tilde\sigma_q( ( 5-p )\mu^2 - (p+1)\tilde\sigma_q )  +
( q\mu^3 + (q-6) \mu \tilde\sigma_q )\sigma \}\nn
&\qquad + \{ 2 N_2 (p,q) - (p-3) N_3(q,p) \}\mu \sigma (\mu -\sigma)
( p\mu +\sigma)\bigr]\nn
&\quad + 8 \delta ((p+1)_0 \mu^2 + q_1 \tilde\sigma_q )^2\bigl[ 6q \mu\sigma^2
(p\sigma + p_1\mu)
+ p\mu \sigma \{ p(p-5)q \mu^2 + (q-6) q_1 \tilde\sigma_q\}\nn
&\qquad  - q_1 \tilde\sigma_q \{ p( p-5 ){\mu }^2 + q_1 {\tilde\sigma_q} \}
\bigr]\ ,\\
G_A &= 2p\alpha_1 \bigl[(p+1) \mu + \sigma\bigr]\nn
&\quad  +8\alpha_4 ( (p+1)\mu + \sigma  ) \bigl[
p_6\mu ^6 + 3p_4(q-1)_2 {\mu }^4 \tilde\sigma_q + 3p_2(q-1)_4 {\mu }^2
\tilde\sigma_q{}^2
+ p(q-1)_6 \tilde\sigma_q{}^3 \bigr]\nn
&\quad + \frac{4p(q-1) \gamma ((p+1)\mu + \sigma) (\mu^2 + \tilde\sigma_q )^2}
{(D-1)^3(D-2)^4}
\bigl[2N_1(q,p)( q\mu ^2 + (q-6) \tilde\sigma_q ) \nn
&\qquad -  \sigma (\sigma + p\mu) ( 2 N_2(p,q) - ( p-3 ) N_3(q,p))\bigr]\nn
&\quad + 8 p\delta ( (p+1)\mu + \sigma) ((p+1)_0 {\mu }^2 + {q_1}\tilde\sigma_q )^2
\bigl[ 6\sigma ( p\mu  + \sigma  )  + (p+1)_0{\mu }^2 +  (q -1)(q -6)
\tilde\sigma_q \bigr] \ ,\\
G_B &=2\alpha_1 \bigl[ (q-1)\sigma ( p\mu  + \sigma  )
- (q-1)_2\tilde\sigma_q \bigr]\nn
&\quad + 8\alpha_4 \bigl[ \sigma ( p\mu+\sigma)(p_5(q-1)\mu^6 + 3p_3(q-1)_3
\mu^4 \tilde\sigma_q
+ 3p_2(q-1)_5\mu ^2 \tilde\sigma_q{}^2 +(q-1)_7\tilde\sigma_q{}^3 )  \nn
&\qquad - \tilde\sigma_q\{ 3(p+1)_0(q-1)_6 \mu^2 \tilde\sigma_q{}^2 + (q-1)_8
\tilde\sigma_q{}^3
+ (q-1)_4 \mu^4( (p+1)_3 \mu^2   + 3p_2(p^2-p- 2) \tilde\sigma_q ) \} \bigr]\nn
&\quad - \frac{4p(q-1) \gamma ( {\mu }^2 + \tilde\sigma_q )^2} {(D-1)^3 (D-2)^4}
\bigl[2N_1(q,p)\{ (p+1) \tilde\sigma_q( (q-2) {\mu }^2 + (q-8) \tilde\sigma_q) \nn
&\qquad + \sigma ( p\mu  + \sigma  )( 2(2p+q-2) {\mu }^2 + (p+2q-13)
\tilde\sigma_q ) \}\nn
&\qquad -  \sigma ( p\mu  + \sigma  ) \{ 2N_2(p,q)\sigma ( p\mu  + \sigma  )
- N_3(q,p) ( ( 1 - p^2 ) {\mu }^2 + (p-3) \sigma ( p\mu  + \sigma  )  ) \}\bigr]\nn
&\quad + 8\delta ((p+1)_0\mu^2 + q_1 \tilde\sigma_q )^2
\bigl[ 6q{\sigma }^2 ( p\mu  + \sigma  )^2  +\sigma ( p\mu  + \sigma)\{
(q-1)(p+1)_0 {\mu }^2 + q_1(q-13) \tilde\sigma_q \}\nn
&\qquad -(p+1)_0(q-1)_2 \mu^2 \tilde\sigma_q - q_1(q-8) (q-1) \tilde\sigma_q{}^2
\bigr]
\end{align}
The condition for the existence of nontrivial solutions~\p{nontri} yields six modes
($\sigma=\sigma_a^{(i)},~a=1,2,\cdots, 6$) for each solution with fixed $\delta$
because we have the new variable
$u_1$ in addition to the variables in \S~\ref{stability1}.
For instance, we have five modes
\begin{align}
\mathrm{ME}6_\pm : \sigma^{(i)} &=( \mp 3.305\,74,\, \mp 2.177\,00,\,
\mp 1.162\,61 \mp 1.516\,84 i,\nn
&\quad \  \mp 1.162\,61 \pm   1.516\,84 i,\, \mp 0.148\,218,\, \pm 0.980\,516)\ ,
  \label{modes_of_ME6}
\end{align}
for $\delta = -0.001$, while
\begin{align}
\mathrm{ME}6_\pm :\sigma^{(i)} &=
  (\mp 1.342\,73 ,\, \mp 0.770\,998 ,\, \mp 0.460\,796 \mp 8.676\,06 i , \nn
&\quad \  \mp 0.460\,796 \pm 8.676\,06 i ,\, \mp 0.150\,595 ,\, \pm 0.421\,134) \ ,
\end{align}
for $\delta = -0.1$.

For $\sigma_p\neq 0$ and $\sigma_q=0$, we have four exact solutions
ME$9_\pm$ -- ME$11_\pm$ obtained in \S~\ref{sec4.2.1}.
Exchanging $\mu, p$ and $\nu, q$, we obtain the following six modes for each
solution:
\begin{align}
  \mathrm{ME}8_\pm : \sigma^{(i)}  &=(\mp4.159\,33,\, \mp3.348\,59,\,
\mp1.721\,40
\mp 2.744\,09 i,\nn
&\quad \ \mp1.721\,40 \pm  2.744\,09i,\, \mp0.094\,2078,\, \pm 0.716\,524)\
,\nn
\mathrm{ME}9_\pm : \sigma^{(i)} &= (\mp3.115\,29 \mp 1.707\,36 i,\,
\mp3.115\,29
\pm  1.707\,36 i,\, \mp2.719\,19,\nn
& \quad \ \mp0.091\,7844,\, \pm 0.304\,322\mp 1.707\,36 i,\, \pm0.304\,322
\pm 1.707\,36 i)\ ,\nn
\mathrm{ME}10_\pm :  \sigma^{(i)}&=(\mp3.304\,07,\, \mp1.429\,35 \mp 0.921\,414
i,\, \mp1.429\,35 \pm  0.921\,414 i,\nn
&\quad\ \mp1.429\,35 \mp 2.731\,49 i,\, \mp1.429\,35 \pm  2.731\,49 i,
\, \pm0.445\,363)
&&\text{for }\delta =  -0.001\ ,\\
\mathrm{ME}9_\pm : \sigma^{(i)}  &=
  ( \mp 1.733\,58 ,\, \mp 1.319\, 40 ,\, \mp 0.703\,698 \mp 15.1267 i ,\nn
&\quad\  \mp 0.703\,698 \pm 15.1267 i ,\, \mp 0.087\,9931 ,\, \mp 0.326\,182)\ ,\nn
 \mathrm{ME}10_\pm:  \sigma^{(i)} &=
( \mp 2.555\,18 ,\, \mp 2.408\,32 \mp 1.076\,03 i ,\, \mp 2.408\,32 \pm 1.076\,03 i ,\nn
&\quad \ \pm 0.209\,864 \mp 1.076\,03 i
 ,\, \pm 0.209\,864 \pm 1.076\,03 i ,\, \pm 0.356\,724)\ ,
&&\text{for }\delta =  -0.1\ .
\end{align}

The numbers of stable and unstable modes for other values of $\delta$ are
summarized in Table~\ref{delta_table2}.
Solution $\mathrm{ME}6_+$ has five stable modes and one unstable mode,
while $\mathrm{ME}7_+$ has four stable modes and two unstable modes.

\section{Solutions in Type-II Superstring}

We can discuss the case of type-II superstring in the same manner as M-theory
if we keep the dilaton field constant and ignore the contributions from other
fields. The low-energy effective actions for type-IIA and IIB superstrings
with tree and one-loop corrections are \cite{TBB}
\begin{align}
S_\mathrm{IIA} &= \frac{1}{2\kappa_{10}{^2}}\int d^{10}x \sqrt{-g}\
 \bigl[ R + \alpha' {^3} \alpha_\mathrm{II} \tilde E_8
 + \alpha' {^3} \gamma_\mathrm{II} L_W\bigr]\ ,\\
S_\mathrm{IIB} &= \frac{1}{2\kappa_{10}{^2}}\int d^{10}x \sqrt{-g}\
 \bigl[ R  + \alpha' {^3} \gamma_\mathrm{II} L_W\bigr]\ ,
\end{align}
where $\alpha_\mathrm{II}$ and $\gamma_\mathrm{II}$ are given by
\begin{align}
\alpha_\mathrm{II} &= \frac{\pi^2 g_\mathrm{s}{^2}}{3^2\cdot 2^8}\ ,&
\gamma_\mathrm{II} &= \frac{\zeta(3)}{2^3}
 + \frac{\pi^2 g_\mathrm{s}{^2}}{3\cdot 2^3}\ .
\end{align}
There is additional tree-level term $L_W$ with coefficient $\zeta(3)/2^3$
in type-IIA superstring which vanishes under the de-compactification limit
and does not exist in the M-theory action (\ref{totaction}).
The one-loop corrections in type-IIA superstring, i.e. $\tilde E_8$ and
the rest of $L_W$, are lifted to $D=11$ in the limit $g_\mathrm{s}\to\infty$ and
in agreement with M-theory corrections (\ref{S4}) and (\ref{4th1}).
We can again rescale $\alpha_\mathrm{II}$ and $\gamma_\mathrm{II}$ as
Eq.~(\ref{nonzeroc}), and have
\begin{align}
\tilde \alpha_\mathrm{II} &= \frac{1}{3\cdot 2^5}\left(1+ \frac{3\zeta(3)}{\pi^2
g_\mathrm{s}{^2}}\right)^{-1}\ ,&
\tilde \gamma_\mathrm{II} &= 1\ .
\end{align}
In the type-IIA superstring theory, we have a parameter
$0<\tilde \alpha_\mathrm{II}<1/(3\cdot 2^5)$
corresponding to $0< g_\mathrm{s} <+\infty$, whereas $\tilde\alpha_\mathrm{II}=0$
in type IIB superstring.
In the limit $g_\mathrm{s}\rightarrow\infty$, the value of
$\tilde \alpha_\mathrm{II}$ is equivalent to $\tilde \alpha_4$ in M-theory.

\begin{figure}[htb]
\begin{minipage}{.475\textwidth}
\includegraphics[width=\linewidth, height=\linewidth]{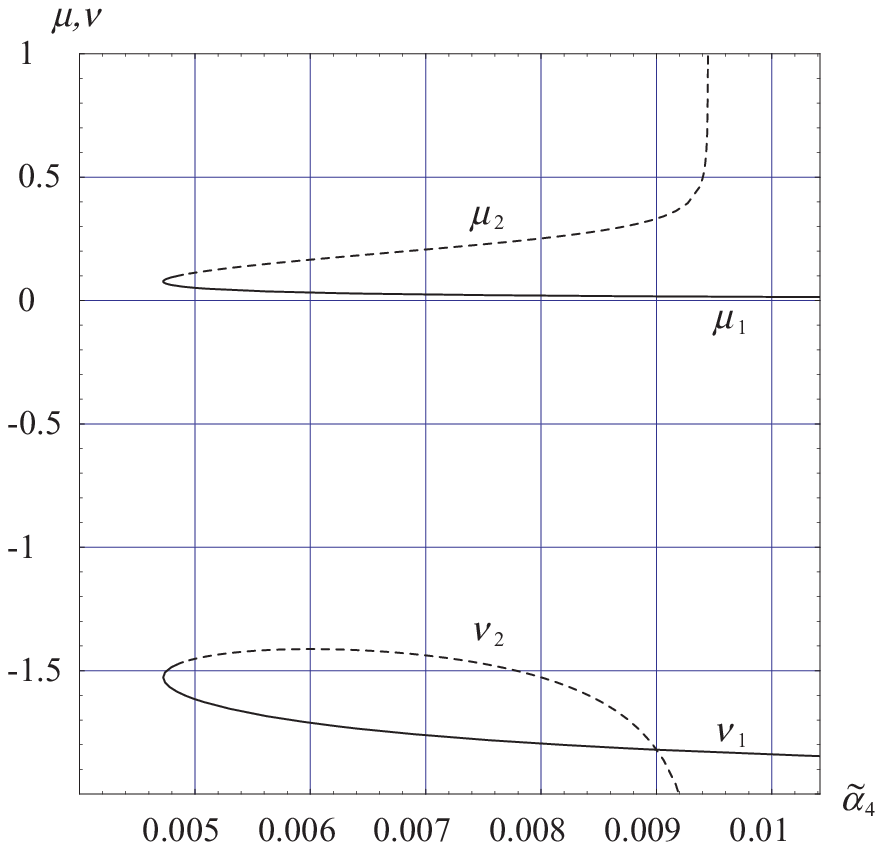}
\caption{\small{Two generalized de Sitter solutions for type II superstring
with $\sigma_p=\sigma_q =0$  with respect to $0\leq \tilde\alpha_4 \leq
 3^{-1}2^{-5}$. $\tilde\alpha_4=0$ corresponds to the type IIB superstring.}}
\label{IIf1}
\end{minipage}\hfill
\begin{minipage}{.475\textwidth}
\includegraphics[width=\linewidth,height=\linewidth]{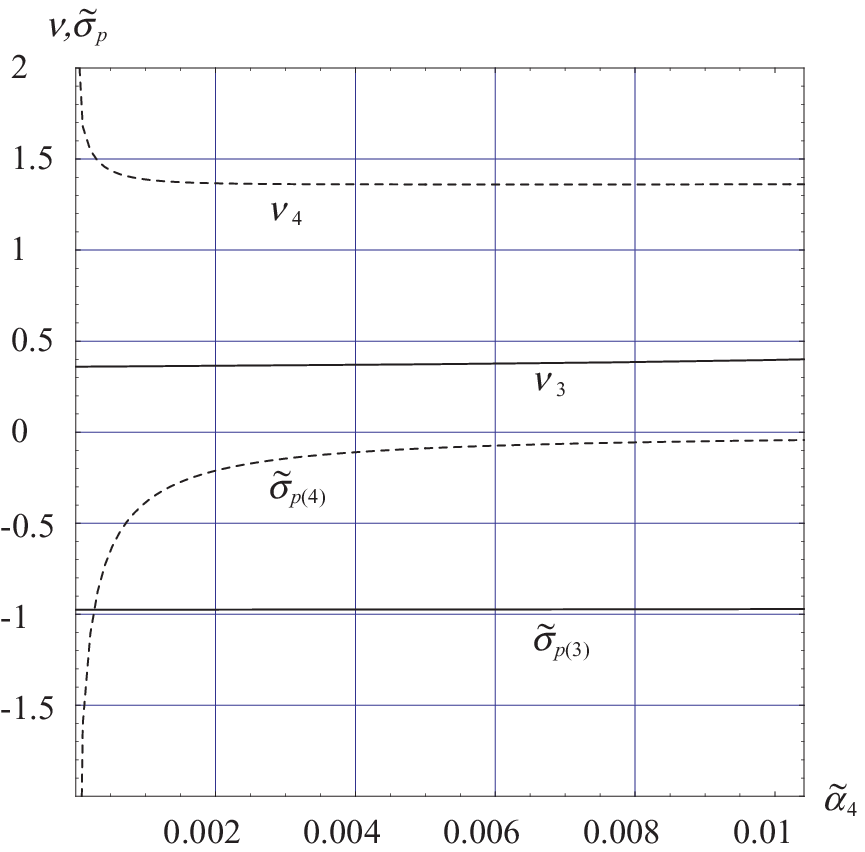}
\caption{\small{Two Generalized de Sitter solutions  for type II superstring
with $\sigma_p\neq0,\ \sigma_q=0$ with respect to
$0\leq \tilde\alpha_4 \leq 3^{-1}2^{-5}$. $\tilde\alpha_4=0$ corresponds
to the type IIB superstring.}}
\label{IIf2}
\end{minipage}
\end{figure}

We can search the exact and asymptotic solutions in a same way in
\S~\ref{sec3},
however we restrict ourselves to exact generalized de Sitter solutions in the
following discussion.
We have found four classes of generalized de Sitter solutions for
$\sigma_p=\sigma_q=0$ and $\sigma_p\neq0,\ \sigma_q=0$
in Type IIA superstring while one solution for  $\sigma_p\neq0,\ \sigma_q=0$
in Type IIB superstring. There is no solution in other cases.

\subsection{$\sigma_p=\sigma_q=0$}

In this case, we have the same Eqs.~(\ref{3.1.1-1}) and (\ref{3.1.1-2})
with $\delta=0$. Exchanging $\alpha_4, \ \gamma$ and
$\tilde \alpha_\mathrm{II}, \ \tilde\gamma_\mathrm{II}$ and setting $q=6$,
we may solve these equations numerically.
In Fig.~\ref{IIf1}, we give numerical solutions $\mathrm{IIE}i_+
(\tilde\alpha_\mathrm{II},\mu_i,\nu_i)\ (i=1,2)$
with $\mu_i>0$ with respect to $\tilde\alpha_\mathrm{II}$, and summarize
their properties in Table~\ref{IIt}.
We have two generalized de Sitter solutions $\mathrm{IIE}1_\pm$ and
$\mathrm{IIE}2_\pm$ for type IIA superstring, whereas no solution for type
IIB superstring, i.e. $\tilde\alpha_\mathrm{II}=0$.

\subsection{$\sigma_p=0,\ \sigma_q\neq0$ (or $\sigma_p\neq0,\ \sigma_q=0$ )}

It is easy to see that there is no exact solution unless $\nu=0$ in which
case we have constant
$A_p=X=\mu^2, \ A_q =\tilde\sigma_q$.
Our basic equations reduce to Eqs.~(\ref{3.2.1-1}) and (\ref{3.2.1-2}) with
$\delta=0$. Exchanging $\alpha_4, \ \gamma$ and
$\tilde \alpha_\mathrm{II}, \ \tilde\gamma_\mathrm{II}$
and setting $p=3,\ q=6$,
we solve these equations numerically and find that there is no solutions
in the range $0\leq\tilde \alpha_\mathrm{II}\leq1/(3\cdot 2^5)$.
Thus we have no solution of this class for both type IIA and type
IIB superstring.

For the case of $\sigma_p\neq 0$ and $\sigma_q=0$, exchanging $\mu,\ p$ and
$\nu,\ q$, we find two solutions $\mathrm{IIE}3_\pm$ and $\mathrm{IIE}4_\pm$.
In Fig.~\ref{IIf2}, we give numerical solutions $\mathrm{IIE}i_+
(\tilde\alpha_\mathrm{II},\mu_i,\nu_i)\ (i=3,4)$
with $\mu_i>0$ with respect to $\tilde\alpha_\mathrm{II}$, and summarize
their properties in Table~\ref{IIt}.
\begin{table}[htb]
\begin{center}
\small{
\caption{\small{Generalized de Sitter Solutions IIE$1_{+}$ -- IIE$4_{+}$
with $\mu_i(\nu_i)\geq 0$ for $0<g_s<+\infty$. Five eigenmodes for linear
perturbations are also shown. ($m$s,$n$u) means that there are $m$ stable
modes and $n$ unstable modes. }}
\vspace{2mm}\label{IIt}
\begin{tabular}{lcccc}
\noalign{\global\arrayrulewidth1pt}
\hline
\noalign{\global\arrayrulewidth.4pt}
Solution & Property &Range & Stability & $3\mu_i +7\nu_i$\\ \hline
IIE$1_{+}$ & $\nu_1<0<\mu_1$ & $-0.0047\,24 < \tilde\alpha_4 \leq 3^{-1}2^{-5}
\ (0.5506 < g_\mathrm{s} < +\infty)$ &(1s,2u) &$-$\\
IIE$2_{+}$ &$\nu_2<0<\mu_2$ &$-0.0049\,49 < \tilde\alpha_4  < 0.009\,447
\ (0.5751 < g_\mathrm{s} < 1.887)$  &(0s,3u)&$-$\\
\hline
IIE$3_{+}$ & $\mu_3 = 0,\ \tilde\sigma_{p(3)}< 0 <\nu_3$
&$0\leq \tilde\alpha_4 \leq 3^{-1}2^{-5}\ (0<g_\mathrm{s}<+\infty)$ &(2s,2u)&$+$\\
IIE$4_{+}$ & $\mu_4 = 0,\ \tilde\sigma_{p(4)}< 0 <\nu_4$
&$0<\tilde\alpha_4 \leq 3^{-1}2^{-5}\ (0<g_\mathrm{s}<+\infty)$ &(3s,1u)&$+$\\
\noalign{\global\arrayrulewidth1pt}
\hline
\noalign{\global\arrayrulewidth.4pt}
\end{tabular}}
\end{center}
\end{table}

We have two generalized de Sitter solutions $\mathrm{IIE}3_\pm$ and
$\mathrm{IIE}4_\pm$ for type IIA
superstring, whereas one solution $\mathrm{IIE}3_\pm$ for type IIB superstring.

\section{Inflationary Scenario in M-theory}
We are ready for discussion about inflation in M-theory.
Since our world is 4-dimensional, we should also
analyze our solutions in 4-dimensional Einstein frame,
in which the Newtonian gravitational constant is constant.
\subsection{Description of solutions in the Einstein frame}
\label{s-and-d}
We have found generalized de Sitter solutions
\begin{align}
 a = a_0 e^{\mu t}\ ,\qquad b=b_0 e^{\nu t}\ ,\qquad {\rm for}\ \epsilon=0\ ,
\label{gdS}
\end{align}
and power-law solutions
\begin{align}
a=a_0 \tau^{\mu}\ , \qquad b=b_0
\tau^{\nu}\ ,\qquad {\rm for}\ \epsilon=1\ .
\label{power1}
\end{align}
These solutions give the power-law expansion and the exponential expansion
in the Einstein frame of our 4-dimensional spacetime~\cite{MO2}.

\begin{itemize}
\item  \textbf{Generalized de Sitter solutions ($\epsilon =0$)}\\
For $\nu \neq 0$, we have the power-law expansion in the Einstein frame.
\begin{align}
  a_\mathrm{E} \equiv \exp \biggl[u_1 + \frac{q}{p-1}u_2\biggr] &\propto t_\mathrm{E}{^\lambda}\ , &
  \lambda &= 1+\frac{(p-1)\mu}{q\nu}\ ,&
t_\mathrm{E} &= t_\mathrm{E}{^{(0)}} \exp\left[\frac{q}{p-1}\nu t\right]\ ,
  \label{lambda-gds}
  \end{align}
where for $\nu >0 $, we have $t_\mathrm{E}{^{(0)}} >0$ and $t\in (-\infty ,\infty)$ is transformed into
$t_\mathrm{E} \in (0,\infty)$, whereas for $\nu <0 $ we have $t_\mathrm{E}{^{(0)}} <0$ and $t\in (-\infty ,\infty)$ is transformed into
$t_\mathrm{E} \in (-\infty , 0)$. For $\nu = 0$, we have the exponential expansion in the Einstein frame
 \begin{align}
    a_\mathrm{E} & \propto \exp[\mu t_\mathrm{E}]\ , &
t_\mathrm{E} &= t\ , &
  \end{align}

  \item  \textbf{Power-law solutions ($\epsilon =1$)}\\
 For $\nu \neq -(p-1)/q$, we have the power-law expansion in the Einstein frame
\begin{align}
  a_\mathrm{E} &\propto t_\mathrm{E}{^\lambda}\ , &
  \lambda &= \frac{(p-1)\mu + q\nu}{(p-1)+ q\nu}\ ,&
t_\mathrm{E} &= t_\mathrm{E}{^{(0)}} \exp\left[\left(1+\frac{q}{p-1}\nu\right) t\right]\ ,
  \label{lambda-pl}
  \end{align}
where for $\nu > -(p-1)/q$, we have $t_\mathrm{E}{^{(0)}} >0$ and
$t\in (-\infty ,\infty)$ is transformed into
$t_\mathrm{E} \in (0,\infty)$, whereas for $\nu < -(p-1)/q$, we have
$t_\mathrm{E}{^{(0)}} <0$ and $t\in (-\infty ,\infty)$ is transformed into
$t_\mathrm{E} \in (-\infty , 0)$. For $\nu = -(p-1)/q$, we have the
exponential expansion in the Einstein frame
 \begin{align}
    a_\mathrm{E} & \propto \exp[(\mu-1) t_\mathrm{E}]\ , &
t_\mathrm{E} &= t\ ,
  \end{align}
\end{itemize}

We list the behavior of a scale factor and show the condition
for inflation in the
Einstein frame in Table~\ref{einsteinframe}.
Note that the values of $\mu$ and $\nu$ in generalized de Sitter solutions
(\ref{gdS}) depend on the choice of the unit. In the M-theory,
we use the unit of $|\gamma|=1$, i.e.
$m_{11}=6^{-1/2}(4\pi)^{-5/9}\sim 0.1818176$.
If we set $m_{11}=1$, the values of $\mu$ and $\nu$ in the following
tables should be multiplied by the factor
$6^{1/2}(4\pi)^{5/9}\sim 5.5$.
On the other hand, the power exponent $\mu$ and $\nu$ in the power-law
solutions (\ref{power1}) or $\lambda$ in Eqs.~(\ref{lambda-gds}) and
(\ref{lambda-pl}) are dimensionless and
they do not depend on the choice of the unit.

\begin{table}[htb]
\begin{center}
\caption{\small{Behavior of solutions in the Einstein frame; ``Condition''
means condition for causing inflation in the
  Einstein frame, while the super-inflationary solution behaves as
 $|t_\mathrm{E}|^{-|\lambda|}$
  for $t_\mathrm{E}\rightarrow 0_-$.}}
\label{einsteinframe}
\vspace{2mm}
\small{
\begin{tabular}{c|llccc}
\noalign{\global\arrayrulewidth1pt}
\hline
\noalign{\global\arrayrulewidth.4pt}
   &  & Scale Factor & Condition & Range of $t_\mathrm{E}$ & Type of inflation \\
\hline
 & $\nu > 0$ & $a_\mathrm{E} \propto t_\mathrm{E}{^\lambda}$ & $\mu / \nu>0$ &
$(0,\infty)$ & power-law \\
  $\epsilon =0 $ & $\nu = 0$ & $a_\mathrm{E} \propto \exp[\mu t_\mathrm{E}]$ &
$\mu>0$ & $(-\infty,\infty)$ & exponential \\
     & $\nu < 0$ & $a_\mathrm{E} \propto t_\mathrm{E}{^\lambda}$ & $\lambda<0$
 & $(-\infty,0)$& superinflation \\ \hline
 & $\nu > -(p-1)/q$ & $a_\mathrm{E} \propto t_\mathrm{E}{^\lambda}$ & $\mu>1$
& $(0,\infty)$ & power-law \\
 $\epsilon = 1$  & $\nu = -(p-1)/q$ & $a_\mathrm{E} \propto \exp[(\mu-1)
 t_\mathrm{E}]$ & $\mu>1$ & $(-\infty,\infty)$ &
  exponential \\
  & $\nu < -(p-1)/q$ & $a_\mathrm{E} \propto t_\mathrm{E}{^\lambda}$ &
$\lambda <0$ & $(-\infty,0)$ & superinflation \\
\noalign{\global\arrayrulewidth1pt}
\hline
\noalign{\global\arrayrulewidth.4pt}
\end{tabular}}
\end{center}
\end{table}

\subsection{Conditions for successful
inflation and some preferable solutions}
\label{coso}
Before we apply our solutions to cosmology, we have to specify what kind of
solutions we need. We list necessary conditions for successful
inflation  in our model below.

\begin{itemize}
\item[(1)] $\mu > \nu$ and $\mu \geq 0$:\\
Our four-dimensional universe makes sense only if it is much larger than the
internal space, so the external space should expand faster than the internal
space. Its expansion may not be necessarily inflationary, but at
least the external space must be expanding in the whole space.
Some solutions give an inflation in the Einstein frame but the external space
shrinks in the original higher dimensions. Such solutions are not suitable
for a good cosmological model.
\item[(2)] 60 e-foldings of inflationary expansion:\\
We need at least 60 e-foldings of inflationary expansion in the Einstein
frame. This may give some constraint on the power exponent for a power-law
inflation, that is the power exponent should be significantly larger than unity.
Specifically, for generalized de Sitter solution and power-law solution, we find
that the number of e-foldings $N_\mathrm{E}$ in the Einstein frame is given by
\begin{align}
N_\mathrm{E} &= \ln \frac{a_\mathrm{E}(t_\mathrm{fin})}{a_\mathrm{E}
(t_\mathrm{ini})} = \left(\mu +
\frac{q\nu}{p-1}\right)(t_\mathrm{fin} - t_\mathrm{ini}) \ ,
\end{align}
where $t_\mathrm{ini}$ is the beginning time of inflation in the original
frame whereas $t_\mathrm{fin}$ is the ending time.
\item[(3)] (semi)stability against the linear perturbation:\\
Solutions which have many unstable modes may not be generic. The most
preferable solutions in this model are those with only stable modes, which means
that the solution is stable against linear perturbations. However, this kind of
solution predicts that inflation never ends because it is stable.
For a realistic cosmological model, the solution must have small instability
for the inflation to end. On the other hand, we also want such a solution to be
rather generic which requires some sort of stability. This would be achieved
if the solution contains only one small unstable mode and many other stable modes,
and then the generic spacetime may first approach this solution and gradually
leave it.
\end{itemize}

We summarize our solutions in the Einstein frame
as well as those in the original frame
in Appendix \ref{summary}.
Using the tables in Appendix \ref{summary},
we shall pick up the preferable solutions for inflation.


For $\delta=0$, although we find exponential expansion of the external space
in the original frame, this gives non-inflationary power-law expansion
in the Einstein frame as described above.
There are power-law inflations (MP1, MP11, MP12 and MP13) in the past regime.
However, the power exponent of them are 1.3 - 2.3,
which may be too small to solve the flatness and horizon problems, because
we do not expect the expansion in these solutions continues so long.
Thus, these solutions are excluded by the condition (2).

For $\delta=-0.001$, we have six candidates [ME$2_+$, ME$3_+$ (MF1),
 ME$6_+$ (MF2),
MP6 and MP7]. Among these, the condition (1) exclude solutions
ME$3_+$ (MF1) and MP7 for the internal space expands at the same rate as
the external space. Solutions ME$2_+$ and MP6 give super-inflation in
the Einstein frame. Especially, the solution
ME$2_+$ has only one unstable mode, and fulfills the condition (3).
In this case, we come close to the singularity at $t_\mathrm{E}=0$, but
we hope that stringy effects renders the singularity harmless when
the curvature becomes large.
For the remaining solution ME$6_+$, we find an exponential
expansion of the external space both in the original and the Einstein
frames, and the internal space is static, viz. modulus is fixed.
{}From the Table \ref{delta_table2}, we also find that this solution fulfills
the condition (3) and exists for wide range of $\delta$.
The radius of the external space is arbitrary whereas that of the internal
space is $b_0 = 1.893 = 0.3442\ m_{11}{^{-1}}$.

The above solution ME$6_+$ has one unstable mode so this is
the most interesting cosmological solutions in our criterion.
But the instability
$\sigma_5^{(i)}$ is the same order of magnitude as other eigenvalues of
stable modes as seen from Eq.~(\ref{modes_of_ME6}) and appears a little too
large to give enough expansion. Specifically, the time scale in which this
unstable mode becomes important is evaluated as
$t_\mathrm{us} \sim (\sigma_5^{(6)})^{-1} \sim 1.020$.
The number of e-foldings will be given by
$N_\mathrm{E}= \mu t_\mathrm{us} \sim 0.7905$, except for some
fine-tuned initial conditions.

If the eigenvalue of the unstable mode is much smaller than those of other
four stable modes, however, a preferable solution is naturally obtained for
a wide range of initial conditions. We may not need a fine-tuning.
Although we do not find
any value of $\delta$ which gives enough small
eigenvalue of an unstable mode,
we note that our starting Lagrangian has some ambiguity, that is,
the forth-order correction term $S_W$ is fixed up to
the Ricci curvature tensors. We have included additional quartic Ricci scalar
term~(\ref{R4}) in order to take this ambiguity into account,
but we may still have another kind of correction term including
the Ricci curvature tensors. To further improve the solutions, we have
the possibility of finding more interesting solutions with these appropriate
extra corrections.

Another interesting possibility is the following. We have seen that
many of our exact solutions have unstable modes as well as stable modes.
But it is not immediately obvious what happens to any inflationary solutions
after the solutions decay into other solutions. Our above analysis appears to
indicate that we cannot obtain big enough e-foldings before the exact solutions
decay. However there is a possibility that we may obtain enough e-foldings
after the decay if we can follow the evolution of our solutions.
Indeed we will show in the next section that we obtain some numerical
solutions which first approach ME$6_+$ and give enough e-foldings
if we fine-tune the initial data.

\subsection{Numerical analysis for generic initial conditions}

To study whether we obtain a sufficient e-foldings
in the present model, we have performed the numerical calculation around the
solution ME$6_+$ for $u_0 = 0,\ \delta=-0.1$ and $\sigma_q \neq 0$.
In this case, Eqs.~(\ref{basic1}) and (\ref{basic2}) give evolution equations
for $\dddot u_1$ and $\dddot u_2$, and
Eq.~(\ref{basic0}) gives a constraint equation for seven variables
$\dddot u_1,\ \dddot u_2,\ \ddot u_1,\ \ddot u_2,\ \dot u_1,\ \dot u_2$ and $u_2$,
which we used to check the numerical error. Thus, we can give six independent
initial values and the remainder is given through the constraint equation.
Under separate five sets of initial conditions, we have performed the numerical
calculations for seven dynamical variables.

\begin{figure}[htb]
\begin{minipage}{.475\textwidth}
\includegraphics[width=\linewidth, height=\linewidth]{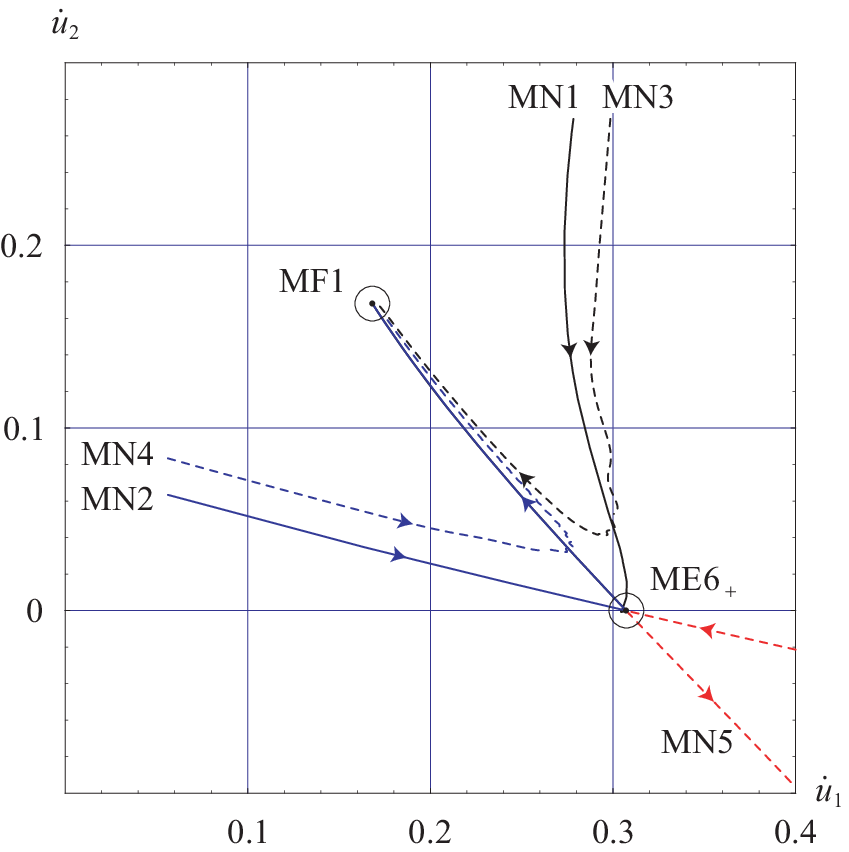}
\caption{\small{Numerical results around ME$6_+$ in the $\dot u_1$-$\dot u_2$
plane for $\delta=-0.1$. Solutions ME$6_+$ and MF1 are indicated by a point
in a circle. }}
\label{u1u2}
\end{minipage}\hfill
\begin{minipage}{.475\textwidth}
\includegraphics[width=\linewidth,height=\linewidth]{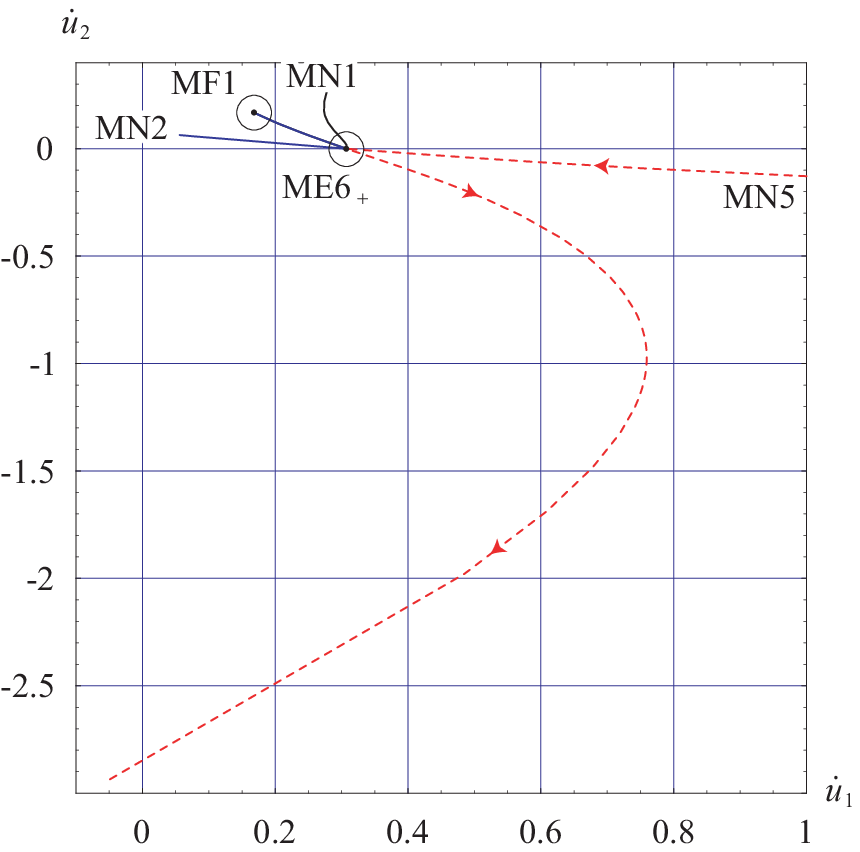}
\caption{\small{Numerical results around ME$6_+$ in the $\dot u_1$-$\dot u_2$
plane. The solution MN5 runs away to $\dot u_1 , \dot u_2 \rightarrow -\infty$
in the last phase.}}
\label{u1u22}
\end{minipage}\\
\\
\small{We give the initial values as follows:
$b=3.7,\ \dot u_1 = 0.28,\ \dot u_2 = 0.27,\ \ddot u_1 = -0.060,\ \ddot u_2 = -0.46,
\ \dddot u_1 = 0.47$ for MN1,
$b=4.4,\ \dot u_1 = 0.056,\ \dot u_2 = 0.063,\ \ddot u_1 = 0.24,\ \ddot u_2 = 0.068,
\ \dddot u_1 = -0.15$ for MN2, same as MN1 except for $\dot u_1 =0.30$ for MN3,
same as MN2 except for $\dot u_2 = 0.083$ for MN4,
$b=5.0,\ \dot u_1 = 1.0,\ \dot u_2 = -0.13,\ \ddot u_1 = -1.6,\ \ddot u_2 = 0.20,
\ \dddot u_1 = 4.7$ for MN5.}
\end{figure}

\begin{figure}[htb]
\begin{minipage}{.475\textwidth}
\includegraphics[width=\linewidth, height=\linewidth]{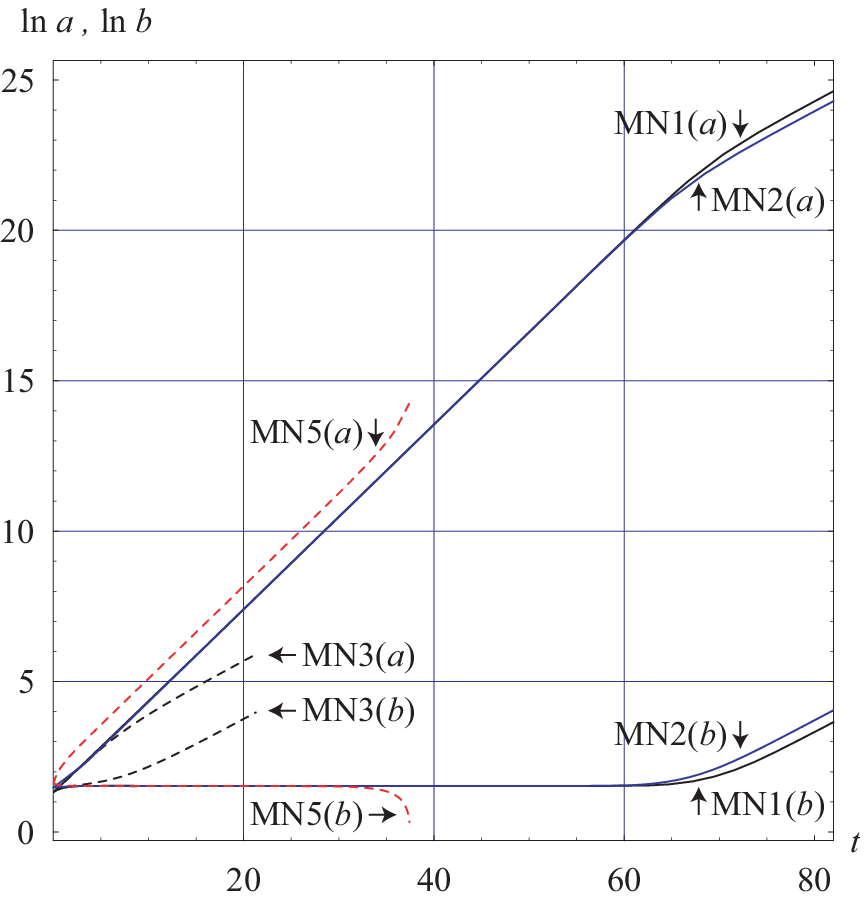}
\caption{\small{Numerical results in the original frame for $\delta=-0.1$. MN$i(a)$
means the plot of $\ln a$ of the solution MN$i$. We omit the solution MN4.}}
\label{tu1u2}
\end{minipage}\hfill
\begin{minipage}{.475\textwidth}
\includegraphics[width=\linewidth,height=\linewidth]{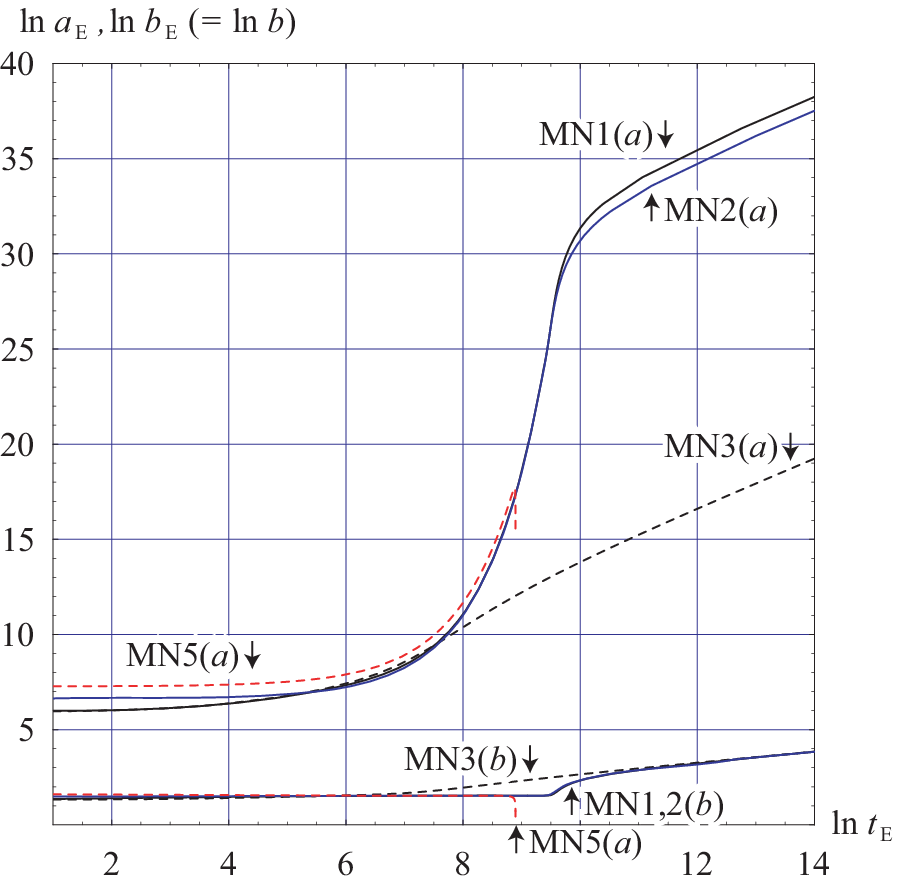}
\caption{\small{Numerical results in the Einstein frame for $\delta=-0.1$.
MN$i(a)$ means the plot of $\ln a_\mathrm{E}$ of the solution MN$i$.
We omit the solution MN4.}}
\label{tEaEbE}
\end{minipage}
\end{figure}

In Figs.~\ref{u1u2} -- \ref{tEaEbE}, we depict five numerical solutions MN1 -- MN5
whose initial values lie in the vicinity of ME6$_+$. In Figs.~\ref{u1u2} and
\ref{u1u22}, we show the behavior of solutions in the $\dot u_1$-$\dot u_2$
plane. Generalized de Sitter solutions are expressed as a point
$(\dot u_1, \dot u_2) = (\mu,\nu)$ in this plane.
In these figures, we have the exact solution ME$6_+$ in Eq.~(\ref{4.2.1-2}) and
the future asymptotic solution MF1 in Eq.~(\ref{futuregds4}) (MF1 is exact
for $\sigma_q=0$). Every solution approaches the ME$6_+$ in the early phase
whereas, in the last phase, solutions MN1 -- MN4 approach the future asymptotic
solution MF1 and the solution MN5 goes to the singularity
$\dot u_1 , \dot u_2 \rightarrow -\infty$ for finite lengths of time as shown
in Fig.~\ref{u1u22}.

In Figs~\ref{tu1u2} and \ref{tEaEbE}, we show the behavior of the scale factors
in the original frame ($a,\ b$) and in the Einstein frame
($a_\mathrm{E},\ b_\mathrm{E}$), respectively. Here we set $a_0=b_0$ and
$t_\mathrm{E}(t=0)=1$. In the early phase, every
solution gives exponential expansion $a_\mathrm{E}\propto e^{0.1682 t_\mathrm{E}}$
arising out of ME$6_+$ both in the original frame and Einstein frame, while
solutions MN1 -- MN4 give power-law expansion $a_\mathrm{E}\propto
t_\mathrm{E}{}^{1.286}$ arising out of MF1 in the Einstein frame in the late phase.
The solution MN4 is not shown in the figures since it does not lead to
interesting result.

Despite the analysis in \S~\ref{coso}, we have enough e-foldings around the
solution ME$6_+$ with appropriate initial values. Especially, numerical
solutions MN1 and MN2 have interesting features.
In the early phase, they give the exponential expansion in the external space,
and the power-law expansion in the late
phase as shown in Fig.~\ref{tEaEbE}. They do not give enough e-foldings around
the solution ME6$_+$ ($N_\mathrm{E}\sim 30$).
However the solution MF1, which is a stable solution,
does also give inflation.
Then after leaving the solution ME6$_+$, a spacetime approaches
to the solution MF1 in the late phase,
which totally gives 60 e-foldings
($N_\mathrm{E}=60$) at
$\ln t_\mathrm{E} = 35.50\ (t=119.7)$ for MN1 and $\ln t_\mathrm{E} =
36.62\ (t=119.2)$ for MN2. The scale of
internal space becomes $R_0 = b = 2.182 \times 10^4 = 3967\ m_{11}{^{-1}}$
for MN1 and
$R_0 = 2.995 \times 10^4 = 5446\ m_{11}{^{-1}}$ for MN2 both in the original
and Einstein frames.
After inflation, if the internal space settles down to static one, the present
radius of extra dimensions is given by $R_0$.
Since this is slightly larger than the fundamental scale length, we may adopt
the model of large extra dimensions, which
was first proposed as a brane world by Arkani-Hamed et al.
\cite{Arkani-Hamed:1998rs}. In this model, the 4D Planck mass is given by
$m_4 {^2} = R_0{^7}m_{11}{^9}$. We then find
\begin{align}
m_{11} &= 2.543 \times 10^{-13}\  m_4  = 619.5\ \mathrm{TeV}\ ,
 & \text{for MN1}\ ,\\
m_{11} &= 8.392 \times 10^{-14}\  m_4  = 204.4\ \mathrm{TeV}\ ,
 & \text{for MN2}\ .
\end{align}
This is our fundamental energy scale. The scale of extra dimensions is
$R_0=3967 \ m_{11}{^{-1}} = 6.403\ \mathrm{TeV}^{-1}$ for MN1 and
$R_0=5446 \ m_{11}{^{-1}} = 26.64\ \mathrm{TeV}^{-1}$ for MN2.

Although numerical solutions MN1 and MN2 give interesting scenario, their
behaviors highly  depend on choice of initial values.
For example, numerical solutions MN3 and MN4 have almost the same initial
conditions as those of MN1 and MN2 (see Fig.~\ref{u1u2}), but their
final behaviors are very different (Fig.~\ref{tEaEbE}).
We can find  60 e-foldings
at $\ln t_\mathrm{E} = 49.97\ (t=83.47)$ for MN3,
and the scale of internal space becomes $R_0 = 1.349 \times 10^6 = 2.463 \times 10^5\ m_{11}{^{-1}}$.
Thus, if the extra dimension is so large,
the 11D Planck mass turns out to be
 $m_{11}=0.3285 \ \mathrm{GeV}$, which is excluded
by high-energy experiments.

\section{Concluding remarks}

In this paper, we have examined cosmological
solutions of the effective theories
of M-theory and superstrings with special attention to the
de Sitter-like and
power-law expansions. In order to evade the no-go theorem in this
setting, we have included higher order quantum corrections. We have also taken
the ambiguity arising from the field redefinition into account.
We have found that there is an interesting solution ME$6_+$
in which the external
space is expanding whereas the size of the internal space is static.
This is interesting in another respect that this may be regarded as ``moduli
stabilization'' by the higher order corrections.
It is true that this solution does not realize the usual ``moduli stabilization''
where all moduli are fixed and stable because we have one unstable mode
in the fluctuations around this solution. But if the solution were completely
stable, the inflation would not end. On the other hand, it is necessary that
the size of the internal space does not grow much during inflation.
So the feature of this solution is desirable for giving successful inflation.
Naively it may be expected
that there are many inflationary solutions. It is somewhat surprising
to find that we have found few solutions with desirable features.

Quite interestingly,
we find some numerical solutions which first approach an exact inflationary
solution and then give enough e-foldings if we fine-tune the initial data.
The key to understand this result is that we cannot see how the solution
evolves after the exact solution decays if we are looking only at the analytic
solutions. We were able to follow the evolution by the numerical analysis.
Even though it appears that this may need some special initial conditions,
we find that this opens another possibility of achieving successful inflation
and is a very interesting result.

There is also still ambiguity in other terms involving
Ricci tensors and scalar curvatures in the effective theory, which we have not
analyzed. So rather than taking our results literally, we should understand
these results as an indication that such inflationary solutions are possible in
this direction. It is certainly an interesting problem to see if our results
may be further improved by including other possible quantum corrections.
We also have to look at the effects of fluxes, which play important roles
in inflationary models with an effective potential~\cite{KKLMMT}.
These may lead to successful inflation insensitive to initial conditions.

\section*{Acknowledgments}

We would like to thank N. Sakai, Y. Tanii and A. Tseytlin for useful
correspondence, N. Deruelle, R. Kallosh and S. Mukohyama for
valuable discussions.
This work was partially supported by the Grant-in-Aid for Scientific Research
Fund of the JSPS (Nos. 16540250 and  17540268) and another for the
Japan-U.K. Research Cooperative Program,
and by the Waseda University
Grant for Special Research Projects and  for The 21st Century
COE Program (Holistic Research and Education Center for Physics
Self-organization Systems) at Waseda University.

\newpage
\appendix
\section{Explicit form of actions}
\label{appendix_A}
We use the following notation throughout this paper.
\bea
(p-m)_n&\equiv& (p-m)(p-m-1)(p-m-2)\cdots (p-n),
\nonumber \\
(q-m)_n&\equiv&(q-m)(q-m-1)(q-m-2)\cdots (q-n),
\nonumber \\
A_p&\equiv&\dot{u}_1^2+\sigma_p e^{2(u_0-u_1)}, \quad
A_q \equiv \dot{u}_2^2+\sigma_q e^{2(u_0-u_2)},\nn
X &\equiv& \ddot u_1 - \dot u_0 \dot u_1 +\dot u_1^2, \quad
Y \equiv \ddot u_2 - \dot u_0 \dot u_2 +\dot u_2^2.
\label{xy}
\ena

Inputting our ansatz (\ref{met1}) into actions (\ref{eh})-(\ref{j0}),
we can write down the Lagrangians in terms of $u_0$, $u_1$ and $u_2$.
The explicit forms of actions are listed here, where actions
(\ref{eh2}) -- (\ref{ll2}) are integrated by parts.\\

\noindent
{\bf (1) Einstein-Hilbert action ($n=1$)}
\begin{align}
\tilde E_2 &=  e^{-2 u_0}\bigl[p_1A_p + q_1A_q - 2(p_1\dot u_1{^2} +
pq\dot u_1\dot u_2 + q_1\dot u_2{^2} )\bigr]\ .\label{eh2}
\end{align}
{\bf (2) Gauss-Bonnet action ($n=2$)}
\begin{align}
\tilde E_4  &=  e^{-4u_0}\bigl[p_3A_p{^2} + 2p_1q_1A_qA_p + q_3A_q{^2} \nn
&\quad -4A_p(p_3\dot u_1{^2}+p_2q\dot u_1\dot u_2 + p_1q_1\dot u_2{^2} )
 -4A_q(p_1q_1\dot u_1{^2}+pq_2\dot u_1\dot u_2 + q_3\dot u_2{^2})\nn
&\quad  +\frac{4}{3}(2p_3\dot u_1{^4} + 2p_2q \dot u_1{^3}\dot u_2
+ 3p_1q_1\dot u_1{^2}\dot u_2{^2} +2pq_2\dot u_1\dot u_2{^3} + 2
q_3\dot u_2{^4} )\bigr]\ .
\end{align}
{\bf (3) Lovelock action ($n=3,\ 4$)}
\begin{align}
\tilde E_6 &=  e^{-6u_0}\bigl[p_5A_q{^3} + 3p_3q_1A_p{^2}A_q + 3p_1q_3A_pA_q{^2}
+ q_5A_q{^3}
 -6A_p{^2}(p_5\dot u_1{^2} + p_4q\dot u_1\dot u_2 + p_3q_1\dot u_2{^2} )\nn
&\quad -6A_q{^2}(p_1q_3\dot u_1{^2} + pq_4\dot u_1\dot u_2 + q_5\dot u_2{^2} )
 -12A_pA_q(p_3q_1\dot u_1{^2} + p_2q_2\dot u_1\dot u_2 + p_1q_3\dot u_2{^2} )\nn
&\quad + 4A_p(2p_5\dot u_1{^4} + 2p_4q\dot u_1{^3}\dot u_2
+ 3p_3q_1 \dot u_1{^2}\dot u_2{^2} + 2p_2q_2\dot u_1\dot u_2{^3} +
2p_1q_3\dot u_2{^4} )\nn
&\quad + 4A_q(2p_3q_1\dot u_1{^4} + 2p_2q_2\dot u_1{^3}\dot u_2
+ 3p_1q_3 \dot u_1{^2}\dot u_2{^2} + 2pq_4\dot u_1\dot u_2{^3} +
2q_5\dot u_2{^4} )\nn
&\quad  + \frac{8}{5}(2p_5\dot u_1{^6} + 2p_4q \dot u_1{^5}\dot u_2
+ 5p_3q_1 \dot u_1{^4} \dot u_2{^2} + 5p_2q_2\dot u_1{^3}\dot
u_2{^3} + 5p_1q_3\dot u_1{^2} \dot u_2{^4} +  2pq_4 \dot u_1\dot u_2{^5}
+ 2q_5\dot u_2{^6}) \bigr]\ ,\\[0.5em]
\tilde E_8 &= e^{-8u_0}\bigl[ p_7A_p{^4} +4p_5q_1A_p{^3}A_q
+ 6p_3q_3 A_p{^2}A_q{^2}+ 4p_1q_5A_pA_q{^3}+ q_7A_q{^7} \nn
& \quad -8A_p{^3}(p_7\dot u_1{^2} + p_6q\dot u_1\dot u_2 + p_5q_1\dot u_2{^2} )
 -8A_q{^3}(p_1q_5\dot u_1{^2} + pq_6\dot u_1\dot u_2 + q_7\dot u_2{^2} )\nn
& \quad -24A_p{^2}A_q(p_5q_1\dot u_1{^2} + p_4q_2\dot u_1\dot u_2
+ p_3q_3\dot u_2{^2} ) -24A_pA_q{^2}(p_3q_3\dot u_1{^2}
+ p_2q_4\dot u_1\dot u_2 + p_1q_5\dot u_2{^2} )\nn
&\quad + 8A_p{^2}(2p_7\dot u_1{^4} + 2p_6q\dot u_1{^3}\dot u_2
+ 3p_5q_1 \dot u_1{^2}\dot u_2{^2} + 2p_4q_2\dot u_1\dot u_2{^3} +
2p_3q_3\dot u_2{^4} )\nn
&\quad + 8A_q{^2}(2p_3q_3\dot u_1{^4} + 2p_2q_4\dot u_1{^3}\dot u_2
+ 3p_1q_5 \dot u_1{^2}\dot u_2{^2}
+ 2pq_6\dot u_1\dot u_2{^3} + 2q_7\dot u_2{^4} )\nn
&\quad + 16A_pA_q(2p_5q_1\dot u_1{^4} + 2p_4q_2\dot u_1{^3}\dot u_2
+ 3p_3q_3 \dot u_1{^2}\dot u_2{^2} + 2p_2q_4\dot u_1\dot u_2{^3}
+2p_1q_5\dot u_2{^4}) \nn
&\quad
 -\frac{32}{5}A_p(2p_7\dot u_1{^6} + 2p_6q \dot u_1{^5}\dot u_2
 + 5p_5q_1 \dot u_1{^4} \dot u_2{^2} + 5p_4q_2\dot u_1{^3}\dot u_2{^3}
+ 5p_3q_3\dot u_1{^2} \dot u_2{^4}  + 2p_2q_4 \dot u_1\dot u_2{^5}
+ 2p_1q_5\dot u_2{^6}  )\nn
&\quad  -\frac{32}{5}A_q(2p_5q_1\dot u_1{^6} + 2p_4q_2 \dot u_1{^5}\dot u_2
+ 5p_3q_3 \dot u_1{^4} \dot u_2{^2}
+ 5p_2q_4\dot u_1{^3}\dot u_2{^3}+ 5p_5q_3\dot u_1{^2} \dot u_2{^4}
+ 2pq_6 \dot u_1\dot u_2{^5} + 2q_7\dot u_2{^6} )\nn
&\quad +\frac{16}{35}(8p_7\dot u_1{^8} + 8p_6q\dot u_1{^7}\dot u_2
+ 28p_5q_1\dot u_1{^6}\dot u_2{^2}
+ 28p_4q_2\dot u_1{^5}\dot u_2{^3} + 35p_3q_3\dot u_1{^4}\dot u_2{^4}\nn
&\quad+ 28p_2q_4\dot u_1{^3}\dot u_2{^5} + 28p_1q_5\dot u_1{^2}\dot u_2{^6}
+ 8pq_6\dot u_1\dot u_2{^7}+ 8q_7\dot u_2{^8}) \bigr]\ .\label{ll2}
\end{align}
{}\\

\noindent
{\bf (4) $S_W$ action}\\
Components of the Weyl tensor defined in~\p{weyl} are given by
\bea
C^t{}_{itj} &=& \frac{e^{-2u_0}}{(D-1)(D-2)} q B_1 g_{ij}, \nn
C^t{}_{atb} &=& -\frac{e^{-2u_0}}{(D-1)(D-2)} p B_1 g_{ab}, \nn
C^i{}_{jkl} &=& \frac{e^{-2u_0}}{(D-1)(D-2)}qB_2(g^i{}_k g_{jl}-g^i{}_l g_{jk}),\nn
C^a{}_{bcd} &=& \frac{e^{-2u_0}}{(D-1)(D-2)}pB_3(g^a{}_c g_{bd}-g^a{}_d g_{bc}),\nn
C^i{}_{ajb} &=& \frac{e^{-2u_0}}{(D-1)(D-2)} B_4 g^i{}_j g_{ab},
\label{c_ijkl}
\ena
where $i,j, \ldots$ run over external space, and $a,b, \ldots$ over internal
space, respectively, and we have defined
\begin{align}
 B_1 &= (D-3)(X-Y) - (p-1)A_p + (q-1) A_q + (p-q) \dot u_1 \dot u_2\ , \nn
B_2 &= -2(X-Y) + (q+1)A_p + (q-1) A_q -2q \dot u_1 \dot u_2\ , \nn
B_3 &= 2(X-Y) + (p-1)A_p + (p+1) A_q -2 p \dot u_1 \dot u_2\ , \nn
B_4 &= (p-q)(X-Y) - (p-1)q A_p - p(q-1) A_q + (2pq-p-q) \dot u_1 \dot u_2\ .
\end{align}
Although we have four quantities $B_1,\ B_2,\ B_3$ and $B_4$, only two of them
are independent. Actually, the tracelessness of the Weyl tensor gives
the following relations.
\begin{align}
B_1 + (p-1) B_2 + B_4 &= 0\ , &
-B_1 + (q-1) B_3 + B_4 &= 0\ .
\end{align}
Taking independent variables as $B_2$ and $B_3$ in order to consider the case $p=1$
or $q=1$ at the same time, we find that the other variables $B_1$ and $B_4$
are given as the following with respect to $B_2$ and $B_3$.
\begin{align}
B_1 &= -\frac{1}{2}[(p-1)B_2-(q-1)B_3]\ , &
B_4 &= -\frac{1}{2}[(p-1)B_2+(q-1)B_3] \label{B1,B4}
\end{align}
Substituting Eqs.~(\ref{B1,B4}) and (\ref{c_ijkl}) into the action Eq.~(\ref{4th1}),
we have the Lagrangian with respect to $B_2$ and $B_3$ as follows.
\begin{align}
L_W = \frac{pq\ e^{-7u_0 +pu_1 + qu_2}}{16(D-1)^4(D-2)^4}
\bigl[n_{1pq} B_2{^4} - 4n_{2pq}B_2{^3} B_3 + 2n_{3pq} B_2{^2} B_3{^2}
- 4n_{2qp} B_2B_3{^3} +n_{1qp} B_3{^4}\bigr]
\end{align}
where we define $n_{1pq}$, $n_{2pq}$ and $n_{3pq}$ as
\begin{align}
n_{1pq} &= (p^2-1) \bigl[-(p-1)^4(p-2) + (p-1)^3 (p^2 + 3p-7) q \nn
        &\quad +2(p-1)^2(p^2-4p+7) + (p^3-13p^2 +59p-71) q^3\bigr]\ ,\label{1pq} \\
n_{2pq} &=(p-1)^2(q-1) \bigl[-3p^4+3p^3+p^2+p-2 + (p-1)(p^3+2p^2+5)q \nn
        &\quad +2(p(p-1)^2-4)q^2 + (p-3)(p-5)q^3\bigr]\ ,\label{2pq} \\
n_{3pq} &= (p-1) (q-1)\bigl[ p\bigl\{p^3(3q^2-18q+23) - p^2(9q^2-16q+13)-p(11p+1)
-3 \bigr\} \nn
        &\quad +q\bigl\{q^3(3p^2-18p+23) - q^2(9p^2-16p+13)-q(11q+1)-3\bigr\} \nn
        &\quad +2\bigl\{3(pq)^3 + 5(pq)^2 + 8pq -3\bigr\} \bigr]\label{3pq} \ ,
\end{align}
while $n_{1qp}$ and $n_{2qp}$ are given by changing $p$ for $q$ in
Eqs.~(\ref{1pq}) and (\ref{2pq}), respectively.\\

\noindent
{\bf (5) $R^4$ action}
\begin{align}
L_{R^4} &=  e^{-7{u_0} + pu_1 +qu_2}\bigl[ 2pX + 2qY + p_1{A_p} + q_1{A_q}
+ 2pq\dot u_1 \dot u_2\bigr]^4\ . \label{Sr4}
\end{align}

\section{Field equations}

Taking variation of the actions (\ref{eh2})-(\ref{Sr4}), we find the basic
equations~\p{basic0} -- \p{basic2}, where each term is summarized here
according to which action it originates from.
The explicit forms of each term in the field equations are listed here:\\
{\bf (1) EH action ($n=1$) term}
\begin{align}
\label{eh1}
F_1&=\alpha_1 e^{-u_0}\bigl[p_1 A_p+q_1 A_q+2pq\dot{u}_1\dot{u}_2\bigr]\ ,\\
f_1^{(p)}&=\alpha_1 e^{-u_0}\bigl[ (p-1)_2A_p+q_1 A_q+2(p-1)q\dot{u}_1
\dot{u}_2\bigr]\ , \\
f_1^{(q)}&=\alpha_1 e^{-u_0}\bigl[p_1 A_p+ (q-1)_2A_q+2p(q-1)\dot{u}_1
\dot{u}_2\bigr]\ ,\\
g_1^{(p)}&= 2(p-1)\alpha_1 e^{-u_0}\ , \qquad g_1^{(q)}=2(q-1)\alpha_1
e^{-u_0}\ ,\\
h_1^{(p)}& =2q\alpha_1 e^{-u_0}\ ,\qquad  h_1^{(q)}=2p\alpha_1 e^{-u_0}\ .
\end{align}
{\bf (2) Lovelock action ($n=4$) term}
\bea
F_4&=&
\alpha_4 e^{-7u_0} \bigl[
p_7 A_p^4+4p_5q_1 A_p^3A_q+6p_3q_3A_p^2A_q^2
+4p_1q_5 A_pA_q^3+q_7 A_q^4
\nonumber \\
&&
+8\dot{u}_1\dot{u}_2
(p_6 q A_p^3+3p_4q_2A_p^2A_q+3p_2q_4A_pA_q^2 +p q_6 A_q^3)
\nonumber
\\
&&
+24\dot{u}_1^2\dot{u}_2^2(p_5q_1A_p^2+2p_3q_3A_pA_q+p_1q_5A_q^2)
+32\dot{u}_1^3\dot{u}_2^3(p_4q_2A_p+p_2q_4A_q)
+16p_3q_3\dot{u}_1^4\dot{u}_2^4 \bigr],
\nonumber \\
~~
\\
f_4^{(p)}&=&\alpha_4 e^{-7u_0}\bigl[ (p-1)_8 A_p^4+4(p-1)_6q_1
A_p^3A_q+6(p-1)_4q_3A_p^2A_q^2 +4(p-1)_2q_5A_pA_q^3+q_7 A_q^4
\nonumber \\
&&
+8\dot{u}_1\dot{u}_2 \left((p-1)_7qA_p^3+3(p-1)_5q_2A_p^2A_q
+3(p-1)_3q_4A_pA_q^2+(p-1)q_6 A_q^3\right)
\nonumber \\
&&
+24\dot{u}_1^2\dot{u}_2^2\left((p-1)_6q_1A_p^2+2(p-1)_4q_3A_pA_q
+(p-1)_2q_5A_q^2\right)
\nonumber \\
&&
+32\dot{u}_1^3\dot{u}_2^3\left( (p-1)_5q_2A_p+(p-1)_3q_4 A_q\right)
+16(p-1)_4q_3\dot{u}_1^4\dot{u}_2^4\bigr],
\\
f_4^{(q)}&=&\alpha_4 e^{-7u_0}\bigl[ (q-1)_8 A_q^4+4(q-1)_6p_1
A_q^3A_p+6(q-1)_4p_3A_p^2A_q^2 +4(q-1)_2p_5A_qA_p^3+p_7 A_p^4
\nonumber \\
&&
+8\dot{u}_1\dot{u}_2 \left((q-1)_7pA_q^3+3(q-1)_5p_2A_q^2A_p
+3(q-1)_3p_4A_qA_p^2+(q-1)p_6 A_p^3\right)
\nonumber \\
&&
+24\dot{u}_1^2\dot{u}_2^2\left((q-1)_6p_1A_q^2+2(q-1)_4p_3A_pA_q
+(q-1)_2p_5A_p^2\right)
\nonumber \\
&&
+32\dot{u}_1^3\dot{u}_2^3\left( (q-1)_5p_2A_q+(q-1)_3p_4 A_p\right)
+16(q-1)_4p_3\dot{u}_1^4\dot{u}_2^4 \bigr],
\\
g_4^{(p)}&=&8(p-1)\alpha_4 e^{-7u_0}\bigl[
(p-2)_7A_p^3+3(p-2)_5q_1A_p^2A_q+3(p-2)_3q_3A_pA_q^2+q_5A_q^3
\nonumber\\
&&
+6\dot{u}_1\dot{u}_2\left( (p-2)_6qA_p^2+2(p-2)_4q_2A_pA_q
+(p-2)q_4A_q^2\right)
\nonumber\\
&&
+12\dot{u}_1^2\dot{u}_2^2\left( (p-2)_5q_1A_p+(p-2)_3q_3A_q\right)
+8(p-2)_4q_2\dot{u}_1^3\dot{u}_2^3\bigr],
\\
g_4^{(q)}&=&8(q-1)\alpha_4 e^{-7u_0}\bigl[
(q-2)_7A_q^3+3(q-2)_5p_1A_q^2A_p+3(q-2)_3p_3A_qA_p^2+p_5A_p^3
\nonumber\\
&&
+6\dot{u}_1\dot{u}_2\left( (q-2)_6pA_q^2+2(q-2)_4p_2A_pA_q
+(q-2)p_4A_p^2\right)
\nonumber\\
&&
+12\dot{u}_1^2\dot{u}_2^2\left( (q-2)_5p_1A_q+(q-2)_3p_3A_p\right)
+8(q-2)_4p_2\dot{u}_1^3\dot{u}_2^3\bigr],
\\
h_4^{(p)}&=&8q\alpha_4 e^{-7u_0}\bigl[
(p-1)_6A_p^3+3(p-1)_4(q-1)_2A_p^2A_q+3(p-1)_2(q-1)_4A_pA_q^2
+(q-1)_6A_q^3
\nonumber\\
&&
+6\dot{u}_1\dot{u}_2\left( (p-1)_5(q-1)A_p^2+2(p-1)_3(q-1)_3A_pA_q
+(p-1)(q-1)_5A_q^2\right)
\nonumber \\
&&
+12\dot{u}_1^2\dot{u}_2^2\left( (p-1)_4(q-1)_2A_p+(p-1)_2(q-1)_4A_q\right)
+8(p-1)_3(q-1)_3\dot{u}_1^3\dot{u}_2^3\bigr],
\\
h_4^{(q)}&=&8p\alpha_4 e^{-7u_0}\bigl[
(q-1)_6A_q^3+3(q-1)_4(p-1)_2A_q^2A_p+3(q-1)_2(p-1)_4A_qA_p^2 +(p-1)_6A_p^3
\nonumber\\
&&
+6\dot{u}_1\dot{u}_2\left( (q-1)_5(p-1)A_q^2+2(p-1)_3(q-1)_3A_pA_q
+(q-1)(p-1)_5A_p^2\right)
\nonumber \\
&&
+12\dot{u}_1^2\dot{u}_2^2\left( (q-1)_4(p-1)_2A_q+(q-1)_2(p-1)_4A_p\right)
+8(p-1)_3(q-1)_3\dot{u}_1^3\dot{u}_2^3\bigr].
\ena
{\bf (3) $S_W$ action term}
\begin{align}
F_W &= \gamma  e^{-pu_1-qu_2} \Big[ -7 L_W
       + 2\sigma_p e^{2(u_0-u_1)}
\Bigl((q+1)\frac{\partial L_W}{\partial B_2}
       + (p-1) \frac{\partial L_W}{\partial B_3}\Bigr)\nn
    &\quad +2\sigma_q e^{2(u_0-u_2)}\Bigl((q-1)
\frac{\partial L_W}{\partial B_2}
       + (p+1) \frac{\partial L_W}{\partial B_3}\Bigr)
       - 2\frac{d}{dt}\Big\{ (\dot u_1-\dot u_2)
\Big(  \frac{\pa L_W}{\pa B_2}
       - \frac{\pa L_W}{\pa B_3} \Big) \Big\} \Big]\ , \\
pF_W^{(p)} &= \gamma e^{-pu_1-qu_2} \Big[ p L_W
-2 \s_p e^{2(u_0-u_1)} \Big\{ (q+1) \frac{\pa L_W}{\pa B_2} + (p-1)
 \frac{\pa L_W}{\pa B_3}\Big\} \nn
&\quad -2\frac{d}{dt}\Big\{  \Big( \dot u_0 +(q-1) \dot u_1 -q \dot u_2 \Big)
\frac{\pa L_W}{\pa B_2}
+ \Big( -\dot u_0 +(p+1) \dot u_1 -p \dot u_2 \Big) \frac{\pa L_W}{\pa B_3}
\Big\}\nn
&\quad   -2\frac{d^2}{dt^2}\Big\{ \frac{\pa L_W}{\pa B_2}-
 \frac{\pa L_W}{\pa B_3}\Big\} \Big]\ , \\
q F_W^{(q)} &= \gamma e^{-pu_1-qu_2} \Big[ q L_W -2\sigma_q e^{2(u_0-u_2)}
 \Big\{ (q-1) \frac{\pa L_W}{\pa B_2} + (p+1) \frac{\pa L_W}{\pa B_3}
 \Big\} \nn
&\quad -2\frac{d}{dt}\Big\{
 \Big( -\dot u_0 -q \dot u_1 +(q+1) \dot u_2 \Big) \frac{\pa L_W}{\pa B_2}
 +\Big( \dot u_0 -p \dot u_1 +(p-1) \dot u_2 \Big) \frac{\pa L_W}{\pa B_3}
\Big\} \nn
&\quad +2\frac{d^2}{dt^2}\Big\{  \frac{\pa L_W}{\pa B_2} - \frac{\pa L_W}{\pa B_3}\Big\} \Big]\ ,\label{gbl}
\end{align}
where
\begin{align}
\frac{ \partial L_W}{\partial B_2} &= \frac{pq\ e^{-7u_0 + pu_1 + qu_2}}{4(D-1)^4(D-2)^4}
  \bigl[ n_{1pq}B_2{^3} -3n_{2pq}B_2{^2} B_3 + n_{3pq} B_2B_3{^2} - n_{2qp} B_3{^3}\bigr] \ ,\\
\frac{ \partial L_W}{\partial B_3} &= \frac{pq\ e^{-7u_0 + pu_1 + qu_2}}{4(D-1)^4(D-2)^4}
  \bigl[-n_{2pq}B_2{^3} + n_{3pq}B_2{^2} B_3 -3 n_{2qp} B_2B_3{^2} + n_{1qp} B_3{^3}\bigr] \ .
\end{align}
{\bf (4) $R^4$ action term}
\begin{align}
 F_{R^4} &=\delta e^{-pu_1-qu_2} \left[-7L_{R^4}+2\sigma_p e^{2(u_0-u_1)}{\partial
L_{R^4}\over \partial A_p}+2\sigma_q e^{2(u_0-u_2)}{\partial L_{R^4}\over
\partial A_q}+{d\over dt}\left(\dot{u}_1{\partial L_{R^4}\over
\partial X}+\dot{u}_2{\partial L_{R^4}\over \partial Y}\right)\right]\,, \\
 pF_S^{(p)} &=\delta e^{-pu_1-qu_2} \left[
pL_{R^4}-2\sigma_p e^{2(u_0-u_1)}{\partial L_{R^4}\over \partial A_p}
+{d\over dt}\left((\dot{u}_0-2\dot{u}_1){\partial L_{R^4}\over \partial
X}-2\dot{u}_1{\partial L_{R^4}\over \partial A_p}-{\partial L_{R^4}\over \partial
\dot{u}_1}\right)+  {d^2\over dt^2}\left({\partial L_{R^4}\over
 \partial X}\right) \right]\,,\\
 qF_S^{(q)} &=\delta e^{-pu_1-qu_2} \left[
qL_{R^4}-2\sigma_q e^{2(u_0-u_2)}{\partial L_{R^4}\over \partial A_q}
+{d\over dt}\left((\dot{u}_0-2\dot{u}_2){\partial L_{R^4}\over \partial
Y}-2\dot{u}_2{\partial L_{R^4}\over
\partial A_q}-{\partial L_{R^4}\over \partial \dot{u}_2}\right)
+ {d^2\over dt^2}\left({\partial L_{R^4}\over\partial Y}\right)
\right]\,,
\end{align}
where
\begin{gather}
\begin{aligned}
\frac{\partial L_{R^4}}{\partial X} &=  e^{-7u_0+pu_1+qu_2} 8p \tilde R^3 \ ,&
\frac{\partial L_{R^4}}{\partial Y}&=  e^{-7u_0+pu_1+qu_2} 8q \tilde R^3 \ , \\
\frac{\partial L_{R^4}}{\partial A_p}&=  e^{-7u_0+pu_1+qu_2} 4p_1 \tilde R^3 \ , &
\frac{\partial L_{R^4}}{\partial A_q}&=  e^{-7u_0+pu_1+qu_2} 4q_1 \tilde R^3 \ , \\
\frac{\partial L_{R^4}}{\partial \dot u_1}&=  e^{-7u_0+pu_1+qu_2}
8pq\dot u_2\tilde R^3 \ , &
\frac{\partial L_{R^4}}{\partial \dot u_2}&=  e^{-7u_0+pu_1+qu_2} 8pq\dot u_1
\tilde R^3 \ .
\end{aligned}\\
\tilde R =2pX + 2qY + p_1{A_p} + q_1{A_q} + 2pq\dot u_1 \dot u_2
\end{gather}

\section{Inputting our ansatz into solutions}

In order to find solutions, we assume
\begin{align}
u_0=\epsilon t\ ,\qquad  u_1=\mu t  + \ln a_0\ ,\qquad  u_2=\nu t  + \ln b_0 \ .
\label{2.20}
\end{align}
Inserting this form into the above equations (Eqs.~\p{eh1} -- \p{gbl})
and setting
\begin{gather}
A_p =\mu^2+\tilde{\sigma}_p e^{2(\epsilon-\mu)t}\ , \qquad
A_q =\nu^2+\tilde{\sigma}_q e^{2(\epsilon-\nu)t}\ ,\qquad
\tilde\s_p =\frac{\s_p}{a_0^2}\ ,\qquad
\tilde\s_q =\frac{\s_q}{b_0^2}\ ,
\label{A_pA_q}\\
X=\mu(\mu-\epsilon)\ , \qquad Y=\nu(\nu-\epsilon)\ ,
\label{XY}
\end{gather}
we obtain the following explicit equations:\\[1em]
{\bf (1) EH action ($n=1$) term}
\bea
F_1&=&\alpha_1 e^{-\epsilon t}\left[p_1A_p+q_1A_q+2pq \mu\nu\right],
\nn
f_1^{(p)}&=&\alpha_1 e^{-\epsilon t}\left[(p-1)_2A_p+q_1A_q+2(p-1)q
\mu\nu\right],
\nn
f_1^{(q)}&=&\alpha_1 e^{-\epsilon t}\left[p_1A_p+(q-1)_2A_q+2p(q-1)
\mu\nu\right],
\nn
g_1^{(p)}&=&2(p-1)\alpha_1 e^{-\epsilon t}
, ~~~g_1^{(q)}~=~2(q-1)\alpha_1 e^{-\epsilon t},
\nn
h_1^{(p)}&=&2q\alpha_1 e^{-\epsilon t}
, ~~~~~~~~~~~
h_1^{(q)}~=~2p\alpha_1 e^{-\epsilon t},
\label{emn-EH}
\ena
{\bf (2) Lovelock action ($n=4$) term}
\bea
F_4&=&\alpha_4 e^{-7\epsilon t}
\left[ p_7 A_p^4+4p_5q_1 A_p^3A_q+6p_3q_3A_p^2A_q^2
+4p_1q_5 A_pA_q^3+q_7 A_q^4 \right. \nn
&&
+8\mu\nu (p_6 q A_p^3+3p_4q_2A_p^2A_q+3p_2q_4A_pA_q^2 +p q_6 A_q^3)
+24\mu^2\nu^2(p_5q_1A_p^2\nn
&&\left.
+2p_3q_3A_pA_q+p_1q_5A_q^2)
+32\mu^3\nu^3(p_4q_2A_p+p_2q_4A_q)
+16p_3q_3\mu^4\nu^4 \right],
\label{M:F_4}
\\
f_4^{(p)}&=&\alpha_4 e^{-7\epsilon t}\Big[
(p-1)_8 A_p^4+4(p-1)_6q_1 A_p^3A_q+6(p-1)_4q_3A_p^2A_q^2
+4(p-1)_2q_5A_pA_q^3 \nn
&&
+q_7 A_q^4 +8\mu\nu\left\{(p-1)_7qA_p^3+3(p-1)_5q_2A_p^2A_q
+3(p-1)_3q_4A_pA_q^2 \right. \nn
&& \left. +(p-1)q_6 A_q^3\right\}
+24\mu^2\nu^2\left\{(p-1)_6q_1A_p^2+2(p-1)_4q_3A_pA_q
+(p-1)_2q_5A_q^2\right\} \nn
&& +32\mu^3\nu^3\left\{(p-1)_5q_2A_p+(p-1)_3q_4 A_q\right\}
+16(p-1)_4q_3\mu^4\nu^4 \Big],
\label{M:f_4p}
\\
g_4^{(p)}&=&8(p-1)\alpha_4 e^{-7\epsilon t}\left[
(p-2)_7A_p^3+3(p-2)_5q_1A_p^2A_q+3(p-2)_3q_3A_pA_q^2+q_5A_q^3
\right.\nn
&&
\left.
+6\mu\nu\left\{ (p-2)_6qA_p^2+2(p-2)_4q_2A_pA_q
+(p-2)q_4A_q^2\right\}\right. \nn
&& \left. +12\mu^2\nu^2\left\{ (p-2)_5q_1A_p+(p-2)_3q_3A_q\right\}
+8(p-2)_4q_2\mu^3\nu^3\right],
\label{M:g_4p}
\\
h_4^{(p)}&=&8q\alpha_4 e^{-7\epsilon t}\left[
(p-1)_6A_p^3+3(p-1)_4(q-1)_2A_p^2A_q+3(p-1)_2(q-1)_4A_pA_q^2 \right. \nn
&& +(q-1)_6A_q^3 + 6\mu\nu\left\{ (p-1)_5(q-1)A_p^2+2(p-1)_3(q-1)_3A_pA_q
\right.\nn
&& \left. +(p-1)(q-1)_5A_q^2\right\} +12\mu^2\nu^2\left\{
(p-1)_4(q-1)_2A_p+(p-1)_2(q-1)_4A_q\right\}\nn
&& +8(p-1)_3(q-1)_3\mu^3\nu^3\Big],
\label{M:h_4p}
\\
f_4^{(q)}&=&\alpha_4 e^{-7\epsilon t}\Big[ (q-1)_8 A_q^4+4(q-1)_6p_1
A_q^3A_p+6(q-1)_4p_3A_p^2A_q^2 +4(q-1)_2p_5A_qA_p^3 \nn
&&
+p_7 A_p^4+8\mu\nu
\left\{(q-1)_7pA_q^3+3(q-1)_5p_2A_q^2A_p+3(q-1)_3p_4A_qA_p^2 \right.\nn
&& \left. +(q-1)p_6 A_p^3\right\}+24\mu^2\nu^2\left\{(q-1)_6p_1A_q^2
+2(q-1)_4p_3A_pA_q +(q-1)_2p_5A_p^2\right\} \nn
&&
\left.
+32\mu^3\nu^3\left\{
(q-1)_5p_2A_q+(q-1)_3p_4 A_p\right\}
+16(q-1)_4p_3\mu^4\nu^4 \right],
\label{M:f_4q}
\\
g_4^{(q)}&=&8(q-1)\alpha_4 e^{-7\epsilon t}\left[
(q-2)_7A_q^3+3(q-2)_5p_1A_q^2A_p+3(q-2)_3p_3A_qA_p^2+p_5A_p^3
\right.\nonumber\\
&&
+6\mu\nu\left\{ (q-2)_6pA_q^2+2(q-2)_4p_2A_pA_q
+(q-2)p_4A_p^2\right\} \nn
&&
\left.
+12\mu^2\nu^2\left\{ (q-2)_5p_1A_q+(q-2)_3p_3A_p\right\}
+8(q-2)_4p_2\mu^3\nu^3\right],
\label{M:g_4q}
\\
h_4^{(q)}&=&8p\alpha_4 e^{-7\epsilon t}\Big[
(q-1)_6A_q^3+3(q-1)_4(p-1)_2A_q^2A_p+3(q-1)_2(p-1)_4A_qA_p^2 \nn
&& +(p-1)_6A_p^3 +6\mu\nu\left\{ (q-1)_5(p-1)A_q^2+2(p-1)_3(q-1)_3A_pA_q
\right. \nn
&& \left. +(q-1)(p-1)_5A_p^2\right\}
+12\mu^2\nu^2\left\{(q-1)_4(p-1)_2A_q+(q-1)_2(p-1)_4A_p\right\} \nn
&& +8(p-1)_3(q-1)_3\mu^3\nu^3\Big],
\ena
{\bf (3) $S_W$ action term}
\begin{align}
F_W &=  \frac{\gamma pq\  e^{-7\epsilon t}}{16(D-1)^4(D-2)^4}
  \bigl[-7\tilde L_W -2(\mu-\nu)(-7\epsilon+p\mu+q\nu)N_- \nn
  &\quad + 2\tilde\sigma_p e^{2(\epsilon -\mu)t} \bigl\{N_+ + N_-
  + (\epsilon -\mu)\bigl((q+1)N_{-,2}+(p-1)N_{-,3}\bigr)\bigr\} \nn
  &\quad + 2\tilde\sigma_q e^{2(\epsilon -\nu)t} \bigl\{N_+ - N_-
  + (\epsilon -\nu)\bigl((q-1)N_{-,2}+(p+1)N_{-,3}\bigr)\bigr\}\bigr]\ .\\
F_W{^{(p)}} &=  \frac{\gamma q\ e^{-7\epsilon t}}{16(D-1)^4(D-2)^4}
  \bigl[p\tilde L_W -2 (-7\epsilon +p\mu +q\nu)\bigl((-6\epsilon +(p-1)\mu
+ q\nu)N_- + (\mu-\nu)N_+\bigr)\nn
  &\quad -2\tilde\sigma_p e^{2(\epsilon -\mu)t} \bigl\{N_+ + N_-
  +2(\epsilon -\mu)\bigl((-11\epsilon
+ (2p-3)\mu+2q\nu)((q+1)N_{-,2}+(p-1)N_{-,3}) \nn
    &\qquad +(\mu-\nu)((q+1)N_{+,2}+(p-1)N_{+,3})\bigr)\bigr\}\nn
  &\quad -4(\epsilon -\nu)\tilde\sigma_q e^{2(\epsilon -\nu)t}
    \bigl\{(-11\epsilon + (2p-1)\mu+2(q-1)\nu)((q-1)N_{-,2}+(p+1)N_{-,3})
\nn
    &\qquad +(\mu-\nu)((q-1)N_{+,2}+(p+1)N_{+,3})\bigr\}\nn
  &\quad - 8(\epsilon -\mu)^2\tilde\sigma_p{^2}e^{4(\epsilon -\mu)t}\bigl
\{(q+1)^2N_{-,22} + 2(p-1)(q+1)N_{-,23}
    + (p-1)^2N_{-,33} \bigr\} \nn
  &\quad - 16(\epsilon -\mu)(\epsilon -\nu)\tilde\sigma_p \tilde\sigma_q
e^{2(2\epsilon -\mu-\nu)t}
    \bigl\{(q^2-1)N_{-,22} + 2(pq+1)N_{-,23} + (p+1)^2N_{-,33} \bigr\} \nn
  &\quad - 8(\epsilon -\nu)^2\tilde\sigma_q{^2}e^{4(\epsilon -\nu)t}\bigl
\{(q-1)^2N_{-,22} + 2(p+1)(q-1)N_{-,23}
    + (p-1)^2N_{-,33} \bigr\}  \bigr]\ ,\\
  F_W{^{(q)}} &=  \frac{\gamma p\  e^{-7\epsilon t}}{16(D-1)^4(D-2)^4}
  \bigl[q\tilde L_W + 2 (-7\epsilon +p\mu +q\nu)\bigl((-6\epsilon +p\mu
+ (q-1)\nu)N_- + (\mu-\nu)N_+ \bigr)\nn
  &\quad + 4(\epsilon -\mu)\tilde\sigma_p e^{2(\epsilon -\mu)t} \bigl\{
    (-11\epsilon + 2(p-1)\mu+(2q-1)\nu)((q+1)N_{-,2}+(p-1)N_{-,3}) \nn
    &\qquad + (\mu-\nu)((q+1)N_{+,2} + (p-1)N_{+,3})\bigr\}\nn
  &\quad - 2\tilde\sigma_q e^{2(\epsilon -\nu)t}
    \bigl\{N_+ -N_- -2(\epsilon -\nu)\bigl((-11\epsilon + 2p\mu+(2q-3)\nu)
((q-1)N_{-,2}+(p+1)N_{-,3}) \nn
    &\qquad + (\mu-\nu)((q-1)N_{+,2}+(p+1)N_{+,3})\bigr)\bigr\}\nn
  &\quad + 8(\epsilon -\mu)^2\tilde\sigma_p{^2}e^{4(\epsilon -\mu)t}\bigl
\{(q+1)^2N_{-,22} + 2(p-1)(q+1)N_{-,23}
    + (p-1)^2N_{-,33} \bigr\} \nn
  &\quad + 16(\epsilon -\mu)(\epsilon -\nu)\tilde\sigma_p \tilde\sigma_q
e^{2(2\epsilon -\mu-\nu)t}
    \bigl\{(q^2-1)N_{-,22} + 2(pq+1)N_{-,23} + (p+1)^2N_{-,33} \bigr\} \nn
  &\quad + 8(\epsilon -\nu)^2\tilde\sigma_q{^2}e^{4(\epsilon -\nu)t}\bigl
\{(q-1)^2N_{-,22} + 2(p+1)(q-1)N_{-,23}
    + (p-1)^2N_{-,33} \bigr\}  \bigr]\ .
\end{align}
where
\begin{align}
\tilde L_W &=n_{1pq} B_2{^4} - 4n_{2pq}B_2{^3} B_3 + 2n_{3pq} B_2{^2} B_3{^2}
             - 4n_{2qp} B_2B_3{^3} +n_{1qp} B_3{^4}\ ,\\
M_2 &= 4  \bigl[ n_{1pq}B_2{^3} -3n_{2pq}B_2{^2} B_3 + n_{3pq} B_2B_3{^2}
- n_{2qp} B_3{^3}\bigr]\ ,\\
M_3 &=4 \bigl[-n_{2pq}B_2{^3} + n_{3pq}B_2{^2} B_3 -3 n_{2qp} B_2B_3{^2}
+ n_{1qp} B_3{^3}\bigr] \ ,\\
N_+ &= qM_2 + p M_3 \nn
    &= 4\bigl[(qn_{1pq} - p n_{2pq})B_2{^3} - (3q n_{2pq} - pn_{3pq})
B_2{^2}B_3 \nn
    &\quad + (qn_{3pq}-3pn_{2qp})B_2B_3{^2} - (qn_{2qp} -pn_{1qp})B_3{^3}
\bigr]\ , \\
N_- &= M_2 -M_3 \nn
    &= 4\bigl[(n_{1pq} + n_{2pq})B_2{^3} - (3n_{2pq}+n_{3pq})B_2{^2}B_3
      + (n_{3pq} + 3n_{2qp})B_2B_3{^2}-(n_{2qp} + n_{1qp})B_3{^3} \bigr]\ ,\\
  N_{+,2} &= 4\bigl[3(qn_{1pq}-pn_{2pq}) B_2{^2} - 2(3q n_{2pq} - pn_{3pq})
B_2B_3
      + (qn_{3pq}-3pn_{2qp})B_3{^2}\bigr]\ ,\\
  N_{+,3} &= 4\bigl[ -(3q n_{2pq} - pn_{3pq})B_2{^2} + 2(qn_{3pq}-3pn_{2qp})
B_2B_3
  - 3(qn_{2qp} -pn_{1pq})B_3{^2} \bigr]\ ,\\
  N_{-,2} &= 4\bigl[3(n_{1pq}+n_{2pq})B_2{^2} - 2(3n_{2pq}+n_{3pq})B_2 B_3
+ (n_{3pq} + 3n_{2qp})B_3{^2}\bigr]\ ,\\
  N_{-,3} &= 4\bigl[-(3n_{2pq}+n_{3pq})B_2{^2} + 2(n_{3pq}+3n_{2qp})B_2 B_3
- 3(n_{2qp}+n_{1qp})B_3{^2} \bigr]\ ,\\
    N_{-,22} &= 8\bigl[3(n_{1pq}+n_{2pq})B_2 - (3n_{2pq}+n_{3pq}) B_3\bigr]\
,\\
    N_{-,23} &= 8\bigl[ - (3n_{2pq}+n_{3pq})B_2
+ (n_{3pq} + 3n_{2qp})B_3\bigr]\ ,\\
    N_{-,33} &= 8\bigl[ (n_{3pq}+3n_{2qp})B_2 - 3(n_{2qp}+n_{1qp})B_3{^2}
\bigr]\ .
\end{align}
{\bf (4) $R^4$ action term}
\begin{align}
F_{R^4} &= \delta e^{-7\epsilon t} \tilde R^2 \bigl[-7\tilde R^2
+ 8(-7\epsilon +p\mu +q\nu)(p\mu +q\nu)\tilde R
+ 16\tilde\sigma_p e^{2(\epsilon -\mu) t}\bigl\{p\tilde R
+ 6p_1(\epsilon - \mu)(p\mu +q\nu)\bigr\}\nn
& \quad +16\tilde\sigma_q e^{2(\epsilon -\nu) t}\bigl\{q\tilde R
+ 6q_1(\epsilon - \nu)(p\mu +q\nu)\bigr\}\bigr]\ ,\\
F_{R^4}{^{(p)} } &= \delta e^{-7\epsilon t } \tilde R \, \bigl[
\tilde R ^3 -8(-7\epsilon + p\mu+q\nu)(6\epsilon+\mu)\tilde R^2\nn
&\quad + 8(p-1)\tilde \sigma_p e^{2(\epsilon -\mu)t} \bigl\{-\tilde R^2
+ 6 p(\epsilon -\mu)(-13\epsilon + (p-1)\mu +q\nu)\tilde R
+ 24 p (\epsilon - \mu)^2 \bigr\}\nn
&\quad + 46 q_1(\epsilon -\nu) \tilde \sigma_q e^{2(\epsilon -\nu)t}
\bigl\{(-13\epsilon + (p-1)\mu +q\nu)\tilde R + 4  (\epsilon - \nu) \bigr\}
+ 192 p_1{^2} (\epsilon -\mu)^2 \tilde \sigma_p{^2}e^{4(\epsilon -\mu)t}
\tilde R\nn
&\quad   + 384 p_1q_1  (\epsilon -\mu)(\epsilon -\nu) \tilde \sigma_p
\tilde \sigma_q e^{2(2\epsilon -\mu-\nu)t}\tilde R
+ 192 q_1{^2} (\epsilon -\nu)^2 \tilde \sigma_q{^2}e^{4(\epsilon -\nu)t}
\tilde R
\bigr]\ ,\\
F_{R^4}{^{(q)} } &=  \delta e^{-7\epsilon t } \tilde R \, \bigl[
\tilde R ^3 -8(-7\epsilon + p\mu+q\nu)(6\epsilon+\nu)\tilde R^2\nn
&\quad + 46 p_1(\epsilon -\mu) \tilde \sigma_p e^{2(\epsilon -\mu)t}
\bigl\{(-13\epsilon + p\mu + (q-1)\nu)\tilde R + 4  (\epsilon - \mu) \bigr\}
\nn
&\quad + 8(q-1)\tilde \sigma_q e^{2(\epsilon -\nu)t} \bigl\{-\tilde R^2
+ 6 q(\epsilon -\nu)(-13\epsilon + p\mu +(q-1)\nu)\tilde R
+ 24 q (\epsilon - \nu)^2 \bigr\}\nn
&\quad+ 192 p_1{^2} (\epsilon -\mu)^2 \tilde \sigma_p{^2}e^{4(\epsilon -\mu)t}
\tilde R + 384 p_1q_1 (\epsilon -\mu)(\epsilon -\nu) \tilde \sigma_p
\tilde \sigma_q e^{2(2\epsilon -\mu-\nu)t}\tilde R\nn
&\quad
+ 192 q_1{^2} (\epsilon -\nu)^2 \tilde \sigma_q{^2}e^{4(\epsilon -\nu)t}
\tilde R
\bigr]\ ,\label{emn-S_W}
\end{align}
where
\begin{align}
\tilde R &= 2pX+2qY + p_1 A_p + q_1 A_q +2pq \mu\nu\ .
\end{align}

\section{Summary Tables of solutions in M-theory}
\label{summary}
Here we summarize our solutions for $\delta=0, -0.001$ and $-0.1$
in tables.
In the last columns of the tables, we include the type of two spaces
($ds_p^2,
ds_q^2$).
K means the kinetic dominant space, in which the curvature term
($\sigma_p$, or $\sigma_q$) is either zero or can be asymptotically ignored.
M denotes the Milne-type space, which is described by
$ds^2=-dt^2+t^2 ds_p^2+\cdots$ with $\sigma_p=-1$, or
$ds^2=-dt^2+\cdots+t^2 ds_q^2$ with $\sigma_q=-1$.
Similarly, we define a constant curvature space C by $\sigma_p=1$ or
$\sigma_q=1$, and
S$_0$ and S$_\pm$ are static spaces with zero curvature and positive (or
negative) curvature, respectively.

\begin{table}[htb]
\begin{center}
\caption{\small{Exact solutions for $\delta=0$. $\epsilon=0$ and 1 correspond
to generalized de Sitter solutions ($a\sim e^{\mu t}$, $b\sim e^{\nu t}$) and
power law ones ($a\sim \tau^{\mu}$, $b\sim \tau^{\nu}$), respectively.
$\lambda$ is a power exponent of power law solutions in the Einstein frame
($a_E \sim t_E^{\lambda}$).
K, $\mathrm{S}_\pm$,
$\mathrm{S}_0$, and M mean a kinetic dominance, a static space with positive
(or negative) curvature, a flat static space, and a Milne-type space,
respectively.
There is no inflationary solution in the Einstein frame.}}
\label{sum1}
\vspace{2mm}
\small{
\begin{tabular}{lcccccccccc}
\noalign{\global\arrayrulewidth1pt}
\hline
\noalign{\global\arrayrulewidth.4pt}
Solution$\!\!\!$ & $\epsilon$ & $\sigma_p$ & $\sigma_q$ & $\mu$ & $\nu$ & $a_0$ & $b_0$
& $\lambda$ & $\phi_1$ & Type\\ \hline
ME$1_{\pm}$ & 0 & 0 & 0 & $\pm 0.1047$ & $\mp 0.9367$ & $\cdots$ & $\cdots$
& 0.9681 & $\mp0.2676$ & K K \\
  \hline
ME12 & 1 & 0 & $-1$ & 0 & 1 & $\cdots$ & 1 & $0.7778$ & $0.7778$ &
$\mathrm{S}_0$ M\\
ME13 & 1 & $-1$ & 0 & 1 & 0 & 1 & $\cdots$ & 1 & 0 &M $\mathrm{S}_0$ \\
\noalign{\global\arrayrulewidth1pt}
\hline
\noalign{\global\arrayrulewidth.4pt}
\end{tabular}}
\end{center}
%
\begin{center}
\caption{\small{Future asymptotic solutions ($t\rightarrow\infty$) for
$\delta=0$. There is no inflationary solution in the Einstein frame.}}
\label{sum2}
\vspace{2mm}
\small{
\begin{tabular}{lccccccccccc}
\noalign{\global\arrayrulewidth1pt}
\hline
\noalign{\global\arrayrulewidth.4pt}
Solution$\!\!\!$ & $\epsilon$ & $\sigma_p$ & $\sigma_q$ & $\mu$ & $\nu$ & $a_0$ & $b_0$
& $\lambda$ & $\phi_1$ & $t_E$ & Type
\\   \hline
MF6 & 1 & 0 & 0 & $0.5583$ & $-0.0964$ & $\cdots$ & $\cdots$ & $0.3333$ &
$-0.1455$ & $\rightarrow\infty$ &Kasner\\
MF7 & 1 & 0 & 0 & $-0.3583$ & $0.2964$ & $\cdots$ & $\cdots$ & $0.3333$
& $0.1455$ & $\rightarrow\infty$ &Kasner\\
MF8 & 1 & $-1$ & $-1$ & 1 & 1 &  $0.4714$ & $0.8165$ & 1 & $0.2222$ &
$\rightarrow\infty$ & M M\\
\noalign{\global\arrayrulewidth1pt}
\hline
\noalign{\global\arrayrulewidth.4pt}
\end{tabular}}
\end{center}
%
\begin{center}
\caption{\small{Past asymptotic solutions ($t\rightarrow-\infty$)
for $\delta=0$.
MP1, MP4, MP5, MP11, MP12 and MP13 are inflationary solutions in the Einstein
 frame.}}
\label{sum3}
\vspace{2mm}
\small{
\begin{tabular}{lccccccccccc}
\noalign{\global\arrayrulewidth1pt}
\hline
\noalign{\global\arrayrulewidth.4pt}
Solution$\!\!\!$ & $\epsilon$ & $\sigma_p$ & $\sigma_q$ & $\mu$ & $\nu$ & $a_0$ & $b_0$
& $\lambda$ & $\phi_1$ & $t_E$ & Type\\
\hline
MP1 & 1 & $0$ & $0,\,\pm1$ & $1.588$ & $0.3193$ & $\cdots$ & $\cdots$ & 1.278
& 0.1508 &$\sim 0$& K K\\
MP2 & 1 & $0,\,\pm1$ & $0,\,\pm1$ & $0.7336$ & $0.082\,88$ & $\cdots$ &
$\cdots$ & 0.7935 & 0.064\,25&$\sim 0$& K K\\
MP3 & 1 & $0,\,\pm1$ & $0,\,\pm1$ & $0.7225$ & $-0.1667$ & $\cdots$ & $\cdots$
& 0.3335 & -0.4004&$\sim 0$& K K\\
MP4 & 1 & $0,\,\pm1$ & $0,\,\pm1$ & $0.6221$ & $-0.4004$ & $\cdots$ & $\cdots$
& 1.942 & 0.9975&$\sim 0$& K K\\
MP5 & 1 & $0,\,\pm1$ & $0,\,\pm1$ & $0.1003$ & $-1.701$ & $\cdots$ & $\cdots$
& 1.182 & 0.3434&$\sim 0$& K K\\
MP6 & 1 & $0,\,\pm1$ & $0,\,\pm1$ & $0.022\,04$ & $0.9906$ & $\cdots$ & $\cdots$
& 0.7811 & 0.2218&$\sim 0$& K K\\
MP7 & 1 & $0,\,\pm1$ &  $0,\,\pm1$  & $-0.030\,14$ & $0.6209$ & $\cdots$
& $\cdots$ & 0.6754 & 0.1957&$\sim 0$& K K\\
MP8 & 1 & $0,\,\pm1$ & $0,\,\pm1$  & $-0.3353$ & $0.8502$ & $\cdots$ & $\cdots$
& 0.6641 & 0.2139&$\sim 0$& K K\\
MP9 & 1 & $0,\,\pm1$ & $0,\,\pm1$  & $-0.6685$ & $0.6343$ & $\cdots$ & $\cdots$
& 0.4818 & 0.1970&$\sim 0$& K K\\
MP10 & 1 & $0,\,\pm1$ & $0$  & $-0.9380$ & $2.573$ & $\cdots$ & $\cdots$ &
0.8063 &0.2571&$\sim 0$& K K\\
  \hline
MP11 & 1 & $0$ & $-1$  & $6.680$ & 1& $\cdots$ & $\cdots$ & 2.262 & 0.2222&
$\sim 0$& K M\\
MP12 & 1 & $0$ & $-1$  & $6.086$ & 1 & $\cdots$ & $\cdots$ &2.1302 & 0.2222 &
$\sim 0$& K M\\
MP13 & 1 & $0$ & $-1$  & $2$ & 1 & $\cdots$ & $\cdots$ & 1.222 & 0.2222&
$\sim 0$& K M\\
MP14 & 1 & $0,\,\pm 1$ & $-1$  & $0.4818$ & 1 & $\cdots$ & $\cdots$ & 0.8848
& 0.2222&$\sim 0$& K M\\
MP15 & 1 & $0,\,\pm 1$ & $-1$  & $0.072\,89$ & 1 & $\cdots$ & $\cdots$ &
0.7940 & 0.2222&$\sim 0$& K M\\
MP16 & 1 & 1 & $0,\,\pm 1$  & 1 & $0.2682$ & $\cdots$ & $\cdots$ & 1 &0.1383
&$\sim 0$& M K\\
MP17 & 1 & 1 & $0,\,\pm 1$  & 1 & $-9.178$ & $\cdots$ & $\cdots$ & 1 & 0.2949
&$\sim 0$& M K\\
MP18 & 1 & $-1$ & $0,\,\pm 1$  & 1 & $0.9318$ & $\cdots$ & $\cdots$ & 1 &0.2187
&$\sim 0$& M K\\
MP19 & 1 & $-1$ & $0,\,\pm 1$  & 1 & $-0.1225$ & $\cdots$ & $\cdots$ & 1
& -0.2144  &$\sim 0$& M K\\
  \hline
MP23 & 1 & $\pm 1$  & $-1$ & 0 & 1 & 1 & $\cdots$ & 0.7778 & 0.2222 &$\sim 0$
& $\mathrm{S}_\pm$ M\\
MP24 & 1 & $-1$  & $\pm 1$ & 1 & 0 & $\cdots$ & 1 &1& 0 &$\sim 0$&M
$\mathrm{S}_\pm$ \\
MP25 & 1 & 1  & $-1$  & 1 & 1 & 0.6495 & 0.8898 &  1& 0.2222 &$\sim 0$&C M\\
MP26 & 1 & $-1$  & $-1$  & 1 & 1 & 0.8413 & 0.091\,06 &  1& 0.2222 &$\sim 0$& M M\\
MP27 & 1 & $-1$  & $-1$  & 1 & 1 & 0.4780 & 2.284 &  1& 0.2222 &$\sim 0$&M M\\
MP28 & 1 & $-1$  & $-1$  & 1 & 1 & 0.4166 & 0.6216 &  1& 0.2222 &$\sim 0$&M M\\
\noalign{\global\arrayrulewidth1pt}
\hline
\noalign{\global\arrayrulewidth.4pt}
\end{tabular}}
\end{center}
\end{table}

\begin{table}[htb]
\begin{center}
\caption{\small{Exact solutions for $\delta=-0.001$.
$\epsilon=0$ and 1 correspond
to generalized de Sitter solutions ($a\sim e^{\mu t}$, $b\sim e^{\nu t}$) and
power law ones ($a\sim \tau^{\mu}$, $b\sim \tau^{\nu}$), respectively.
$\lambda$ is a power exponent of power law solutions in the Einstein frame
($a_E \sim t_E^{\lambda}$).K, $\mathrm{S}_\pm$,
$\mathrm{S}_0$ and M mean a kinetic dominance, a static space with positive
(or negative) curvature, a flat static space, and a Milne-type space,
respectively.
ME3$_\pm$ and ME6$_+$ are inflationary solution in the Einstein frame, but
our 3-space is shrinking in ME3$_-$.}}\label{sum4}
\vspace{2mm}
\small{
\begin{tabular}{lcccccccccc}
\noalign{\global\arrayrulewidth1pt}
\hline
\noalign{\global\arrayrulewidth.4pt}
Solution$\!\!\!$ & $\epsilon$ & $\sigma_p$ & $\sigma_q$ & $\mu$ & $\nu$ & $a_0$ & $b_0$
& $\lambda$ & $\phi_1$ & Type\\ \hline
ME$2_{\pm}$ & 0 & 0 & 0 & $\pm 1.438$ & $\mp 0.067\,07$ & $\cdots$ & $\cdots$
& $-5.126$ & $\pm 0.019\,16$ & K K \\
ME$3_{\pm}$ & 0 & 0 & 0 & $\pm 0.4029$ & $\pm 0.4029$ & $\cdots$ & $\cdots$ &
$1.286$ & $ \pm0.1151$ & K K \\
ME$6_{\pm}$ & 0 & 0 & 1 & $\pm 0.7751$ & 0 & $\cdots$ & $1.893$  & $e^{\mu t_E}$
& $0$ & K $\mathrm{S}_+$\\
ME$8_{\pm}$ & 0 & 1 & 0 & 0 & $\pm0.4918$ & $1.0472$ & $\cdots$ & 1 & $\pm14.05$
& $\mathrm{S}_+$ K\\
ME$9_{\pm}$ & 0 & 1 & 0 &0 &  $\pm 0.4016$ & $1.220$ & $\cdots$ & 1 & $\pm0.1167$
&$\mathrm{S}_+$ K\\
ME$10_{\pm}$ & 0 & $-1$ & 0 & 0 & $\pm0.4084$ & $0.7124$ & $\cdots$ & 1&
$\pm0.1147$ & $\mathrm{S}_-$ K\\
  \hline
ME12 & 1 & 0      & $-1$ & 0 & 1 & $\cdots$ & 1 & $0.7778$ & $0.2222$ &
$\mathrm{S}_0$ M\\
ME13 & 1 & $-1$ & 0      & 1 & 0 & 1 & $\cdots$ & 1 & 0 &M $\mathrm{S}_0$ \\
\noalign{\global\arrayrulewidth1pt}
\hline
\noalign{\global\arrayrulewidth.4pt}
\end{tabular}}
\end{center}
%
\begin{center}
\caption{\small{Future asymptotic solutions ($t\rightarrow\infty$) for
$\delta=-0.001$.
MF1 and MF2 are inflationary solution in the Einstein frame.}}
\label{sum5}
\vspace{2mm}
\small{
\begin{tabular}{lccccccccccc}
\noalign{\global\arrayrulewidth1pt}
\hline
\noalign{\global\arrayrulewidth.4pt}
Solution$\!\!\!$ & $\epsilon$ & $\sigma_p$ & $\sigma_q$ & $\mu$ & $\nu$ & $a_0$ & $b_0$
& $\lambda$ & $\phi_1$ & $t_E$ & Type\\ \hline
MF1 & 0 & $\pm 1$ & $\pm 1$ & $0.4029$ & $0.4029$ & $\cdots$ & $\cdots$ &
$1.286$ & $ 0.1151$ & $\rightarrow\infty$ & ME$3_+$  \\
 MF2 & 0 & $\pm 1$ & 1 & $0.7751$ & 0 & $\cdots$ & $1.893$  & $e^{\mu t_E}$
& $0$ & $\rightarrow\infty$ & ME$6_+$\\
MF3 & 0 & $1$ & $\pm 1$ & 0 & $0.4918$ & $1.0472$ & $\cdots$ & 1 & $14.05$
& $\rightarrow\infty$ & ME$8_+$\\
MF4 & 0 & $1$ & $\pm 1$ &0 &  $0.4016$ & $1.220$ & $\cdots$ & 1 & $0.1167$
& $\rightarrow\infty$ & ME$9_+$\\
MF5 & 0 & $-1$ & $\pm 1$ & 0 & $0.4084$ & $0.7124$ & $\cdots$ & 1&
$0.1147$ & $\rightarrow\infty$ & ME$10_+$\\
 \hline
MF6 & 1 & 0 & 0 & $0.5583$ & $-0.0964$ & $\cdots$ & $\cdots$ & $0.3333$ &
$-0.1455$ & $\rightarrow\infty$ &Kasner\\
MF7 & 1 & 0 & 0 & $-0.3583$ & $0.2964$ & $\cdots$ & $\cdots$ & $0.3333$ &
$0.1455$ & $\rightarrow\infty$ &Kasner\\
MF8 & 1 & $-1$ & $-1$ & 1 & 1 &  $0.4714$ & $0.8165$ & 1 & $0.2222$ &
$\rightarrow\infty$ & M M\\
\noalign{\global\arrayrulewidth1pt}
\hline
\noalign{\global\arrayrulewidth.4pt}
\end{tabular}}
\end{center}
\end{table}

\begin{table}[htb]
\begin{center}
\caption{\small{Past asymptotic solutions $(t\rightarrow -\infty) $ for
$\delta=-0.001$.
MP1, MP2, MP7, MP9, and MP18 are inflationary solution in the Einstein frame,
although our 3-space is shrinking in MP1 and MP2.}}
\label{sum6}
\vspace{2mm}
\small{
\begin{tabular}{lccccccccccc}
\noalign{\global\arrayrulewidth1pt}
\hline
\noalign{\global\arrayrulewidth.4pt}
Solution$\!\!\!$ & $\epsilon$ & $\sigma_p$ & $\sigma_q$ & $\mu$ & $\nu$ & $a_0$ & $b_0$
& $\lambda$ & $\phi_1$ & $t_E$ & Type\\
\hline
 MP1 & 0 & $\pm 1$ & $\pm 1$ & $-0.4029$ & $-0.4029$ & $\cdots$ & $\cdots$ &
$1.286$ & $ -0.1151$ & $\rightarrow -\infty$ & ME$3_+$  \\
 MP2 & 0 & $\pm 1$ & 1 & $-0.7751$ & 0 & $\cdots$ & $1.893$  & $e^{\mu t_E}$
& $0$ & $\rightarrow-\infty$ & ME$6_+$\\
MP3 & 0 & $1$ & $\pm 1$ & 0 & $-0.4918$ & $1.0472$ & $\cdots$ & 1 & $-14.05$
& $\rightarrow-\infty$ & ME$8_+$\\
MP4 & 0 & $1$ & $\pm 1$ &0 &  $-0.4016$ & $1.220$ & $\cdots$ & 1 & $-0.1167$
& $\rightarrow-\infty$ & ME$9_+$\\
MP5 & 0 & $-1$ & $\pm 1$ & 0 & $-0.4084$ & $0.7124$ & $\cdots$ & 1&
$-0.1147$ & $\rightarrow-\infty$ & ME$10_+$\\
\hline
MP6 & 1 & $0$ & $0, \pm1$ & $121.2$ & $-5.488$ & $\cdots$ & $\cdots$ & $-5.603$
 & $0.3014$ & $\sim 0$ &  K K \\
MP7 & 1 & $0$ & $0$ & $27.08$ & $27.08$ & $\cdots$ & $\cdots$ & $ 1.272$ &
$0.2827$ & $\sim 0$ &  K K\\
MP8 & 1 & $0$ & $0, \pm1$ & $26.66$ & $-37.15$ & $\cdots$ & $\cdots$ & $0.8011$
& $0.2879$ & $\sim 0$ &  K K\\
MP9 & 1 & $0$ & $0, \pm1$ & $2.610$ & $-0.1187$ & $\cdots$ & $\cdots$ &
$3.756$ & $ -0.2032$ & $\sim 0$ &  K K\\
MP10 & 1 & $0, \pm1$ & $0, \pm1$ & $0.7376$ & $-0.086\,31$ & $\cdots$ & $\cdots$ &
$0.6240$ & $ -0.1237$ & $\sim 0$ &  K K\\
MP11 & 1 & $0, \pm1$ & $0, \pm1$ & $0.7268$ & $-0.1514$ & $\cdots$ & $\cdots$
& $0.4187$ & $-0.3221$ & $\sim 0$ &  K K\\
MP12 & 1 & $0, \pm1$ &  $0, \pm1$  & $0.1909$ & $0.1395$ & $\cdots$ & $\cdots$ &
$0.4564$ & $0.093\,75$ & $\sim 0$ &  K K \\
MP13 & 1 & $0, \pm1$ & $0, \pm1$  & $0.1548$ & $0.1548$ & $\cdots$ & $\cdots$ &
$0.4519$ & $0.1004$ & $\sim 0$ &  K K \\
 MP14 & 1 & $0, \pm1$ & $0, \pm1$ & $0.1201$ & $0.1699$ & $\cdots$ & $\cdots$ &
$0.4483$ & $0.1066$ & $\sim 0$ &  K K\\
MP15 & 1 & $0, \pm1$ & $0, \pm1$ & $-0.7576$ & $0.6250$ & $\cdots$ & $\cdots$
& $0.4486$ & $0.1961$ & $\sim 0$ &  K K\\
MP16 & 1 & $0, \pm1$ &  $0$  & $-1.162$ & $1.498$ & $\cdots$ & $\cdots$ &
$0.6537$ & $0.2399$ & $\sim 0$ &  K K \\
MP17 & 1 & $0, \pm1$ & $0, \pm1$  & $-2.408$ 9& $0.5986$ & $\cdots$ & $\cdots$ &
$ -0.1009$ & $0.1934$ & $\sim 0$ &  K K \\
  \hline
MP18 & 1 & $0$ & $1$  & $32.50$ & 1 & $\cdots$ & $0.045\,53$ & $8.000$ &
$0.2222$ & $\sim 0$ &  K M\\
MP19 & 1 & $0, \pm1$ & $-1$  & $-47.58$ & 1 & $\cdots$ & $0.076\,13$ & $ -9.795$
& $0.2222$ & $\sim 0$ &  K M\\
MP20 & 1 & 1 & $0$  & 1 & $31.77$ & $0.016\,75$ & $\cdots$ & $1$ & $0.2832$
& $\sim 0$ &  M K\\
MP21 & 1 & 1 & $0$ & 1 & $24.98$ & $0.010\,90$ & $\cdots$ & $1$ & $ 0.2825$
& $\sim 0$ &  M K\\
MP22 & 1 & $-1$ & $0, \pm1$  & 1 & $-0.8418$ & $0.8922$ & $\cdots$ & $1$ &
$0.4325$ & $\sim 0$ & M K\\
 \hline
MP23 & 1 & $\pm 1$  & $-1$ & 0 & 1 & $\cdots$ & 1 & $0.7778$ & $0.2222$  &
$\sim 0$ & $\mathrm{S}_\pm$ M\\
MP24 & 1 & $-1$  & $\pm 1$ & 1 & 0 & 1 & $\cdots$ & 1 & 0 & $\sim 0$ &
M $\mathrm{S}_\pm$ \\
MP25 & 1 & 1  & 1  & 1 & 1 & $0.2210$ & $0.3839$ & 1 & $0.2222$ & $\sim 0$ & C C\\
MP26 & 1 & $-1$  & $1$  & 1 & 1 & $0.069\,37$ & $0.3495$ & 1 & $0.2222$ &
$\sim 0$ & M C\\
MP27 & 1 & $-1$  & $-1$  & 1 & 1  & $1.296$ & $0.6584$ & 1 & $0.2222$ &
$\sim 0$ & M M\\
MP28 & 1 & $-1$  & $-1$  & 1 & 1 & $0.5507$ & $0.8095$ & 1 & $0.2222$ &
$\sim 0$ & M M\\
\noalign{\global\arrayrulewidth1pt}
\hline
\noalign{\global\arrayrulewidth.4pt}
\end{tabular}}
\end{center}
\end{table}

\begin{table}[htb]
\begin{center}
\caption{\small{Exact solutions for $\delta=-0.1$.
$\epsilon=0$ and 1 correspond
to generalized de Sitter solutions ($a\sim e^{\mu t}$, $b\sim e^{\nu t}$) and
power law ones ($a\sim \tau^{\mu}$, $b\sim \tau^{\nu}$), respectively.
$\lambda$ is a power exponent of power law solutions in the Einstein frame
($a_E \sim t_E^{\lambda}$).K, $\mathrm{S}_\pm$,
$\mathrm{S}_0$ and M mean a kinetic dominance, a static space with positive
(or negative) curvature, a flat static space, and a Milne-type space,
respectively.
ME3$_\pm$ and ME6$_+$ are inflationary solution in the Einstein frame, but
our 3-space is shrinking in ME3$_-$.
}}\label{sum7}
\vspace{2mm}
\small{
\begin{tabular}{lcccccccccc}
\noalign{\global\arrayrulewidth1pt}
\hline
\noalign{\global\arrayrulewidth.4pt}
Solution$\!\!\!$ & $\epsilon$ & $\sigma_p$ & $\sigma_q$ & $\mu$ & $\nu$ & $a_0$ & $b_0$
& $\lambda$ & $\phi_1$ & Type\\ \hline
ME$3_{\pm}$ & 0 & 0 & 0 & $\pm 0.1682$ & $\pm 0.1682$ & $\cdots$ & $\cdots$ &
$1.286$ & $\pm 0.2857$ & K K \\
ME$4_{\pm}$ & 0 & 0 & 0 & $\pm 0.7429$ & $\mp 0.4540$ & $\cdots$ & $\cdots$ &
$0.5326$ &  $\pm 0.2857$ & K K \\
ME$6_{\pm}$ & 0 & 0 & 1 & $\pm 0.3072$ & $0$ & $\cdots$ & $4.605$ &
$e^{\mu t}$ & $0$ & K $\mathrm{S}_+$\\
ME$9_{\pm}$ & 0 & 1 & 0 & 0 & $\pm 0.2011$ & $2.658$ & $\cdots$ & $1$
& $\pm 0.2857$ & $\mathrm{S}_+$ K \\
 ME$10_{\pm}$ & 0 & $-1$ & 0 & 0 & $\pm 0.3141$ & $1.149$ & $\cdots$ & $1$
& $\pm 0.2857$ & $\mathrm{S}_-$ K \\
 \hline
ME12 & 1 & 0     & $-1$ & 0 & 1 & $\cdots$ & 1 & $0.7778$ & $0.2222$ &
$\mathrm{S}_0$ M\\
ME13 & 1 & $-1$ & 0      & 1 & 0 & 1 & $\cdots$ & 1 & 0 &M $\mathrm{S}_0$ \\
\noalign{\global\arrayrulewidth1pt}
\hline
\noalign{\global\arrayrulewidth.4pt}
\end{tabular}}
\end{center}
\end{table}

\newcommand{\NP}[1]{Nucl.\ Phys.\ B\ {\bf #1}}
\newcommand{\PL}[1]{Phys.\ Lett.\ B\ {\bf #1}}
\newcommand{\CQG}[1]{Class.\ Quant.\ Grav.\ {\bf #1}}
\newcommand{\CMP}[1]{Comm.\ Math.\ Phys.\ {\bf #1}}
\newcommand{\IJMP}[1]{Int.\ Jour.\ Mod.\ Phys.\ A\ {\bf #1}}
\newcommand{\JHEP}[1]{JHEP\ {\bf #1}}
\newcommand{\PR}[1]{Phys.\ Rev.\ D\ {\bf #1}}
\newcommand{\PRL}[1]{Phys.\ Rev.\ Lett.\ {\bf #1}}
\newcommand{\PRE}[1]{Phys.\ Rep.\ {\bf #1}}
\newcommand{\PTP}[1]{Prog.\ Theor.\ Phys.\ {\bf #1}}
\newcommand{\PTPS}[1]{Prog.\ Theor.\ Phys.\ Suppl.\ {\bf #1}}
\newcommand{\MPL}[1]{Mod.\ Phys.\ Lett.\ {\bf #1}}
\newcommand{\JP}[1]{Jour.\ Phys.\ {\bf #1}}


\end{document}